\documentclass[journal]{rmaa}

\let\gtrsim\undefined
\let\lesssim\undefined
\usepackage{amssymb, graphicx,natbib,hyperref}
\usepackage{paralist}
\usepackage[latin1]{inputenc}
\usepackage{color}
\usepackage{enumerate}
\usepackage{booktabs, threeparttable}

\newcommand{\Reff}       {\mbox {${\rm R_{eff}}$}\,}

\newcommand{\nii}         {\mbox{\rm [N{\small II}]}}
\newcommand{\oiii}        {\mbox{\rm [O{\small III}]}}

\newcommand{\Vlos}          {\mbox{$V_{\rm los}$}}
\newcommand{\Vlosspp}          {\mbox{$V_{\rm los,ssp}$}}
\newcommand{\VlosHa}          {\mbox{$V_{\rm los,H\alpha}$}}
\newcommand{\slos}          {\mbox{$\sigma$}}
\newcommand{\slosSSP}          {\mbox{$\sigma_{\rm ssp}$}}
\newcommand{\slosGAS}          {\mbox{$\sigma_{\rm H\alpha}$}}
\newcommand{\slosGASc}          {\mbox{$\sigma^{corr}_{\rm H\alpha}$}}

\newcommand{\Sstar}       {\mbox{$\Sigma_*$}}
\newcommand{\Mstar}       {\mbox{$\mathrm{M_{\ast}}$}}
\newcommand{\Ssfr}        {\mbox{$\Sigma_{\rm SFR}$}}

\newcommand{\ha}          {\mbox{H$\alpha$}}
\newcommand{\hb}          {\mbox{H$\beta$}}

\newcommand{\msunperpcsqperyr} {\mbox{\rm M$_\odot$~pc$^{-2}$~yr$^{-1}$}}

\newcommand{\avssp}          {\mbox{${\rm Av_{SSP}}$\,}}
\newcommand{\avgas}          {\mbox{${\rm Av_{gas}}$\,}}
\newcommand{\Smol}        {\mbox{$\Sigma_{\rm mol}$}\,}
\newcommand{\SmolAv}        {\mbox{$\Sigma_{\rm mol,Av}$}\,}
\newcommand{\MstarMsun} {\mbox{$\log(\mathrm{M_{\ast}/ M_{\odot}})$}}
\newcommand{\lamssp}        {\mbox{$\lambda_{\rm ssp}$}}
\newcommand{\fgas}          {\mbox{f$_\mathrm{mol}$}}
\DeclareRobustCommand{\ion}[2]{%
\relax\ifmmode
\ifx\testbx\f@series
{\mathbf{#1\,\mathsc{#2}}}\else
{\mathrm{#1\,\mathsc{#2}}}\fi
\else\textup{#1\,{\mdseries\textsc{#2}}}%
\fi}

\title{SDSS-IV MaNGA: The Radial Distribution of Physical Properties within Galaxies in the Nearby Universe}

\author{
  J.~K.~Barrera-Ballesteros\altaffilmark{1},
 S.~F.~S\'anchez\altaffilmark{1},
 C.~Espinosa-Ponce\altaffilmark{1},
 C.~L\'opez-Cob\'a\altaffilmark{2},
 L.~Carigi\altaffilmark{1},
 A.Z.~Lugo-Aranda\altaffilmark{1},
 E.~Lacerda\altaffilmark{1},
 G.~Bruzual\altaffilmark{3},
 H.~Hernandez-Toledo\altaffilmark{1},
 N.~Boardman\altaffilmark{4},
 N.~Drory\altaffilmark{5},
 Richard~R.~Lane\altaffilmark{6}, and
 J.~R.~Brownstein\altaffilmark{7}
 }
\altaffiltext{1}{Universidad Nacional Aut\'onoma de M\'exico, Instituto de Astronom\'ia, AP 70-264, CDMX  04510, M\'exico}
\altaffiltext{2}{Institute of Astronomy and Astrophysics, Academia Sinica, No. 1, Section 4, Roosevelt Road, Taipei 10617, Taiwan}
\altaffiltext{3}{IRyA, Universidad Nacional Aut\'onoma de M\'exico, Campus Morelia, Michoacan, M\'exico C.P. 58089, M\'exico}
\altaffiltext{4}{School of Physics and Astronomy, University of St Andrews, North Haugh, St Andrews KY16 9SS, UK}
\altaffiltext{5}{University of Texas at Austin,McDonald Observatory, 1 University Station, Austin, TX 78712, USA}
\altaffiltext{6}{Centro de Investigaci\'on en Astronom\'ia, Universidad Bernardo O'Higgins, Avenida Viel 1497, Santiago, Chile}
\altaffiltext{7}{Department of Physics and Astronomy, University of Utah, 115 S. 1400 E., Salt Lake City, UT 84112, USA}

\shortauthor{Barrera-Ballesteros J.K., et al.}
\shorttitle{Radial profiles of MaNGA galaxies}

\abstract{Using the largest sample of galaxies observed with an optical integral field unit (IFU, the \mbox{SDSS-IV} MaNGA survey, \mbox{$\sim$10000} targets),  we derive the radial distribution of the physical properties obtained from the stellar continuum and the ionized-gas emission lines. Given the large sample, we are able to explore the impact of the total stellar mass and morphology by averaging those radial distributions for different bins of both global properties. We use a piece-wise analysis to characterize the slopes of the gradients from those properties at different galactocentric distances. In general we find that most of the properties -- derived from both the stellar continuum and the ionized gas emission lines -- exhibit a negative gradient with a secondary impact by global properties such as the total stellar mass or morphology. Our results confirm the intimate interplay between the properties of the stellar component and those of the ionized gas at local (kpc) scales in order to set the observed gradients. Furthermore, the resemblance of the gradients for similar global properties (in particular for the stellar parameters) indicates statistical similar histories of star formation and chemical enrichment with an initial radial gas distribution following the potential of the galaxy.}

\resumen{
Usando la muestra m\'as grande de galaxias con observaciones espacialmente resueltas (el cartografiado \mbox{SDSS-IV} MaNGA, \mbox{$\sim$10000} objetos) estimamos las distribuciones radiales de las propiedades fisicas obtenidas atraves del continuo y de las lineas de emision del espectro en el optico. El gran tama\~no de la muestra nos permite explorar el impacto que parametros estructurales como la masa estelar total y la morfologia tienen en la forma de dichas distribuciones radiales. Para caracterizar las pendientes de dichos perfiles radiales, usamos un analisis a trazos, el cual nos permite medir diferentes pendientes para distintas distancias galactocentricas. De manera general, encontramos que la mayoria de las propiedades -- derivadas para las componentes estelares y del gas ionizado -- muestran un gradiente cuya pendiente es negativa con un impacto secundario de la masa estelar y la morfologia. Nuestros resultados confirman la intima relacion entre las propiedades de la componente estelar con la componente del gas ionizado a escalas locales (kpc), las cuales fijan los perfiles radiales observados. Las similitud entre gradientes para galaxias con propiedades globales similares indican la similitud en las historias de formacion estelar y enriquecimiento quimico con una distribucion inicial de gas que sigue el potencial de la galaxia.
}

\addkeyword{Galaxies: Late-type galaxies}
\addkeyword{Galaxies: Early-type galaxies}
\addkeyword{Spectroscopy: Galaxies}

\begin{document}
\maketitle

\section{Introduction}
\label{sec:intro}

Projected in the sky, galaxies are spatially resolved objects. To truly assess the physical processes that drive the galaxy formation and evolution it is necessary to measure their observables in a spatially-resolved fashion. One of the most well-known techniques to estimate structural properties of a galaxy is through the measurement and fitting of its surface brightness radial profile \citep[e.g.,][]{de_Vaucouleurs_1958, Sersic_1968,Freeman_1970, Kormendy_1977, Kent_1985}. In general, the surface brightness (from broad-band photometry) decreases with radius, indicating that the amount of stars is larger at central regions of galaxies. 

Similarly, there are studies exploring the radial profiles of the \ha\ emission line using narrow-band filters \citep[e.g., ][]{Martin_Kennicutt_2001, Bigiel_2008}. Although, the \ha\ emission usually is small in the outskirts in comparison to central regions, these studies showed that not for all galactocentric distances the \ha\ emission monotonically decreases with radius. Long-slit spectroscopy has also been a powerful tool to measure the radial distributions of physical properties from both the stellar continuum and the ionized gas emission lines \citep[e.g., ][]{Pagel_Edmunds_1981, Peletier_1989}. Through the analysis of different features of the stellar absorption lines and emission lines, such studies have provided a new way to estimate physical properties from each component at different galactocentric distances (e.g., stellar ages, and metallicities; star-formation rates, oxygen abundances, among many others). 

Despite these significant efforts, those studies have been performed in samples of galaxies targeted to explore a specific scientific goal. Furthermore, the above techniques do not fully capture the angular distribution of the observables from both components. These issues has been successfully overcome thanks to large samples of galaxies observed using the Integral Field Spectroscopy (IFS) technique. This observational technique allows to obtain spectra for different positions of a galaxy across its optical extension. Thus, for each of the physical parameter derived from spectroscopy, it is possible to estimate a two-dimensional distribution. Different collaborations have embarked in acquired IFS datacubes for large samples of galaxies \citep[e.g., ][ CALIFA, and SAMI]{Sanchez_2012, Croom_2012}. Of particular interest is the MaNGA survey included in the SDSS-IV collaboration \citep{Bundy_2015}. This collaboration has recently achieved its goal to observe a sample of 10000 galaxies using an integral field unit (IFU). Given the large sample probed by these surveys, they have unveiled the radial distribution of physical properties for a wide demographic range of galaxies in the local universe from both the stellar and ionized gas components. In particular, these surveys allows us to quantify the impact that global and structural parameters such as the total stellar mass or the morphology have in modulating the slopes and absolute values of those radial profiles \citep[see][ and references therein]{Sanchez_2020}.

In this study we explore the radial distribution of physical parameters from the stellar and ionized gas component using the entire MaNGA dataset. Our goal is to quantify the impact that global parameter have in shaping the radial profiles of those properties. Furthermore, to give a more accurate description than a single gradient for those profiles will provide, we make use of a piece-wise analysis which allows us to measure different slopes and breaks for a given radial distribution of a physical property. In Sec.~\ref{sec:data} we provide a brief description of the MaNGA sample, the IFU datacubes, and the analysis pipeline used to estimate the map for a given observable. In this section we describe the criteria to select a set of the closest galaxies with the best spatial coverage and physical spatial resolution from which we will derive their radial profiles (known as {\it Golden Sample}). We also describe the piece-wise analysis to measure the slopes of those profiles. In Secs.~\ref{sec:stellar} and \ref{sec:gas} we present the piece-wise analysis of the radial distribution of the properties derived from the stellar and ionized gas components, respectively. In Sec.~\ref{sec:kin_prop} we present the radial distribution of the line-of-sight velocity and velocity dispersion from both components, as well as comparisons between them. In Sec.~\ref{sec:ALL} we compare the results from Secs.~\ref{sec:stellar} and \ref{sec:gas} with those using the entire sample of MaNGA galaxies (excluding only high-inclined targets, $\sim$ 7000 galaxies). Finally, in  Sec.~\ref{sec:DisConc} we summarize our main results and conclusions.  

\section{Sample, Datacubes, and Analysis}
\label{sec:data}
\begin{figure}
\includegraphics[width=\linewidth]{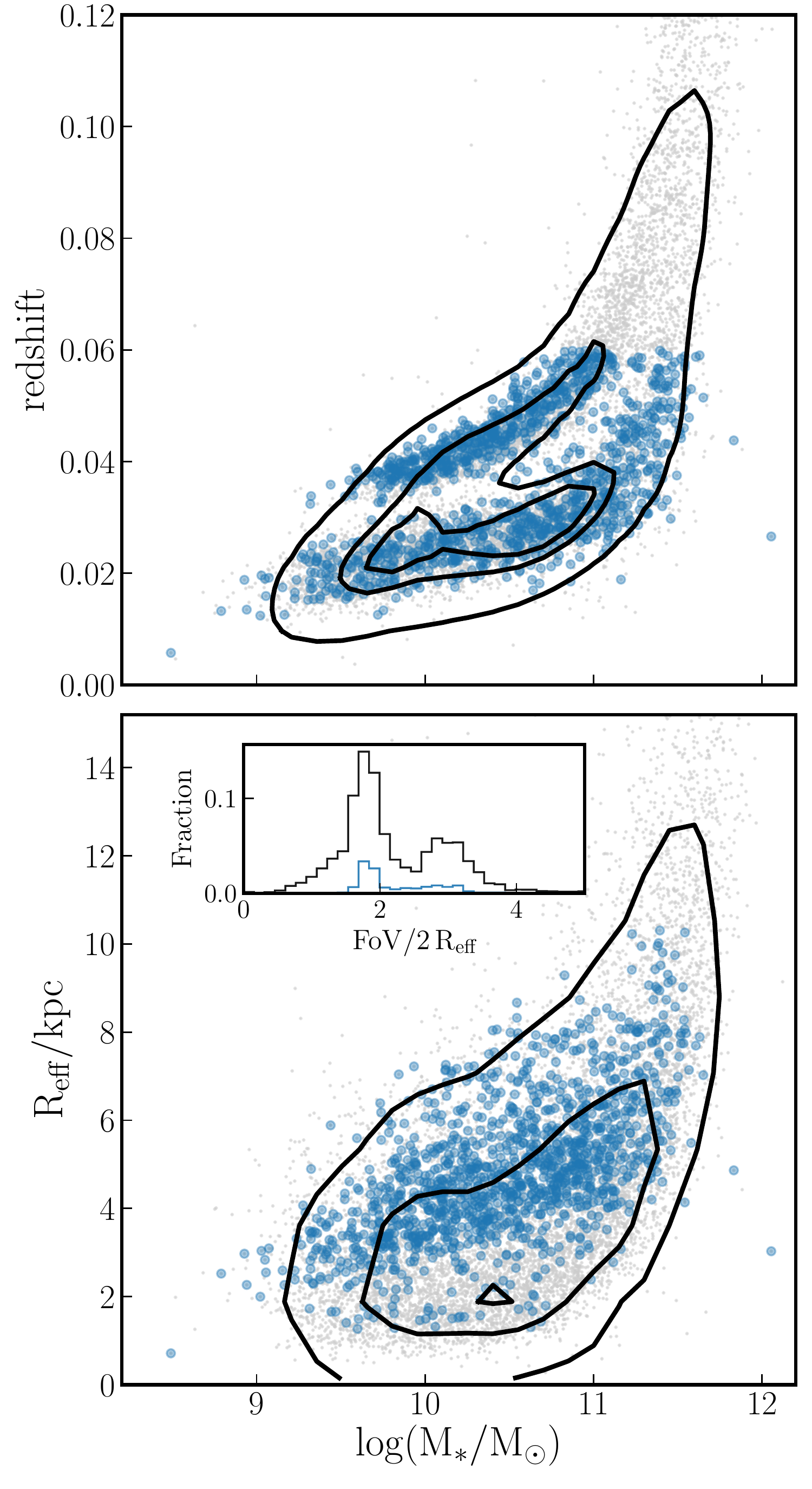}
\caption{Comparing the Golden Sample (GS) with its parent, the MaNGA sample according to their distance (redshift), size (\Reff) and stellar mass, \Mstar (top and bottom panels, respectively). The gray points represent the entire MaNGA survey, while the contours enclose 90\%, 80\%, and 50\% of this sample. The blue circles represent the GS. The inset in the bottom panel shows the distribution of the radial coverage of both samples. The spatial-resolution criteria required to drawn the GS reduces its size to $\sim$10\% the entire MaNGA survey (see details in Sec.~\ref{sec:sample}).}  
\label{fig:GS1}    
\end{figure}
\begin{figure}
\includegraphics[width=\linewidth]{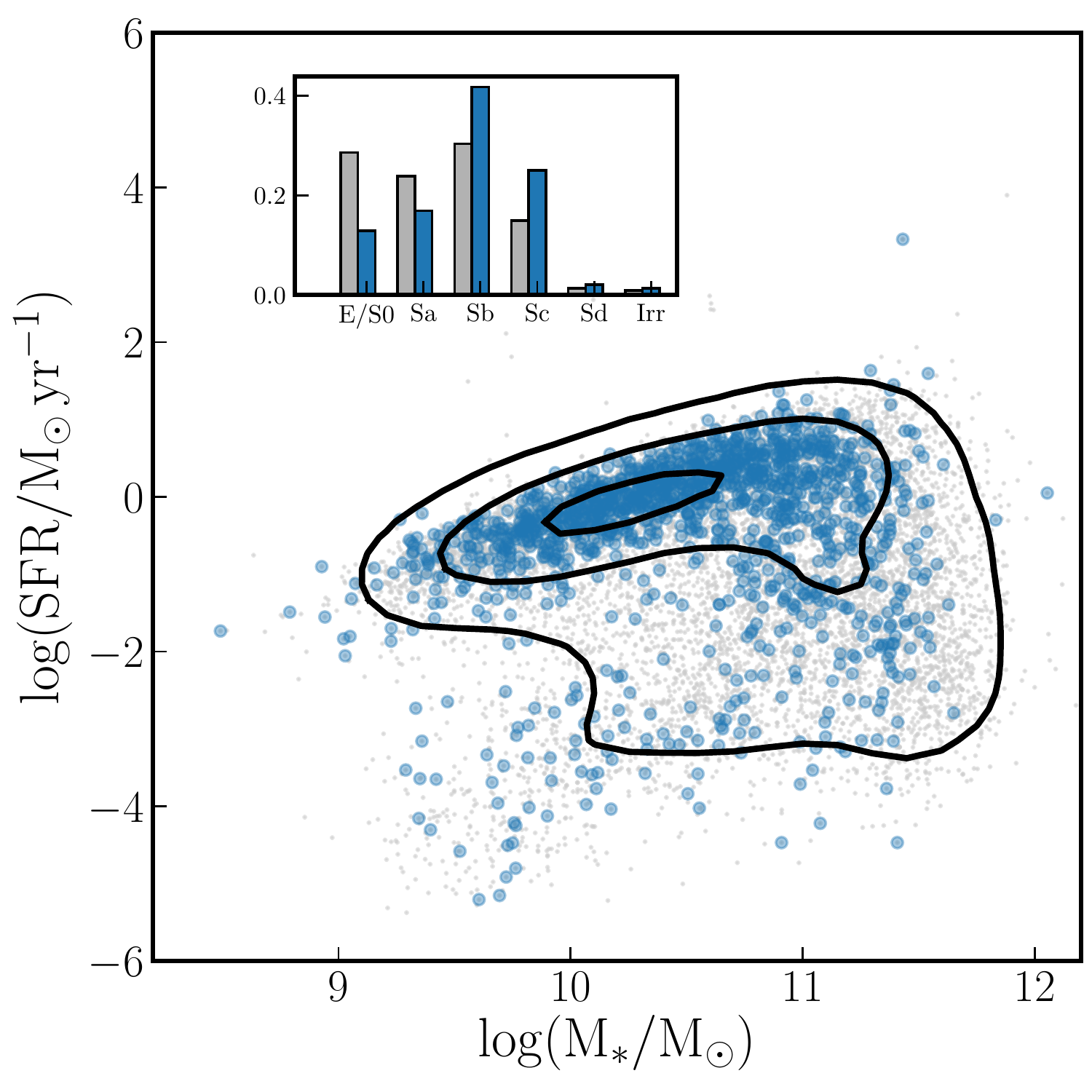}
\caption{Comparison of the Golden Sample and the entire MaNGA sample in the SFR-\Mstar\ plane. As in the previous figure, the blue points represent the golden sample whereas the gray points and the contours represent the MaNGA sample. The blue bars in the histogram in the inset shows the number of galaxies for a given morphology for the golden sample while the gray bars show the morphological distribution for the entire sample. Given our selection criteria from the golden sample it includes a significant fraction of star-forming Sb galaxies in comparison to the MaNGA sample.}  
\label{fig:GS2}    
\end{figure}
\subsection{The MaNGA survey}
\label{sec:manga}

The MaNGA survey \citep{Bundy_2015} was part of the forth generation of surveys included in the Sloan Digital Sky Survey \citep[SDSS-IV, ][]{Blanton_2017}. The goal of this spectroscopic survey was to obtain datacubes in the optical for more than 10000 galaxies via IFU observations. The final sample includes galaxies observed from March of 2014 to September of 2020 (10245 unique datacubes). This sample is  publicly available in the final data release of the SDSS-IV surveys \citep{Abdurrouf_2021_arXiv}. For this study we use the latest MaNGA data release (v3.1.1). In this section we provide a brief description of the most relevant features of this survey. 

Observations for the MaNGA survey took place at the Apache Point Observatory using its 2.5-m telescope \citep{Gunn_2006}. This survey used two spectrographs from the BOSS survey \citep[Baryon Oscillation Spectroscopic Survey][]{Smee_2013} . These spectrographs achieve a nominal spectral resolution of \mbox{$R \equiv \lambda/\Delta\lambda \sim 1900$} covering a large portion of the optical spectra (from 3000 to 10000 \AA). These spectrographs were fed by joined fibers in bundles distributed in a hexagon-like array. The number of fibers in a given bundle varies from 19 to 127. Given that the diameter of each fiber is $\sim$ 2.5\arcsec, thus the Field-of-View (FoV) varies from 12\arcsec to 32\arcsec. A detailed description of the instrumentation of the survey can be found in \citet{Drory_2015}. The reader is refereed to \citet{Law_2016} for a detailed explanation of the data strategy (acquisition, reduction, etc). The MaNGA reduction pipeline includes wavelength calibration, corrections from fiber-to-fiber transmission, subtraction of the sky spectrum and flux calibration \citep{Yan_2016}. The final product is a datacube with $x$ and $y$ coordinates corresponding to the sky coordinates and the $z$-axis corresponds to the wavelength. As result for each datacube, the spaxel size is 0.5\arcsec\ with a spatial resolution of 2.5\arcsec\ corresponding to a mean physical scale of $\sim$ 2 kpc.

In Fig~\ref{fig:GS1} the contours show the distribution of the entire MaNGA sample in three fundamental parameters, their redshift ($z$), size (\Reff), and total stellar mass (\Mstar)\footnote{A full description on the integrated properties derived from the IFU datacubes for the entire MaNGA survey is presented in Sanchez et al. (submitted).}. From the top panel of Fig~\ref{fig:GS1} it is evident that this sample covers both a wide range in redshift and \Mstar (\mbox{$0.01 < z < 0.14$} and \mbox{$9.0 < \log(\mathrm{\Mstar/M_{\odot}}) < 11.5$}, respectively). A wide coverage of redshift implies a large dynamical range in the physical spatial resolution of the survey. It is also clear the strong correlation between these to parameters: massive galaxies are located further away from us than low-mass galaxies. Furthermore, the MaNGA sample shows a bimodality in this plane. This is a result of the intrinsic selection criteria of the survey where $\sim$ 2/3 of the sample is selected in a way that for each galaxy the fiber bundle covers at least 1.5 \Reff, for the remaining fraction of the sample, the bundle covers $\sim$ 2.5 \Reff\ for each galaxy. Thus, to satisfy this criteria, for a given range of \Mstar\ the sample have a fraction of galaxies closer and another further way from us. The bottom panel of Fig~\ref{fig:GS1} shows an expected behavior of \Reff\ with \Mstar: massive galaxies are bigger in comparison to low-mass ones. The distribution of the FoV weighted by \Reff shows the bimodality presented in the $z$-\Mstar\ plane (see inset in bottom panel in Fig~\ref{fig:GS1}). The reader is addressed to \citep{Wake_2017} for a detailed description of the selection criteria for the MaNGA survey.

The statistical strength of this survey allows to explore not only the spatially resolved properties of individual galaxies but also its integrated properties. In Fig.\ref{fig:GS2} we show the integrated star formation rate (SFR) against the \Mstar\ for the entire MaNGA sample (black contours and gray points). The distribution of this sample in the SFR-\Mstar\ plane shows the bimodality observed in larger samples of galaxies using integrated properties. We note that a significant fraction of galaxies are located in the so-called 'Star-Formation Main Sequence'  \citep[SFMS, e.g., ][]{Brinchmann_2004, Renzini_2015}, this is  galaxies that actively form new stars. The remaining fraction of the sample occupy the so-called 'Retired-Galaxies Sequence' (RGS), galaxies with little or none formation of new stars, and the so-called green valley, this is, galaxies located in the middle of the SFMS and the RGS in the  SFR-\Mstar\ plane. Finally when we study the morphological distribution of the MaNGA survey we find that this sample covers a wide range of morphological types from early to late types (see inset in Fig.\ref{fig:GS2}).  In a dedicated article on the integrated properties of the MaNGA survey, Sanchez et al. (submitted) present a detailed description and analysis of the above properties. 

\subsection{Sample Selection: The Golden Sample}
\label{sec:sample}

The main goal of this work is to explore the radial distribution of the properties of galaxies using  the MaNGA survey. Therefore, we require to have a sample in which the fiber bundle of the MaNGA instrumentation provides a good spatial coverage for each of the galaxies in the sample \citep{Ibarra-Medel_2019}. From the MaNGA sample we select a sub-sample of galaxies according to the following criteria: ($i$) galaxies observed with the largest fiber bundles (i.e., 91 and 127 fibers); ($ii$) the diameter of the galaxy (measure by 2 \Reff) has to be larger than 2 times the spatial resolution ($\sim$5\arcsec);  ($iii$) the diameter of the FoV has to cover at least 1.7 times the diameter of the galaxy (i.e., 3.4 \Reff); ($iv$) the major/minor axis ratio has to be smaller than 0.45; and ($v$) the redshift of the galaxy has to be in the range \mbox{$0.005 < z < 0.06$}. The above selection criteria ensure that we select galaxies with a reliable coverage of the galaxy as well as a significant independent data within each galaxy. In particular, the selection criteria ($v$) ensures that within this sample we are considering a consistent physical spatial resolution and that the evolution of the galaxies is similar.  This criteria yields a final sample of 1347 galaxies, representing $\sim$ 13\% of the total MaNGA sample. We are referring to this sample in this article as the {\it Golden Sample (GS)}. 

We overplot the GS in Figs.~\ref{fig:GS1} and \ref{fig:GS2} (blue circles). We note that in general the GS follows similar trends as those observed by the entire MaNGA sample. In Fig.~\ref{fig:GS1}, the GS shows a bimodality in the redshift-\Mstar plane and an increment of \Reff\ as \Mstar\ increases. It is also clear the cut on redshift from the GS due to our selection criteria. In the inset of the bottom panel of Fig.~\ref{fig:GS1} we compare the spatial coverage of the GS with respect to its parent sample. The GS follows the distribution of the MaNGA sample, however given our constrains to have both a reliable spatial coverage and good spatial sampling its size is heavily reduced in comparison to the total sample of the MaNGA survey. In Sec.~\ref{sec:ALL} we explore the impact of deriving the azimuthal-averaged radial distributions using the entire low-inclined MaNGA sample.

\subsection{\textsc{pyPipe3D:} The data-analysis pipeline}
\label{sec:pipe3d}

To analyze the large amount of datacubes provided from the MaNGA survey we use the \textsc{pyPipe3D} analysis pipeline \citep{Lacerda_2022arXiv}. This is an update in python language of the \textsc{Fit3D} and \textsc{Pipe3D} software and analysis pipeline \citep{Sanchez_2015, Sanchez_2016} with significant improvements. Here we highlight the main features of this pipeline, while in Sanchez et al. (submitted) we provide a detailed description of the use of this pipeline in the entire MaNGA dataset.

In a nutshell, the \textsc{pyPipe3D} analysis pipeline disentangle the contributions of the stars and the ionized gas emission in the observed spectrum for each spaxel for each datacube. To obtain the contribution from the stellar component, the pipeline provides the best fit of the continuum using a linear combination of single-stellar populations (SSPs) spectra. To account for the Line-of-sight Velocity Distribution function (LOSVD) first and second moments (i.e., the systemic stellar velocity, \Vlos, and the stellar velocity dispersion, \slos) the set of SSPs is shifted and convolved with a Gaussian function as well as dust attenuated adopting a Cardelli extinction curve \citep{Cardelli_1989}. Details on how \textsc{pyPipe3D} estimates \Vlos\ and \slos\ are described in \citep{Sanchez_2015, Sanchez_2016}. The decomposition of the stellar continuum in different SSPs for a spectrum provides fundamental parameters of the stellar component, thus for each datacube \textsc{pyPipe3D} provides two-dimensional distributions (or maps) of the derived properties as well as their uncertainties. Among the many properties derived from the stellar continuum (see Sec.5 in Sanchez et al., submitted), we use in this study the maps of the following properties: the mass-to-light ratio (M/L), stellar mass density (\Sstar), the luminosity-weighted stellar age and metallicity (Age, [Z/H]), the dust attenuation (\avssp), and the systemic stellar velocity (\Vlos) and velocity dispersion (\slos). Once the stellar continuum has been modeled by the SSPs for each datacube, it is removed from the observed datacube. This continuum-free datacube is thus used to estimated the properties from the emission lines. \textsc{pyPipe3D} pipeline provides a moment analysis to derive the physical properties for a large set of emission lines. For each emission line, the pipeline provides the map for its integrated flux, its systemic velocity, its velocity dispersion, and its equivalent width. We use these properties to build the radial profiles presented in this study for both the stellar and the ionized gas components.  

\subsection{Radial Profiles and Gradients}
\label{sec:grad_profiles}
\begin{figure*}
\includegraphics[width=\linewidth]{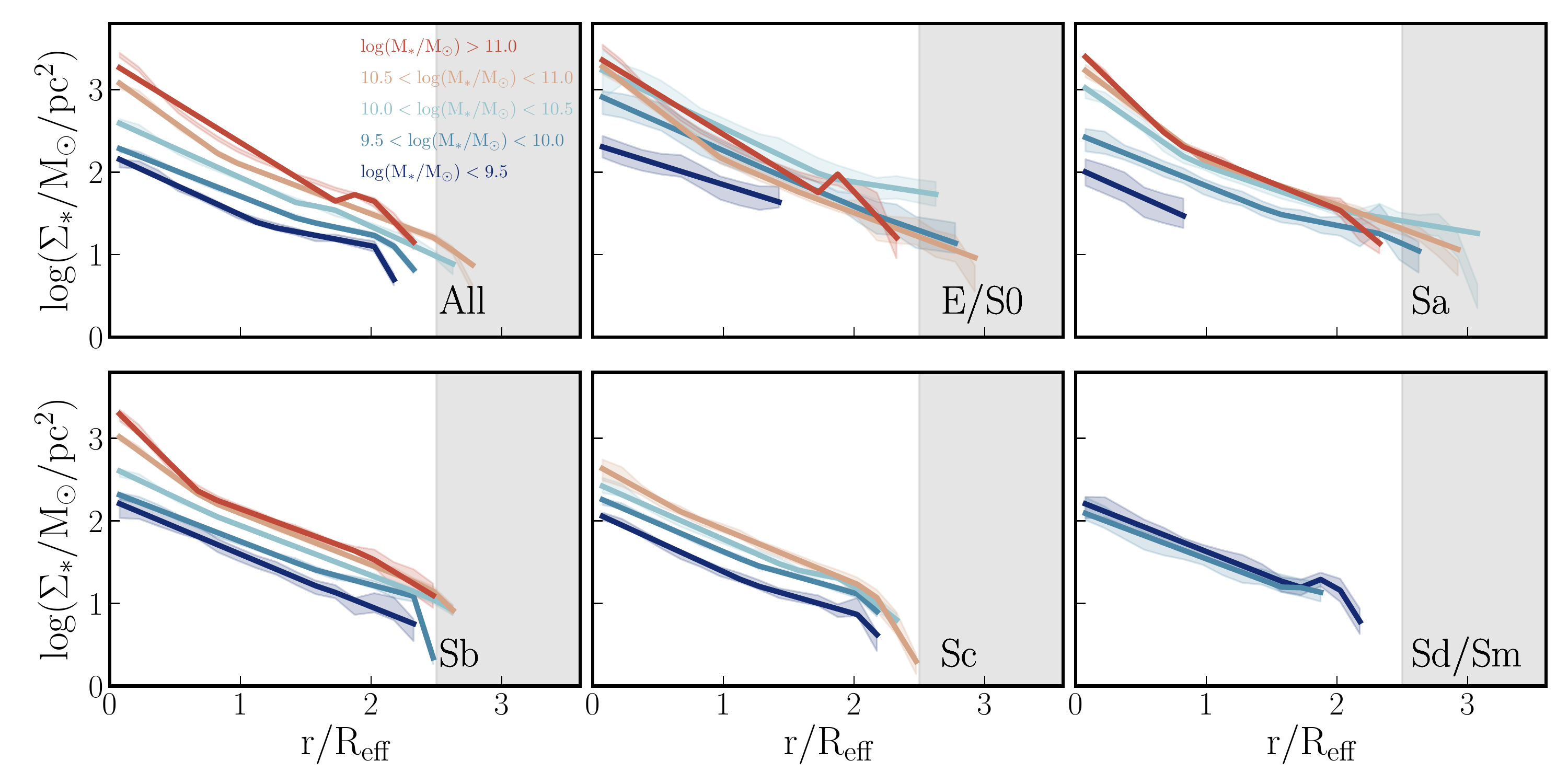}
\caption{The average radial distribution of the stellar mass density, \Sstar, as an example of the radial distribution of the physical properties derived for the GS of the MaNGA sample. The gradients are averaged by morphology (panels from left to right and top to bottom) and total stellar mass bins (shaded colored areas; from blue to red, each color represents a stellar mass bin see the legend at the top-left panel).The gray shaded area in each panel represent the maximum radius covered by the FoV (\mbox{\Reff $\sim$ 2.5}). The solid lines in each panel represent the best fit derived from fitting a piecewise function to the radial distribution (see details in Sec.~\ref{sec:grad_profiles}).} 
\label{fig:Sstar_rad}    
\end{figure*}

From the two-dimensional map of each galaxy's parameter derived by \textsc{pyPipe3D} we build its azimuthal-average radial profile. This procedure is performed for the entire MaNGA survey. For each map in each galaxy we make radial bins of 0.15 \Reff\ width up to $\sim$3 \Reff. Projected in the sky each of these bins is an elliptical annuli centered in the optical (V-band) brightest region with an ellipticity and position angle drawn from the NSA survey \citep{Blanton_2011}. In Fig.~\ref{fig:Sstar_rad} we plot as example the radial distribution of the stellar mass surface density, \Sstar, for the MaNGA GS. Instead of plotting the radial gradients for individual galaxies, for each morphological type (each panel in Fig.~\ref{fig:Sstar_rad}) we plot the median radial distribution for different bins of total stellar mass (each shaded area colored in Fig.~\ref{fig:Sstar_rad}). The borders of those shaded areas show the 1-$\sigma$ distribution for each galactocentric distance. To ensure a reliable estimation of the radial gradient we select those radial bins with good signal-to-noise ratio in the continuum (\mbox{SNR $>$ 3}) for stellar-derived properties, whereas for the emission-line-derived measurements, besides using a SNR cut in the continuum we also impose a SNR cut in the \ha\, line (\mbox{SNR $>$ 2}). Furthermore, when performing the median value for each morphology - stellar mass bin we exclude those radial bins where there are less than ten measurements.

As in other studies exploring the radial profiles of galaxy's properties \citep[e.g.,][]{Sanchez-Menguiano_2018}, we note that a single-slope gradient for a large fraction of radial distributions is not accurate to describe them. In other words, to provide a proper description on how the physical properties of galaxies changes at different galactocentric distances, it is required to use a procedure that fits more than one gradient. To account for these different variations in slopes we fit those radial distributions with a two piece-wise function. We fit the following functional form for a given radial profile $P(r)$:
\begin{equation}
P(r) = \begin{cases} 
      P_0 + k_0 (r/\Reff)   r/\Reff \leq r_0 \\
      P_{\rm ini}+ k_1 (r - r_0 /\Reff)  r_0 \leq r/\Reff \leq r_1 \\
      P_{\rm med}+ k_2 (r- r_1  /\Reff)  r_1 \leq r/\Reff \\
   \label{eq:pw}
   \end{cases}
\end{equation}

where \mbox{$ P_{\rm ini} = P_0 + k_0 (r_0/\Reff)$} and $P_{\rm med} = P_{\rm ini} + k_1 (r_1- r_0 /\Reff)$. The solid lines in Fig.~\ref{fig:Sstar_rad} represent the best fit of Eq.~\ref{eq:pw} for each radial distribution of \Sstar. We constrain this fit to $r/\Reff < 2.5$. It is evident that this functional form provides an accurate representation of the radial distribution for each morphology and stellar mass bin. In Appendix~\ref{app:profiles_grads} we show the radial distribution as well as the best-fit gradients using Eq.~\ref{eq:pw} for the set of parameters explored using the MaNGA GS. 

In the next sections (Secs.~\ref{sec:stellar} and \ref{sec:gas}), we present the best-fit parameters using Eq.~\ref{eq:pw} (i.e., the slopes of each line: $k_0$, $k_1$, $k_2$, and the break radius: $r_0$, $r_1$), the differences between these slopes ($\Delta$ slope), and the value of each parameter at \Reff. To avoid spurious results, in the following analysis we exclude those slopes derived where the range in galactocentric distance is smaller than $r/\Reff < 0.3$ (i.e., \mbox{$r_0 < 0.3$}, and \mbox{$r_0 - r_1 < 0.3$}). This criteria exclude those slopes where the radial range is comparable to the size of the MaNGA PSF. We show for each of these parameters the variation with morphology for the different stellar mass bins (see as example Fig.~\ref{fig:ML}) 

%
\section{Stellar Properties}
\label{sec:stellar}
\subsection{The $M/L$ ratio, and Stellar Mass Surface Density, \Sstar}
\label{sec:Sstar}
\begin{figure}
\includegraphics[width=\linewidth]{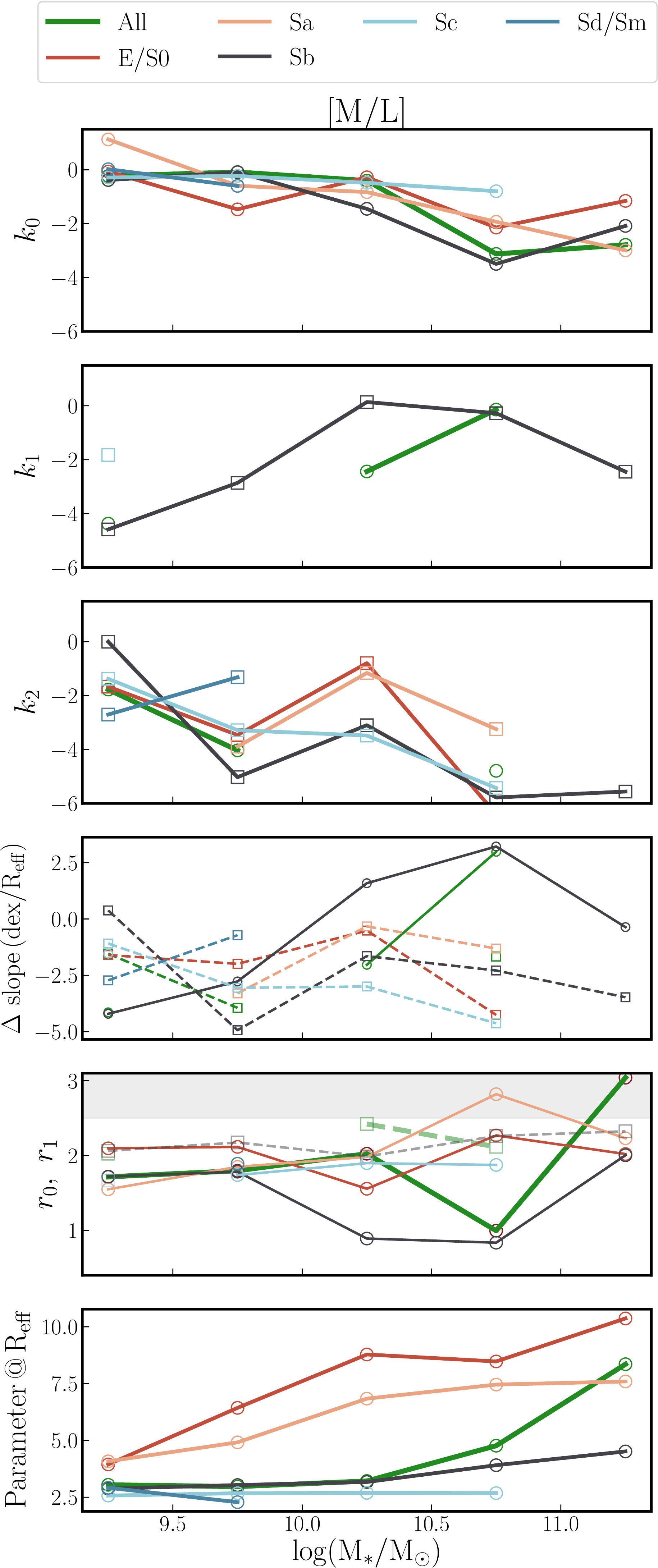}
\caption{Parameters derived from the radial distribution of the M/L ratio for different bins of total stellar mass. The color code of each symbol and line represents a morphological type (see legend). From top to bottom each panel shows: $k_0$, $k_1$, and $k_2$ from Eq.~\ref{eq:pw}; the differences between those slopes (solid and dashed lines represent the  $k_1$ - $k_0$, and $k_2$ - $k_0$ differences, respectively); $r_0$, and $r_1$ from Eq.~\ref{eq:pw} (solid and dashed lines, respectively); and the value of M/L at \Reff.}  
\label{fig:ML}    
\end{figure}

\begin{figure}
\includegraphics[width=\linewidth]{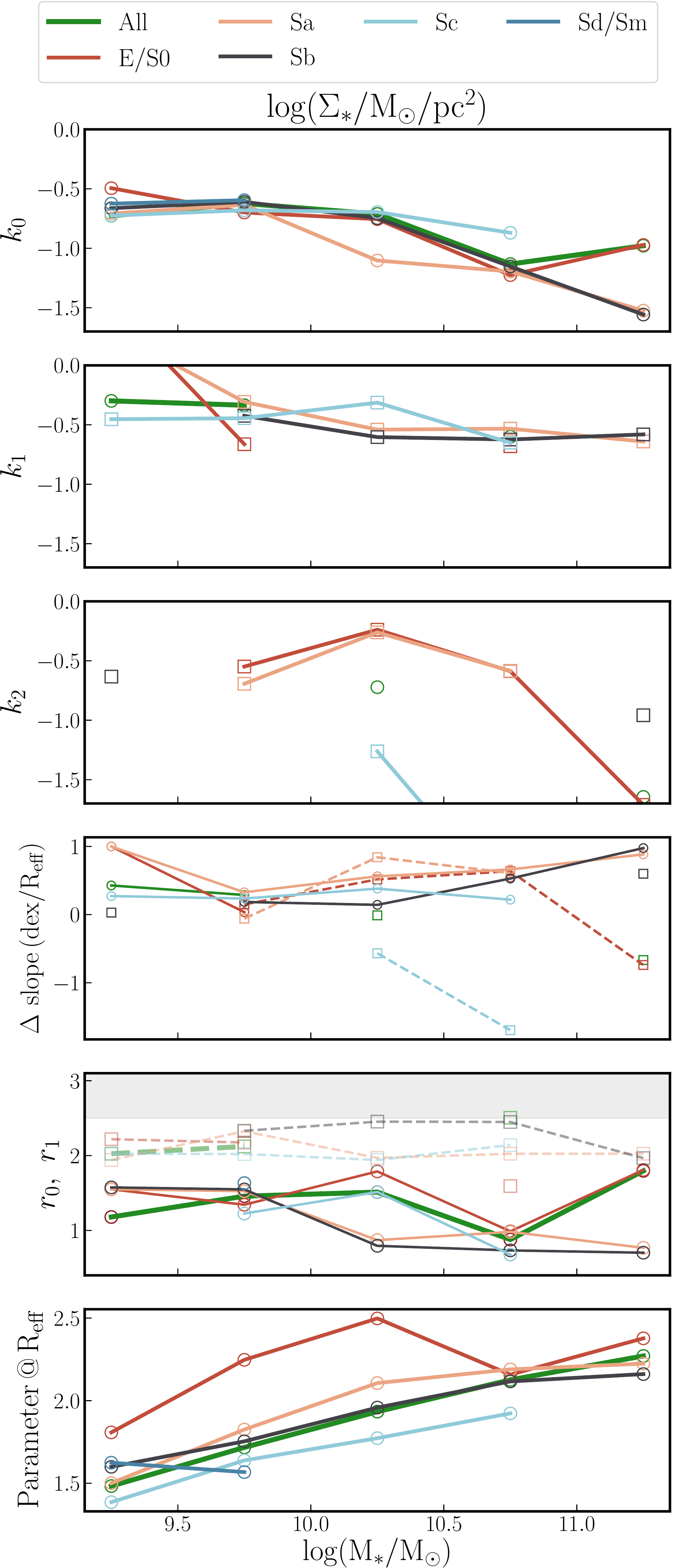}
\caption{Similar to Fig.~\ref{fig:ML}. Parameters derived from the piece-wise analysis for the stellar mass surface density gradients.}  
\label{fig:Sstar}    
\end{figure}
As we mention above, in general galaxies show in the optical a decreasing radial distribution of their surface brightness \citep[e.g.,][]{de_Vaucouleurs_1958,Sersic_1968, Freeman_1970, Kormendy_1977}. These surface brightness profiles are correlated mainly with the amount of stars that produces the observed flux via the mass-to-light ratio, M/L \citep[e.g.,][]{Portinari_Salucci_2010}. In Fig.~\ref{fig:ML_rad} we plot the median values of the M/L ratio for different morphologies and stellar masses, whereas in  Fig.~\ref{fig:ML} we summarize the best-fit parameters derived from the piece-wise analysis. From this analysis we find that the central slope, $k_0$ has mostly negative values and  decreases with the total stellar mass -- with little variation with respect to the morphology. Except for the Sb type, our analysis is not able to detect a significant value of $k_1$. For Sb galaxies the best-fit values of $k_1$ are negative and increase with \Mstar (except for most massive galaxies). For the outermost parts of the galaxies, the gradient of the M/L ratio, $k_2$, is also negative with a significant negative trend with \Mstar\ than $k_0$. These differences are highlighted in the forth panel of Fig.~\ref{fig:ML}, suggesting a sharper drop of the M/L ratio at the outskirts of galaxies. The fifth panel of  Fig.~\ref{fig:ML} indicates that those breaks of the gradients occur between $\sim$ 1.7 and 2.0 \Reff. Finally, we note that for a given stellar mass bin, the M/L ratio at \Reff changes significantly for different morphologies. In average, late type galaxies present mild variations with \Mstar\ whereas early-type galaxies show significant variations of this ratio. This highlights the combined role of \Mstar\ and the morphology in shaping the radial distribution of the M/L ratio. These results reflect the different amount of old stars vs young stars for the different morphological types. Although less bright, old low-mass stars are the main contributor to the stellar mass in galaxies, at least for the mass and morphological ranges explored in this study.
%

One of the more significant parameters that represents a galaxy is its total stellar mass. Similarly, at spatially resolved scales, its local analog, the stellar mass surface density, \Sstar,  has been observed to be a fundamental parameter to understand the angular distributions of different observables \citep[e.g.,][]{Cano-Diaz_2016, Gonzalez_Delgado_2015, Sanchez_2020}. In Fig.~\ref{fig:Sstar_rad} we show the average radial distributions of \Sstar\, for different masses and morphologies as well as the gradients from the piece-wise analysis. In Fig.\ref{fig:Sstar} we show the best-fit parameters of that analysis. We find similar distributions as those reported previously in the literature \citep[e.g.,][]{Gonzalez_Delgado_2015, Sanchez_2020}. In general, regardless the morphological type and the total stellar mass, \Sstar\, decreases with the galactocentric distance with relatively similar slopes. The piece-wise analysis reveals that the slope at small galactocentric distances for \Sstar\ decreases from $ k_0 \sim -0.5$ to $k_0 \sim -1.5$ dex/\Reff, as \Mstar\, increases. We do not find a significant impact of morphology in setting the $k_0$ slope. The $k_1$ slope is rather stable for different masses and morphologies, although we note that it is not derived for low-mass or massive early-type galaxies, suggesting that a single slope is required to estimate the gradient of these galaxies. Except for early-type galaxies, the analysis does not derive $k_2$ slopes, suggesting that for most bins of the \Mstar-morphology parameter space only single or double slope is required to describe the \Sstar\, radial distribution. The difference between slopes suggest that in general the external gradient has a slightly more positive gradient than the inner one (i.e., $k_1 > k_0$). The break where the change of slope occurs is $r_0 \sim$ 0.5 to 2.0 \Reff, depending on the stellar mass and morphology. For those radials distributions with estimations of  $k_2$, $r_1$ varies between 2.0 to 2.5 \Reff. The characteristic value of \Sstar\, at \Reff increases with \Mstar\, with little effect from the morphology (except for early-type E/S0 galaxies where the characteristic value of \Sstar\ is larger in comparison to other morphologies). 

\subsection{Luminosity Weighted Age, Metallicity, and Extinction}
\label{sec:Age_Z_Av}
\begin{figure}
\includegraphics[width=\linewidth]{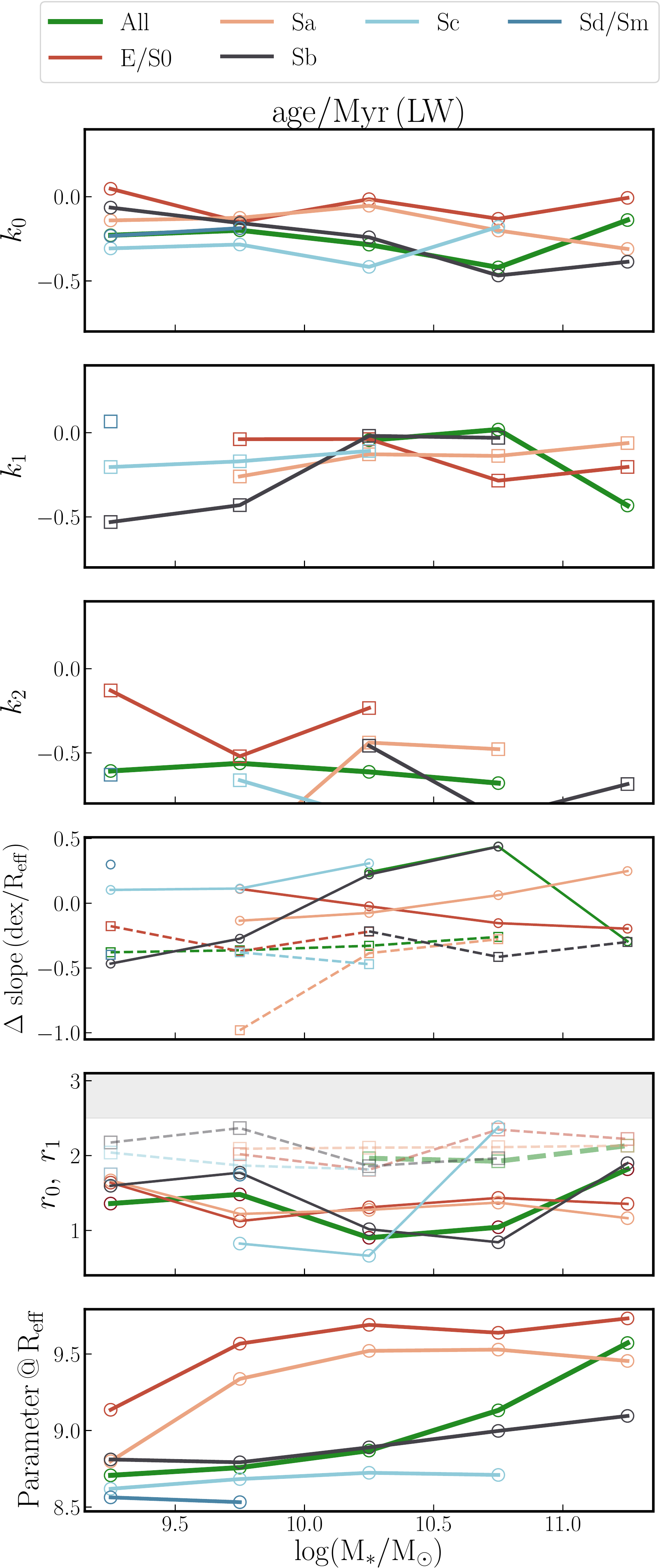}
\caption{Similar to Fig.~\ref{fig:ML}. Parameters derived from the piece-wise analysis for the luminosity-weighted age of the stellar population.}  
\label{fig:age}    
\end{figure}
\begin{figure}
\includegraphics[width=\linewidth]{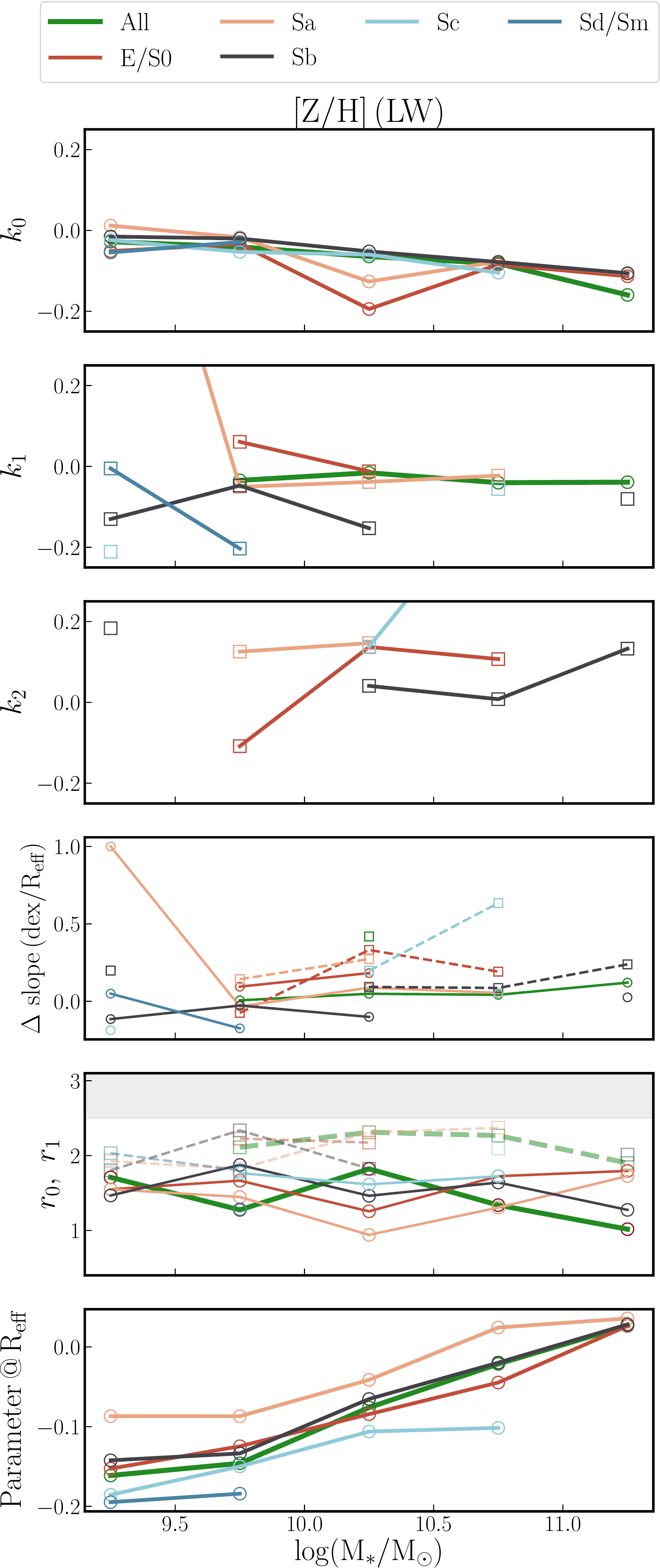}
\caption{Similar to Fig.~\ref{fig:ML}. Parameters derived from the piece-wise analysis for the luminosity-weighted stellar metallicity.}  
\label{fig:ZH}    
\end{figure}
\begin{figure}
\includegraphics[width=\linewidth]{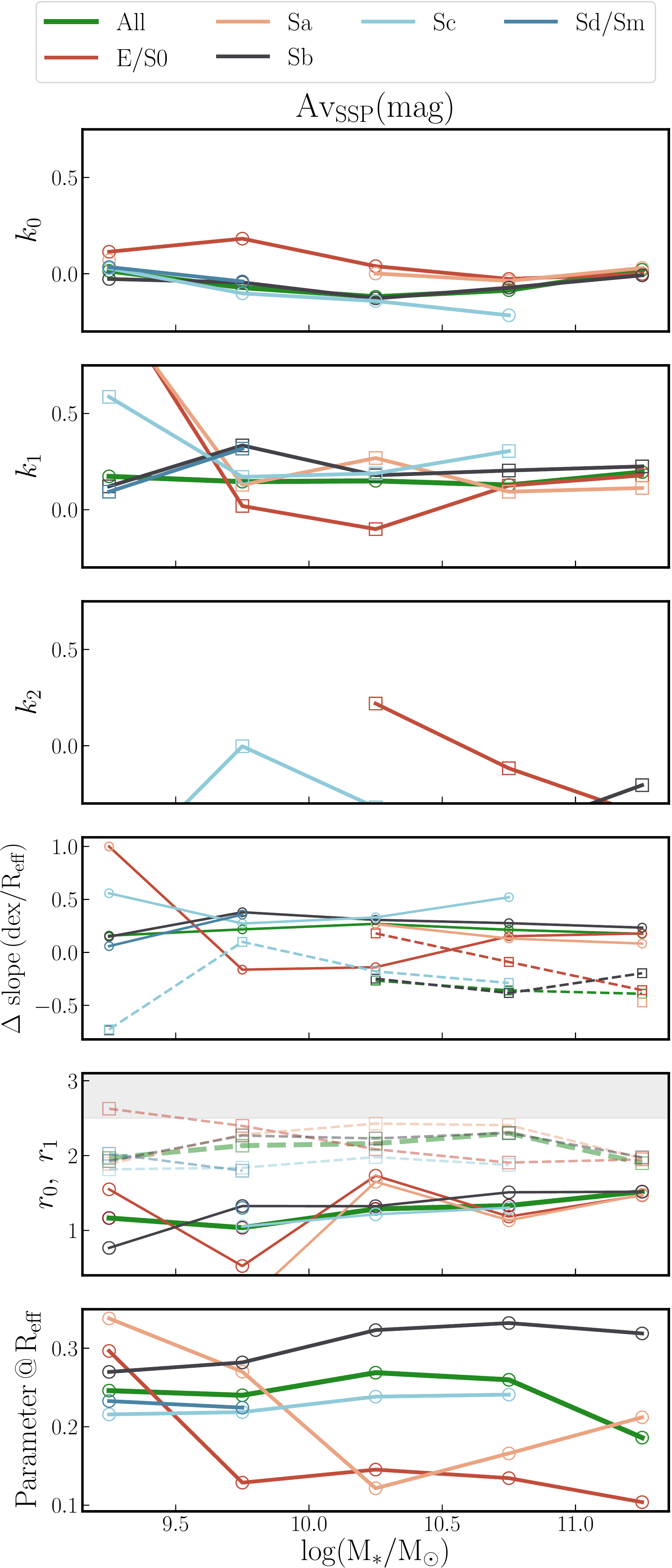}
\caption{Similar to Fig.~\ref{fig:ML}. Parameters derived from the piece-wise analysis for the optical extinction derived from the SSP analysis.}  
\label{fig:avssp}    
\end{figure}
Other than \Sstar, the SSP analysis also provides estimations of the average stellar age and metallicity, [Z/H]. In Fig.~\ref{fig:age} we show the parameters derived from the piece-wise analysis of the radial distribution of the luminosity-weighted stellar age (see Fig.~\ref{fig:age_rad}). The slope of the central gradient is negative ($k_0 \sim$ -0.3 dex/\Reff, regardless the stellar mass). When considering the morphology, this slope appears to decrease for late-type galaxies while early-type galaxies tend to have either flat or mildly positive gradients. In average, the $k_1$ slope is slightly flatter than $k_0$, while the outer gradient ($k_2$) has a significant larger negative value than the central gradient. The radii where the change of slope occurs are relatively well confined ($r_0 \sim$ 1.0 - 1.5 \Reff, and $r_1 \sim$ 2 \Reff). By exploring the average stellar age at \Reff, we find that early-type galaxies show the oldest ages in comparison to other morphological types. Furthermore, the stellar age measured at \Reff increases with \Mstar (green solid line in bottom panel of Fig.~\ref{fig:age_rad}). However morphology may play a more significant role, late-type galaxies are in general younger than early-type ones. These results are in agreement with different studies presented in the literature \citep[e.g., ][]{Sanchez-Blazquez_2014a,Sanchez-Blazquez_2014b, Gonzalez_Delgado_2015, Morelli_2015, Zheng_2017,Parikh_2021}. In order to describe these mild negative gradients, these studies suggest a 'inside-out' growth of the galaxies. We note that contrary to these studies, \citet{Goddard_2017} found positive mass-weighted stellar age gradients for early-type galaxies (using the same dataset that this study) suggesting that an 'outside-in' growth for these galaxies. In this study, we find that in general, the age gradients are -- mildly -- negative, regardless the morphological type suggesting an 'inside-out' growth for all the different morphological types or stellar masses.

In Fig.~\ref{fig:ZH} we show the result of the piece-wise analysis for the stellar metallicity radial profiles for our MaNGA GS (see Fig.~\ref{fig:ZH_rad}). We find that regardless the morphology (except for intermediate-mass early-type galaxies), the gradient of the central radial profile ($k_0$) slightly decreases with \Mstar. For low-mass galaxies, this gradient is nearly flat, whereas most massive galaxies have a mild negative gradient ($k_0 \sim - 0.1$ dex/\Reff). Similar results have been found in the literature assuming a single gradient for the entire radial distribution of metals (e.g., S20). The slopes of the central gradients have been usually interpreted as an inside-out growth more evident in massive galaxies. However, for those galaxies where we are able to measure an external metallicity gradient ($k_1$, and $k_2$), they usually have a flatter behavior than $k_0$. Even more, in some cases (e.g., Sb-Sc galaxies), we detect positive gradients, suggesting a radical change: an increment of metallicity at the outskirts of these galaxies. Since the gradients of the stellar metallicity have been usually described with a single negative slope \citep[e.g., ][]{Gonzalez_Delgado_2015, Zheng_2017,Sanchez_2020}, this could be an artifact due to either the lack of statistics at large galactocentric distances (although as we will explore in Sec.~\ref{sec:ALL}, these trends are also observed when using a much larger sample) and/or the low signal-to-noise from the spectra at the outskirts of galaxies. 
Finally, we note that contrary to the stellar age at \Reff, [Z/H] at \Reff\  depends significantly on \Mstar\ rather than the morphology. The metallicity of the stellar component at this radius increases with \Mstar. Similar results have been found using the same dataset \citep{Zheng_2017}.  

From the stellar continuum fitting, the SSP analysis also allows us to have an estimation of the optical extinction afecting the stellar continuum, \avssp. In Fig.~\ref{fig:avssp} we show the results of the piece-wise analysis to the radial distribution of \avssp (see  Fig.~\ref{fig:av_ssp_rad}). From the piece-wise analysis  we note that in the central portion of the galaxies the radial distribution of \avssp\ is almost flat with a mild negative gradients for most morphological types ($k_0 \sim -0.1 dex/\Reff$), except for early-type galaxies where these gradients are slightly positive. For external regions (\Reff $<$ r $<$ 2 \Reff) the gradients of \avssp\, change their sign, exhibiting a slight increment of \avssp\, as the galactocentric distance increases. Similar radial trends have been described in the literature with  different dataset \citep[e.g.,][]{Gonzalez_Delgado_2015}. From the measurement of \avssp at \Reff we note that their absolute values are rather small (\avssp $<$ 0.3 mag). Contrary to other stellar properties derived from the SSP analysis, for the radial distribution of the optical extinction it is not clear to disentangle the impact of either the morphology or \Mstar. Our analysis suggests that both structural parameters play a similar role in shaping the radial distribution of \avssp. In Sec.~\ref{sec:BD} we show the ratio between \avssp\ and the optical extinction derive from the Balmer decrement. 

\section{Emission Lines Properties}
\label{sec:gas}

As we mention in Sec.~\ref{sec:pipe3d}, the \textsc{pyPipe3D} analysis pipeline provides the angular distribution of the properties of different emission lines observed in the optical (i.e., the integrated flux, equivalent width, velocity, and velocity dispersion) for each of the galaxies included in the MaNGA survey. These properties, including their ratios, has been widely used to explore physical properties of the ionized gas component of the ISM. In this section we explore the radial distribution of the properties derived directly by the pipeline (Secs.~\ref{sec:Flux}), and those derive by the flux ratios of these emission lines (Secs.~\ref{sec:Ratios}, \ref{sec:SFRs}, \ref{sec:Chem}, and \ref{sec:ism}). 

\subsection{Fluxes, and Equivalent widths}
\label{sec:Flux}
\begin{figure*}
\begin{minipage}{.24\textwidth}
\includegraphics[width=\linewidth]{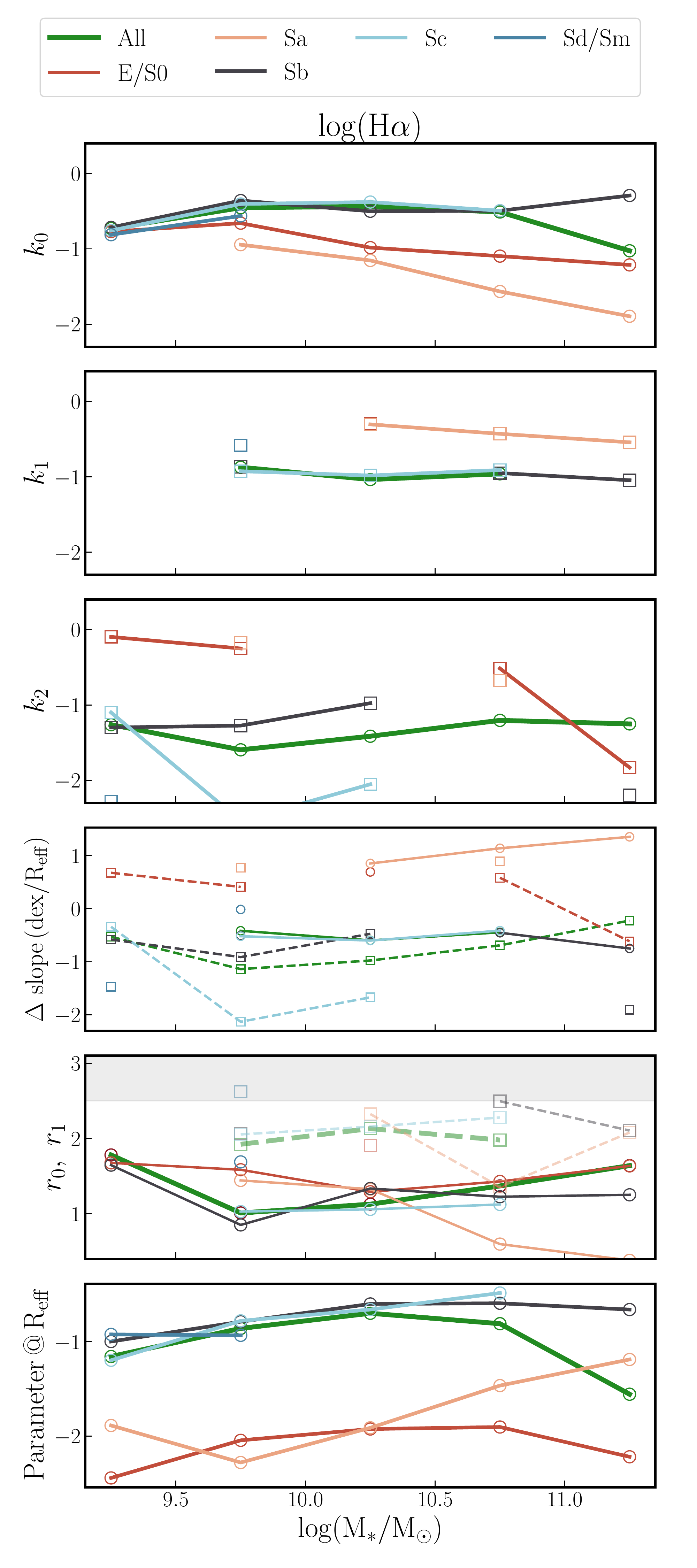}
\end{minipage}
\begin{minipage}{.24\textwidth}
\includegraphics[width=\linewidth]{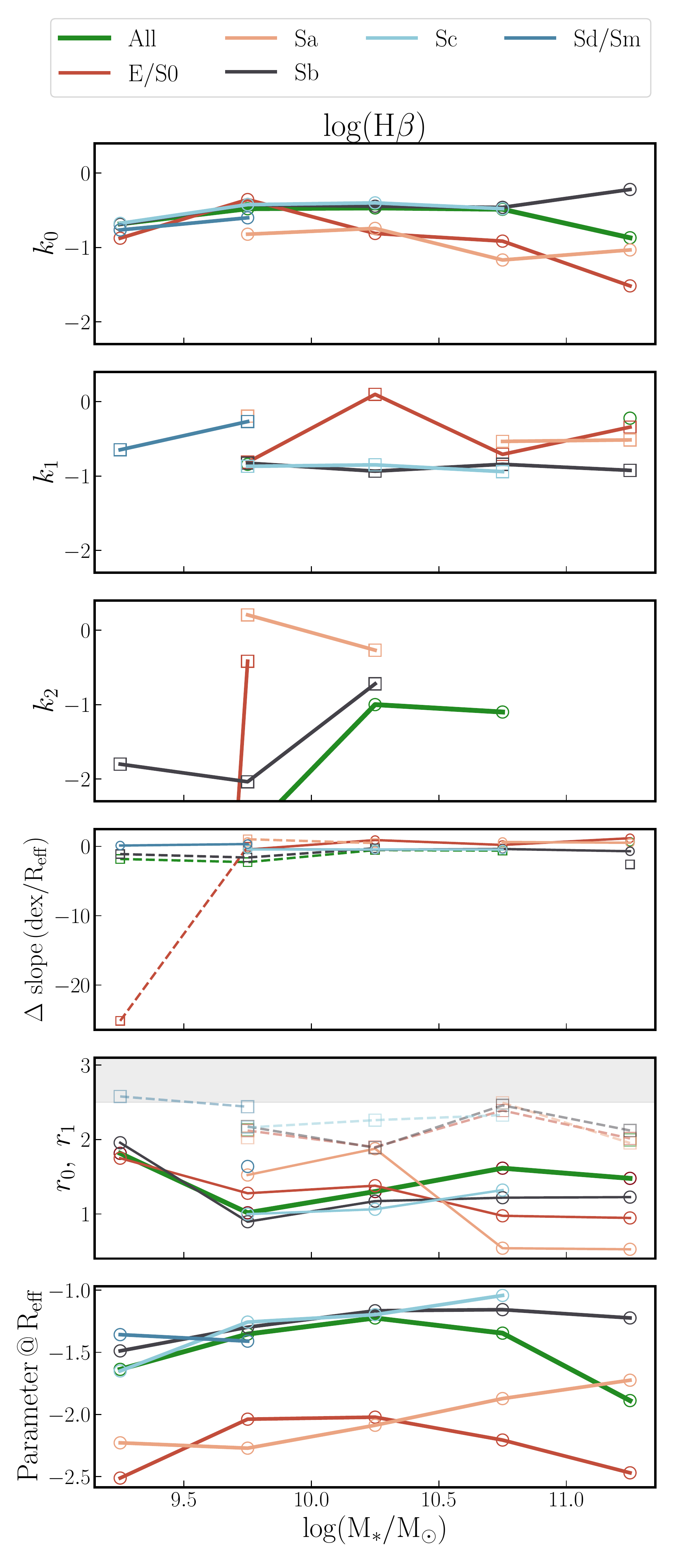}
\end{minipage}
\begin{minipage}{.24\textwidth}
\includegraphics[width=\linewidth]{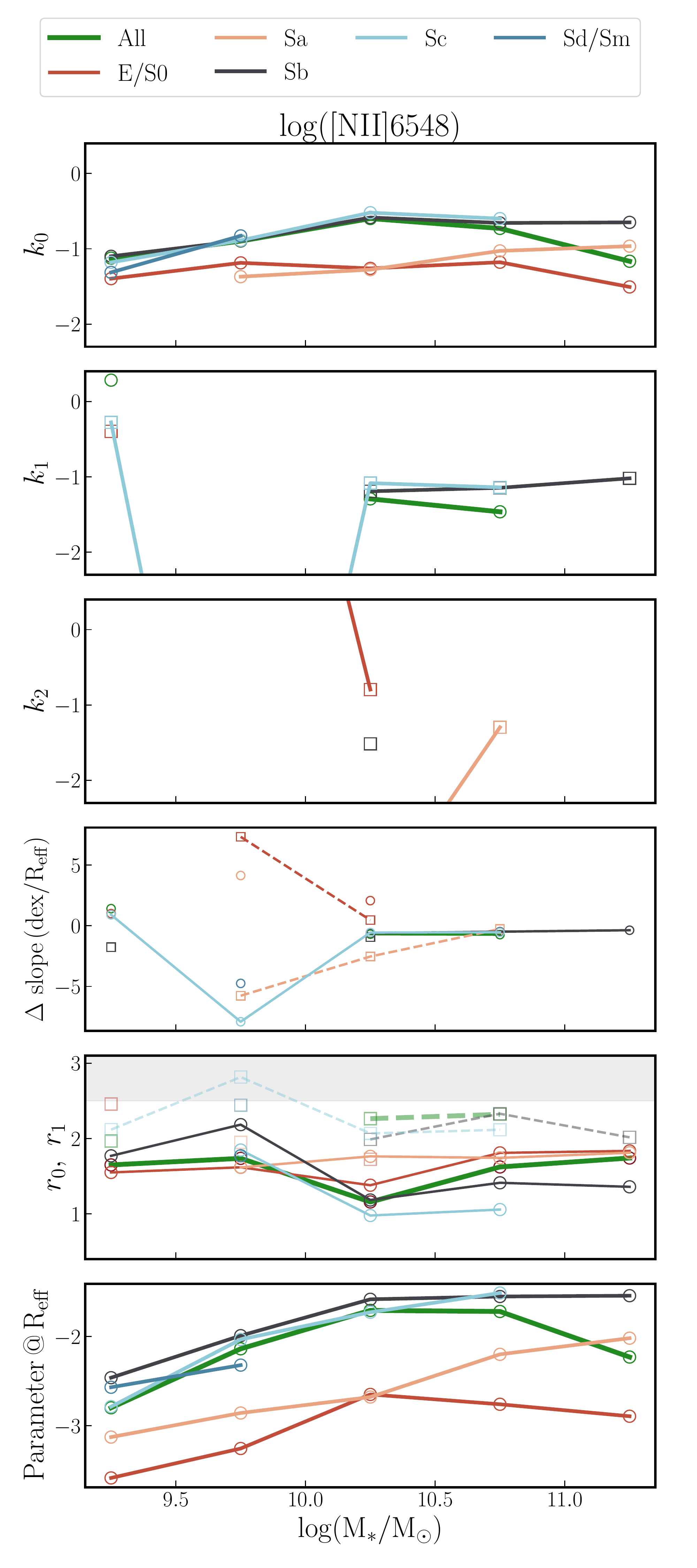}
\end{minipage}
\begin{minipage}{.24\textwidth}
\includegraphics[width=\linewidth]{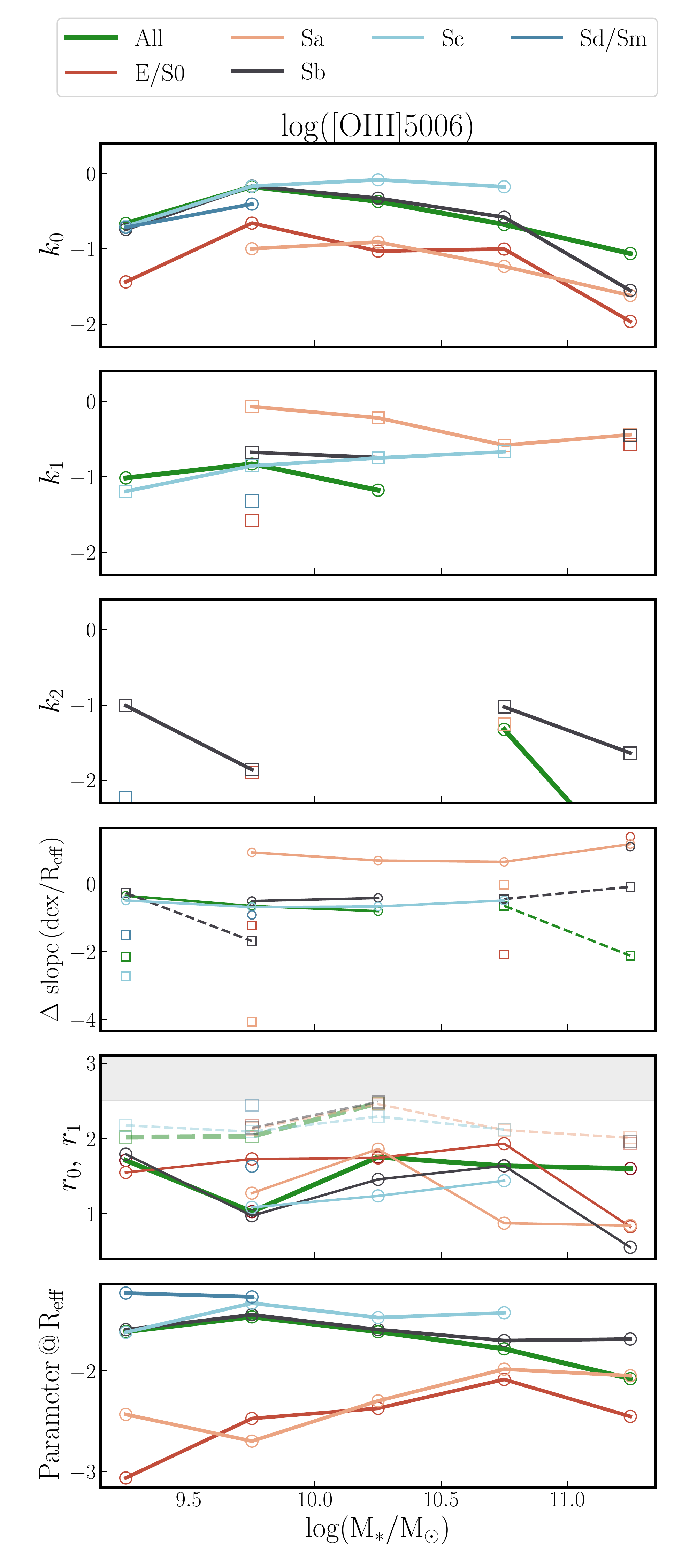}
\end{minipage}
\caption{Parameters derived from the piece-wise analysis for the four most bright emission lines in the optical. From left to right: }
\label{fig:flux}    
\end{figure*}
\begin{figure}
\includegraphics[width=\linewidth]{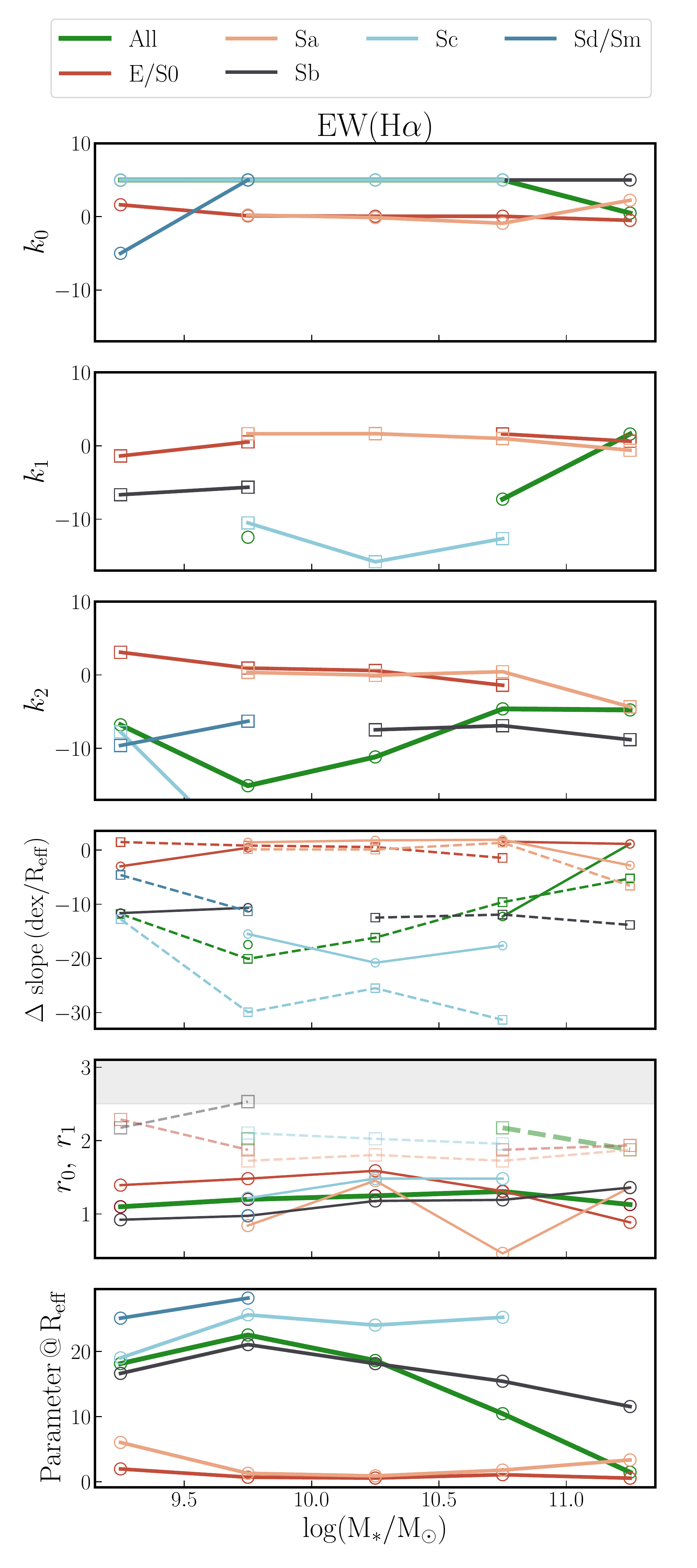}
\caption{Parameters derived from the piece-wise analysis for the \ha\, emission line equivalent width. The layout of the figure is similar as Fig.\ref{fig:ML}.}  
\label{fig:EW}    
\end{figure}
In Fig.~\ref{fig:flux} we show the results of the piece-wise analysis for  the four bright emission lines in the optical regime: \ha, \hb, \nii, and \oiii\footnote{Apart from these emission lines, the \textsc{pyPipe3D} analysis pipeline also allows us to derive the flux from low-brightness emission lines. 
} (see their radial profiles in Figs.~\ref{fig:ha_rad}, \ref{fig:hb_rad}, \ref{fig:n2_rad}, and~\ref{fig:o3_rad}). We first note that due to our SNR selection criteria, in particular the one from the \ha\, emission line, there are some bins of morphology/\Mstar where it is not possible to derive the gradients from these emission lines. This is the case for low-mass early-type (Sa) galaxies. 

For those stellar masses and morphologies where we measure a gradient, the slope in the central region \mbox{$(r < 1.5 \Reff)$} is negative regardless the emission line ($-1 < k_0 < 0$). For the Balmer emission lines, \ha\ and \hb, we find little dependence of $k_0$ with respect to \Mstar. For late-type galaxies, \mbox{$k_0 \sim -0.5$ dex/\Reff}. On the other hand, for early-type galaxies $k_0$ tends to be stepper in comparison to those derived for late type ones at the same stellar mass bin. For those galaxies where we are able to estimate an external gradient ($k_1$), we find steeper negative gradients in comparison to $k_0$ for late-type galaxies whereas for early galaxies $k_1$ $>$ $k_0$, in general.
For the \nii\ emission line $k_0$ increases with as \Mstar\ increases, reaching a constant value of $k_0 \sim -0.5$ dex/\Reff for the most massive bins. Similar to the Balmer lines, the slope of the central gradient for the \nii\, lines for early-type galaxies is steeper in comparison to late-type galaxies for the same stellar mass bin. For this emission line the outer gradients are steeper than those derived in central regions.
On the other hand, for the \oiii\ emission line we find that the central gradient, $k_0$ decreases with the stellar mass, low-mass galaxies have a flatter gradient in comparison to massive ones. Similar to the other emission lines, we estimate negative steeper gradients for the outskirts of galaxies ($k_1$). The sharp drop of the gradients at the outskirts for the different emission lines could be expected given the low signal-to-noise from these lines at those large galactocentric radii. 
For the \ha, \hb, and \nii\, emission lines their fluxes at \Reff\ slightly increases with the stellar mass. For early-type galaxies the flux of these emission lines at \Reff\, is significantly fainter than those derived for late-type galaxies. On the other hand, the flux of the \oiii\ slightly decreases with the stellar mass. 


The equivalent width of the \ha\, emission line, EW(\ha), has been extensively used to explore the star-formation activity of galaxies in both at integrated and spatially resolved scales \citep[e.g.,][]{Sanchez_2012, Lacerda_2018}. Therefore it is quite relevant to understand how this parameter changes with radius. In Fig.~\ref{fig:EW} we present the piece-wise analysis of the radial distribution for the EW(\ha). The radial distributions of EW(\ha) for different bins of stellar mass and morphology are presented in Fig.~\ref{fig:EW_rad}. We find that the slope of the  central gradient of  EW(\ha) is nearly constant for different bins of \Mstar regardless the morphology. However, we note that although late-type galaxies show a positive gradient, early-type galaxies shows a nearly flat gradient. Interestingly, we find a change in the slope of the radial distribution of EW(\ha) for the outskirts of late-type galaxies (i.e., $k_1$, and $k_2$ are negative). This change in the slope occurs at $\sim$ 1.5 \Reff. This may reflect the impact of morphological features on setting the radial distribution of the EW(\ha) in these type of galaxies. Finally, when we explore the value of EW(\ha) at \Reff, we find that low-mass late-type galaxies (Sd/Sm) have the largest values of EW(\ha). For the other late-type morphological bins we note that as \Mstar increases the value of EW(\ha) at \Reff\ tends to decrease. On the other hand, for early-type galaxies, the EW(\ha) measured at \Reff\ is constant regardless \Mstar\ \mbox{(EW(\ha) $<$ 6 \AA)}. This value has been usually used to differentiate star-forming regions from other ionization mechanisms \citep[e.g., ][see dashed line in Fig.~\ref{fig:EW_rad}]{Cid-Fernandes_2011, Lacerda_2018}. As derived in other studies, we find that the star-formation activity is largely affected by the morphology than by the stellar mass \citep[e.g.,][ and references therein]{Sanchez_2020}. In Sec.~\ref{sec:Disc}

\subsection{Lines Ratios}
\label{sec:Ratios}

The ratios between fluxes of different emission lines have been essential to explore physical properties of the ISM. In this section we describe the radial gradients and the best-fit gradients of three emission-line ratios from the most bright lines presented in the previous section (\ha, \hb, \nii, and \oiii).

\subsubsection{Balmer Decrement, \avgas, and \SmolAv}
\label{sec:BD}
\begin{figure}
\includegraphics[width=\linewidth]{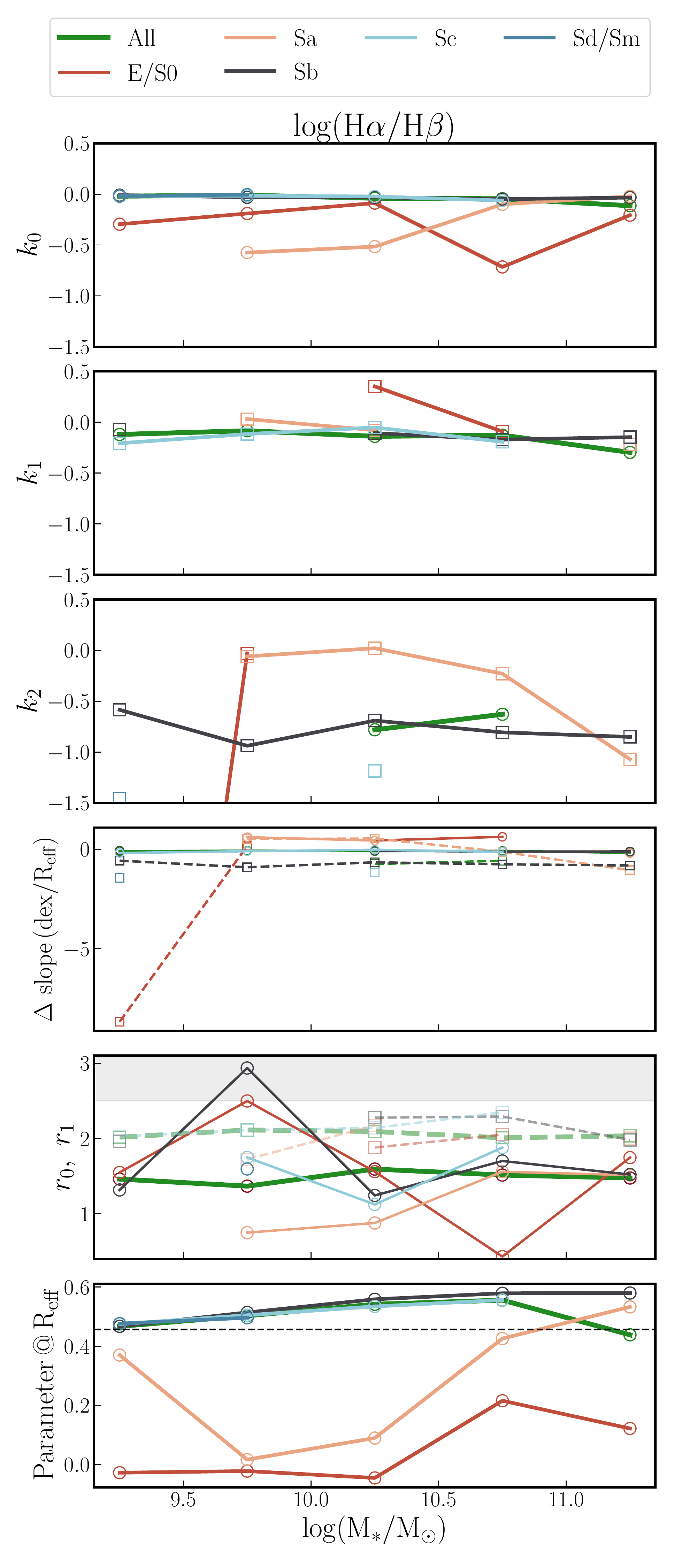}
\caption{Parameters derived from the piece-wise analysis for the Balmer decrement. The layout of the figure is similar as Fig.\ref{fig:ML}. The dashed line in the bottom panel represent the expected value of this ratio from a Case B of recombination (\ha/\hb = 2.86).}  
\label{fig:BD}    
\end{figure}
\begin{figure}
\includegraphics[width=\linewidth]{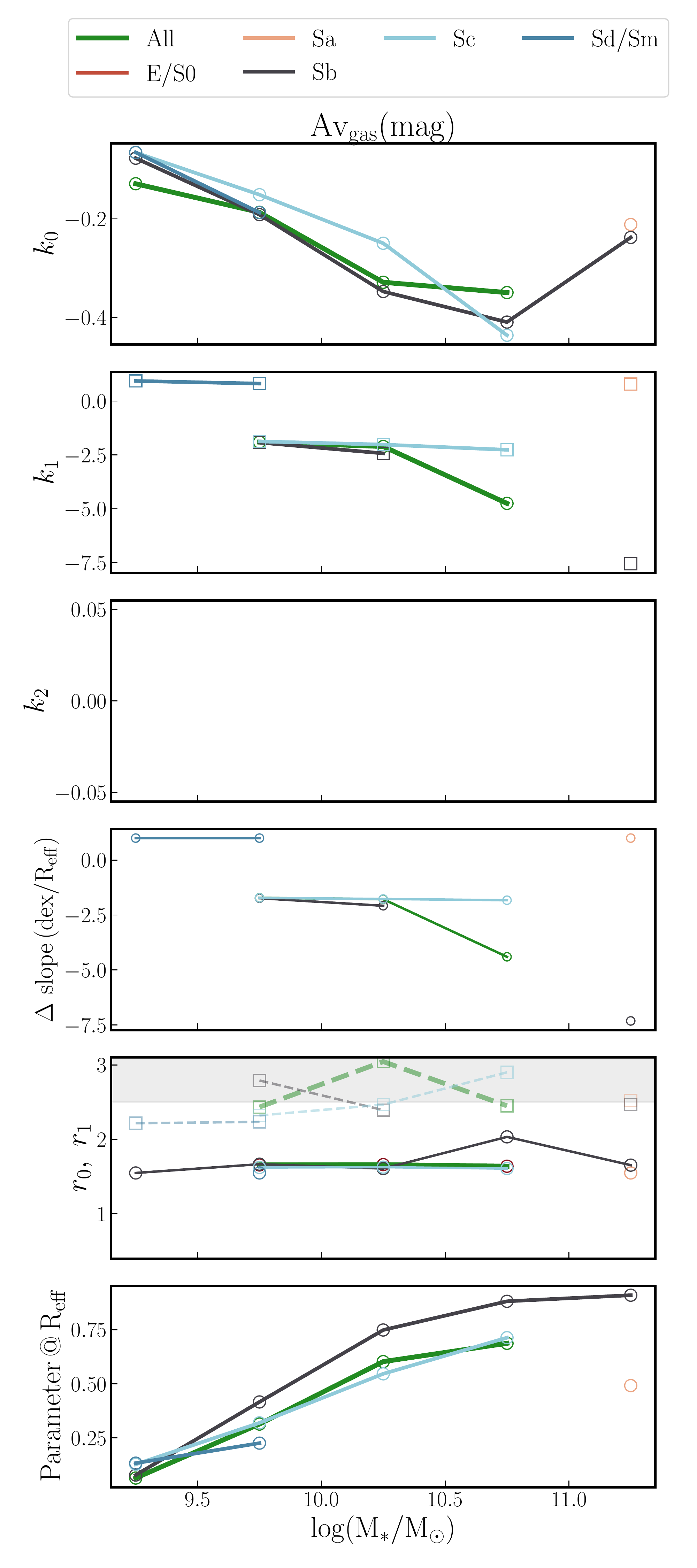}
\caption{Parameters derived from the piece-wise analysis for the optical extinction derived from the Balmer decrement, \avgas. The layout of the figure is similar as Fig.\ref{fig:ML}. }  
\label{fig:avgas}    
\end{figure}
\begin{figure}
\includegraphics[width=\linewidth]{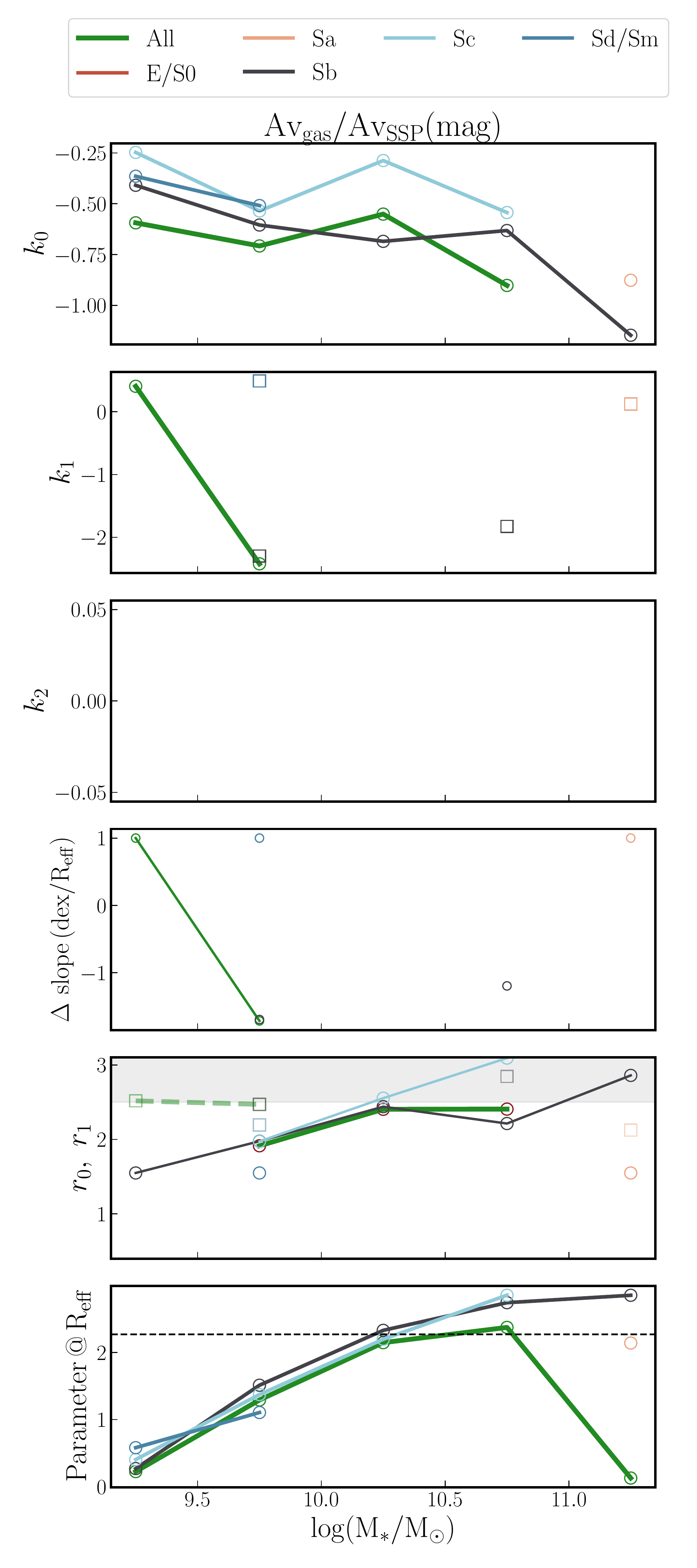}
\caption{Parameters derived from the piece-wise analysis for the \avgas/\avssp ratio. The layout of the figure is similar as Fig.\ref{fig:ML}. The dashed line in the bottom panel shows the value derive from \citet{Calzetti_1997}: \avgas/\avssp $\sim$ 2.27.}  
\label{fig:avgasssp}    
\end{figure}
\begin{figure}
\includegraphics[width=\linewidth]{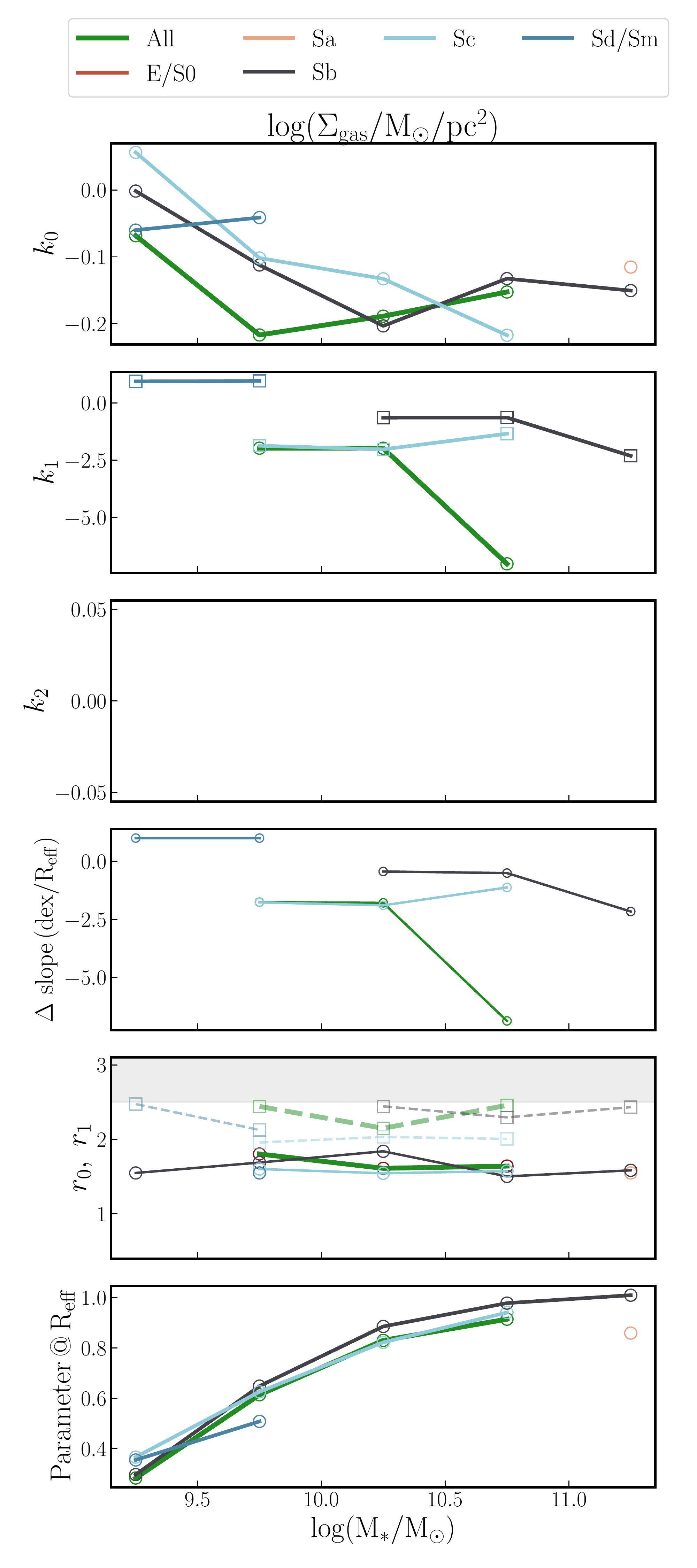}
\caption{Parameters derived from the piece-wise analysis for the molecular gas mass derived from \avgas, \SmolAv. The layout of the figure is similar as Fig.\ref{fig:ML}.}  
\label{fig:mol}    
\end{figure}
\begin{figure}
\includegraphics[width=\linewidth]{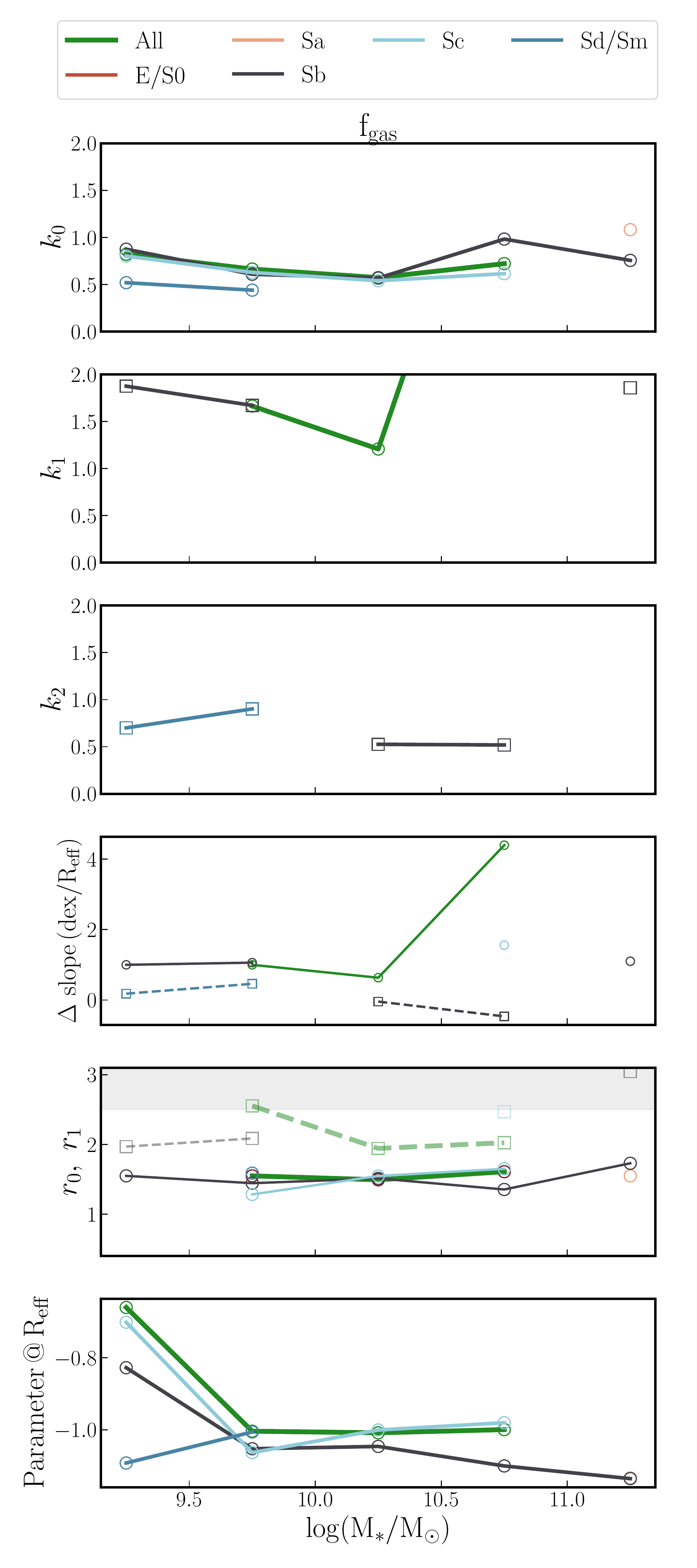}
\caption{Parameters derived from the piece-wise analysis for the gas fraction, \fgas. The layout of the figure is similar as Fig.\ref{fig:ML}.}  
\label{fig:fgas}    
\end{figure}

The \ha/\hb\, flux lines ratio (also known as the Balmer decrement, BD) has been extensively  used to estimate the effect of the dust in the optical extinction \citep[\avgas; e.g., ][]{Kennicutt_1992, Brinchmann_2004, Moustakas_2006, Dominguez_2013}. Recently, in \citet{Barrera-Ballesteros_2020} we provided a proxy between \avgas and the molecular gas surface mass density, \Smol, at kpc scales using the spatially resolved data from the optical and the molecular gas of the EDGE-CALIFA survey.   


In Fig.~\ref{fig:BD} we present the piece-wise analysis of the \ha/\hb\, radial profiles for our MaNGA Golden Sample, whereas in Fig.~\ref{fig:BD_rad} we show the radial distribution of the BD as well as the best-fit gradients derived from this analysis. In Fig.~\ref{fig:BD_rad} we also indicate the theoretical value of this ratio $\sim$ 2.86 \citep[see dashed lines, this is the expected value for the \ha/\hb\, ratio for a temperature of 10$^4$ K and Case B of recombination,][]{Osterbrock_2006, Raga_2015}. From the piece-wise analysis we find that the central gradient ($k_0$) for late-type galaxies is rather constant and flat for different stellar mass bins (i.e., $k_0 \sim$ 0 dex/\Reff). For the early-type galaxies their central gradients are vary significantly between negative and flat for the different bins of probed stellar mass. For those galaxies where we are able to estimate external gradients (i.e., $k_1$ and $k_2$) we find that these are mostly negative. The values of the Balmer decrement measured at \Reff\ shows that this parameter increases with \Mstar for late-type galaxies. Furthermore, we find that for all late-type galaxies -- at least within  1 \Reff\ -- the BD is larger than the expected value from theory (see dashed line in bottom panel of in Fig.~\ref{fig:BD}). On the other hand, only the most massive Sa galaxies have a BD larger than 2.86. For the rest of early-type galaxies the BD measured at \Reff\ is smaller than this value. Furthermore we note that for E/S0 galaxies BD $\sim$ 1, suggesting that the flux from these two lines are close to the values expected from fluctuations within the noise.  
The fact that for most of the late-type galaxies, regardless the stellar mass, the BD distribution is larger than 2.86 suggests that it is possible to have a radial measurement of the optical extinction from this ratio. On the other hand, for early-type galaxies it is not possible to estimate the optical extinction from the \ha/\hb\, ratio since their values are smaller than the one expected from the case B of recombination. 

We follow \citet{Barrera-Ballesteros_2020} and \citet{Catalan-Torrecilla_2015} to estimate the optical extinction for the \ha\, emission line, $A(\ha)$. Assuming a Cardelli extintion curve with $R_V$ = 3.1 \citep{Cardelli_1989}, the optical extinction is given by:  
\begin{equation}
\avgas =  A(\mathrm{H}\alpha) / 0.817. 
\end{equation} 

Thus to estimate \avgas\, we do not consider those radial bins in galaxies where BD $<$ 2.86. In Fig.~\ref{fig:avgas} we present the analysis of the piece-wise analysis of the radial distribution of \avgas\, (see Fig.~\ref{fig:avgas_rad}). As we find in the previous analysis, for early-type galaxies it is not possible to derive \avgas\, from the Balmer decrement because most of the radial bins have a value of the BD smaller than 2.86. Therefore, most of the results presented in Fig.~\ref{fig:avgas} are for late-type galaxies, in particular Sb, and Sc ones. For a wide range of stellar masses (i.e., \mbox{$9.2 < \MstarMsun < 10.7$}), the gradient of \avgas, $k_0$, decreases with \Mstar\, (from \mbox{$k_0 \sim - 0.2~\mathrm{mag}/\Reff$} to  \mbox{$k_0 \sim -0.4~\mathrm{mag}/\Reff$}). For the most massive bin, the gradients are flatter. The radial extension of \avgas\, goes up to $\sim$ 1.5 \Reff. When measuring \avgas\, at \Reff, we find that this extinction increases with \Mstar. For low-mass galaxies \avgas\, $\sim$ 0.2 mag, whereas massive ones show values of \avgas\, $\sim$ 1.0 mag. Our results show the impact that \Mstar\ has in setting the radial distribution of \avgas\ in late-type galaxies: the more massive is the galaxy the steeper is the gradient of \avgas.



Since we have the estimation of the optical extinction from both the stellar continuum (\avssp) and the emission lines (\avgas) we are able to compare the radial variations of the ratio of these values. Different studies have explored this ratio for integrated and angular resolved scales. On the one hand, some studies suggested that this ratio is constant for late-type galaxies in both integrated and spatially-resolved measurements \citep[e.g., ][]{Calzetti_1997,Calzetti_2000,Kreckel_2013}. On the other hand, other studies have suggested variations of this ratio for different local and global properties \citep[e.g.,][]{Wild_2011, Koyama_2015,Koyama_2019, Qin_2019,Lin_2020,Li_2021}. 

In Fig.~\ref{fig:avgasssp} we present the results of the piece-wise analysis of the radial distribution of the \avgas/\avssp\, ratio (see Fig.~\ref{fig:avgasssp_rad}). We first note that the ratio of these two estimations of the optical extinction is not constant with radius. Furthermore the gradient of this ratio varies depending on the stellar mass. As for \avgas, the radial extension of this ratio is limited by the \ha\, SNR selection criteria. Therefore, for most radial distributions, a single gradient ($k_0$) describes well the entire radial distribution. Also similar to \avgas, this ratio is derived mostly for late-type galaxies. We find a similar trend in $k_0$ as the one described by \avgas\, with respect to the stellar mass. The gradient of this ratio decreases with \Mstar; massive galaxies have steeper negative gradients in comparison to low-mass ones. 
The value of this ratio at \Reff, increases with \Mstar; going from values below 1  to 3 from low-mass to massive galaxies, respectively. The bottom panel of Fig.~\ref{fig:avgasssp} shows that only intermediate mass Sb galaxies have a similar ratio as the one expected in the literature \citep{Calzetti_1997}. Our results suggest that the estimation of the optical extinction varies depending on the adopted proxy (i.e., ionized gas or stellar). Furthermore, this ratio shows a significant variation radially as well as with the stellar mass. In contrast to studies that proposed a constant stellar-gas extinction ratio \citep{Calzetti_1997,Calzetti_2000, Kreckel_2013}, we find significant variations of this ratio across the optical extension of the probed galaxies. Different studies have suggested that a possible scenario to explain these differences in this ratio is due mainly to geometrical effects \citep[e.g., ][]{Price_2014, Reddy_2015, Koyama_2019}. Nevertheless, using a sample of galaxies also included in the MaNGA survey, \citet{Lin_2020} and \citep{Li_2021} found that the difference in this ratio is due to different properties from both the stellar and the ionized gas components. On the one hand, \citet{Lin_2020} found that this ratio strongly depends on the oxygen abundance as well as the ionization stage at kpc scales. On the other hand, \citet{Li_2021} found that this ratio also depends on the luminosity-weigthed age, thus the \avgas/\avssp\, ratio. Thus, our results favors the scenario presented by these works where the extinction (i.e., the dust properties) is affect (mainly) by local physical conditions such as the chemical enrichment of the ISM and/or the ionization parameter.



As we mention above, using a sample of galaxies with spatially resolved observations of the molecular gas and the optical properties we estimate a calibrator between the optical extinction derived from the Balmer decrement (\avgas), and the molecular gas surface density, \Smol \citep{Barrera-Ballesteros_2020}. Using this calibration, we present in Fig.\ref{fig:mol} the piece-wise analysis of the radial distribution of the molecular gas density, \SmolAv, derived from the optical extinction, \avgas. As expected, we find similar trends of the gradients and values of \SmolAv at \Reff with respect to the stellar mass and morphology in comparison to those derived for \avgas. The gradient of \SmolAv decreases with respect to \Mstar and is available in general only for late-type galaxies. However, we note that the gradients from \SmolAv are  flatter in comparison to those derived from \avgas. The value of \SmolAv at \Reff increases with the stellar mass. Following this calibrator as a reliable estimation of \Smol at kpc scales, our results suggest that, except for low-mass galaxies, the radial distribution of \Smol slightly decreases with radius. 


Finally, given the estimation of \SmolAv, we can also provide a measurement of the ratio between the molecular and stellar gas surface mass density -- \mbox{\fgas = \SmolAv / \Sstar}. In Fig.~\ref{fig:fgas} we show the piece-wise analysis of the radial distribution of $\log(\fgas)$\, (see Fig.~\ref{fig:fgas_rad}). We find that, in general, regardless the stellar masses and morphologies probed, \fgas\ has a positive constant gradient ($k_0 \sim$ 0.7 dex\Reff).  We also find that ,except for the low mass galaxies, the gas fraction measured at the effective radius is relatively constant for the range of probed stellar masses with \mbox{$\log(\fgas) \sim $ -1.1}. Our results suggest that the gas density, with respect to the stellar mass density, increases at large radii. In other words, in comparison to the radial gradient of \Sstar, the radial distribution of \SmolAv is rather flat. The little variations in both slope and \fgas\ at \Reff\, for different stellar masses, suggests that this radial trend is a ubiquitous property for late-type galaxies.


\subsubsection{Other Lines Ratios}
\label{sec:ORatios}
\begin{figure}
\includegraphics[width=\linewidth]{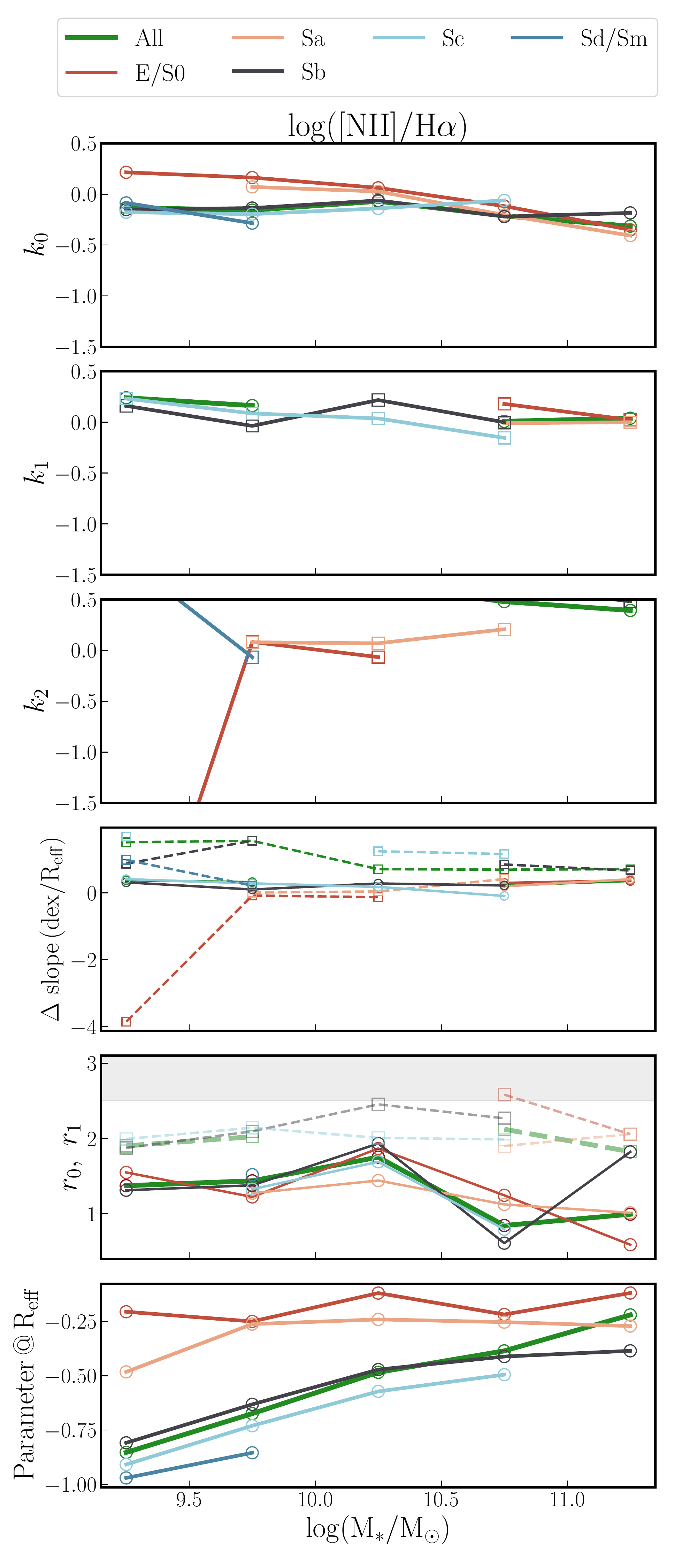}
\caption{Parameters derived from the piece-wise analysis for the  emission lines ratio. The layout of the figure is similar as Fig.\ref{fig:ML}.}  
\label{fig:N2}    
\end{figure}
\begin{figure}
\includegraphics[width=\linewidth]{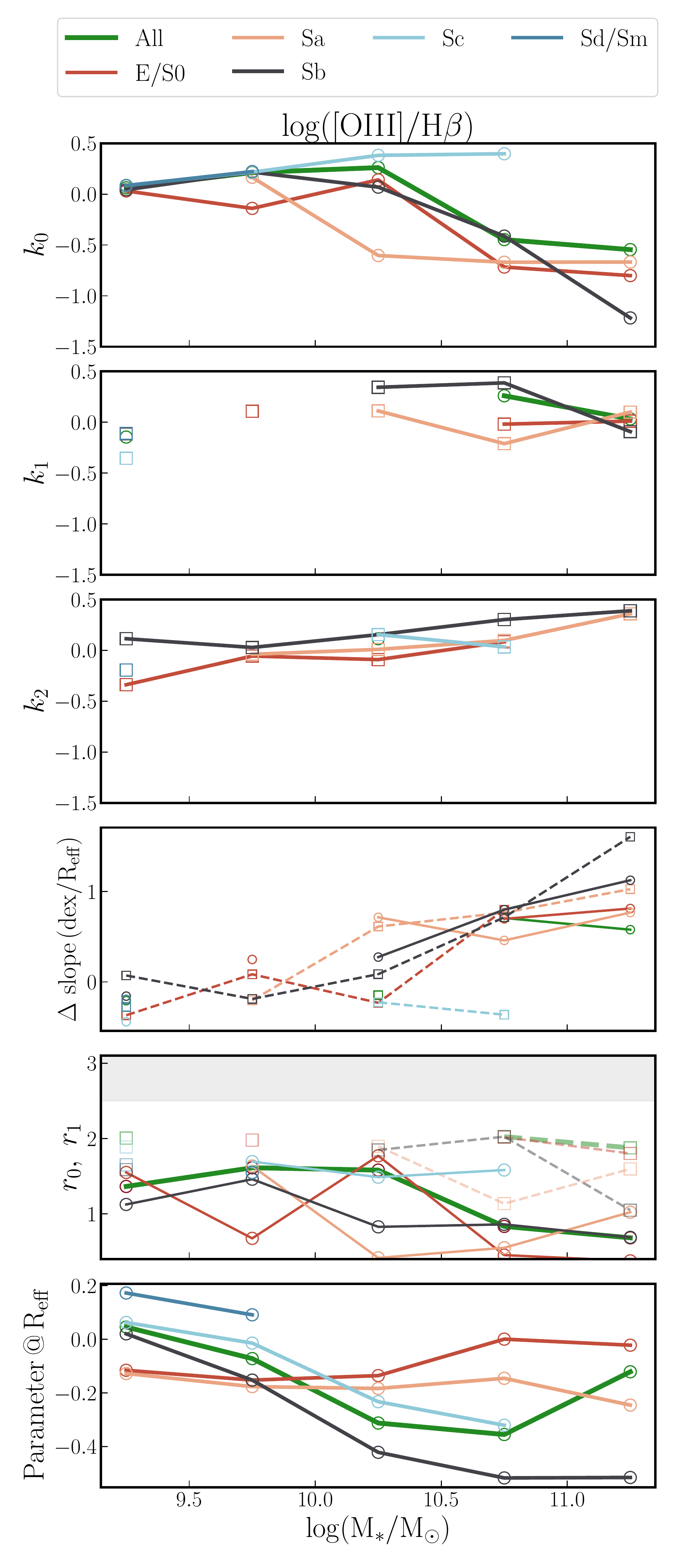}
\caption{Parameters derived from the piece-wise analysis for the emission lines ratio. The layout of the figure is similar as Fig.\ref{fig:ML}.}  
\label{fig:O3}    
\end{figure}

Apart from the \ha/\hb\, ratio, there are other emission line ratios that provide insights on the physical condition of the ISM -- and even the very young stellar population -- in galaxies. This is the case of the \nii/\ha\, ratio. In star-forming galaxies, this ratio has been linked to the fraction of young stellar population \citep[e.g., ][]{Sanchez_2015}. In Fig.~\ref{fig:N2} we present the parameters derived from the piece-wise analysis from the radial distribution of this emission line ratio (see Fig.~\ref{fig:N2_rad}). We find that regardless \Mstar, the central gradient ($k_0$) has negative values with a mild decrement with the stellar mass. On the other hand, for early-type galaxies we find a change from positive to negative gradients as stellar mass increases. For the external part of the galaxies, we note that the piece-wise analysis yields  positive gradients (in some cases sharp ones, e.g.,  Sb galaxies). We consider that those strong radial variations at the outskirts of these galaxies in the \nii/\ha\, ratio could be spurious, induced by low SNR values of both emission lines. At low SNRs, emission lines fluxes tends to be similar to one other, in other words the ratio is close to $\sim$ 1. The \nii/\ha\, ratio measured at \Reff\, increases with \Mstar\, from \mbox{$\log(\nii/\ha) \sim -0.8$ dex} to \mbox{$\log(\nii/\ha) \sim -0.2$ dex}, for late-type galaxies. For early-type galaxies, this ratio measured at \Reff\ does not significantly changes with \Mstar, with a constant value of \mbox{$\log(\nii/\ha) \sim -0.2$ dex}.

In Fig.~\ref{fig:O3} we show the results of the piece-wise analysis from the radial distribution of the \oiii/\hb\, ratio (see Fig.~\ref{fig:O3_rad}). The central gradient ($k_0$) of this ratio has significant variations for both, different morphologies and stellar masses. On the one hand, for late-type galaxies $k_0$ is positive and increases from low to intermediate stellar masses. However for massive galaxies $k_0$ is negative becoming  larger -- negative stepper --  as \Mstar\ increases. On the other hand, for early-type galaxies varies from positive to negative for different bins of stellar mass.  In contrast to $k_0$, $k_1$ and $k_2$ have larger positive gradient indicating the strong radial variation of the \oiii/\hb\, ratio. Given the fact that these variations occur at smaller radius than the \nii/\ha\, ratio ($\sim$ 1 \Reff, and $\sim$ 2 \Reff, respectively), we suggest that these variations may have a physical origin. For instance large central \oiii/\hb\, ratios could be indicating the presence of a hard ionizing source (e.g., an active nucleus), whereas large ratios at the outskirts could be due ionization due to a large star formation activity. Nevertheless, we cannot rule out that, as for the \nii/\ha\, ratio, these variations in gradients are due to low SNR measurements. Contrary to the \nii/\ha\, ratio, we find that both \Mstar\ and the morphology play a significant role in setting the observed \oiii/\hb\, ratio at \Reff. For late-type galaxies, \oiii/\hb\, measured at \Reff decreases with \Mstar\, however for a given stellar mass bin, this ratio decreases from Sd/Sm to Sb galaxies. On the other hand, for early-type galaxies this ratio is rather constant for different bins of \Mstar, furthermore the value of this ratio is similar to the one derived from the \nii/\ha\, ratio for this morphological type. In the next section, using these two line ratios, we explore the possible ionization mechanism that could explain their radial distribution.

\subsubsection{The BPT diagram}
\label{sec:BPT}
\begin{figure*}
\includegraphics[width=\linewidth]{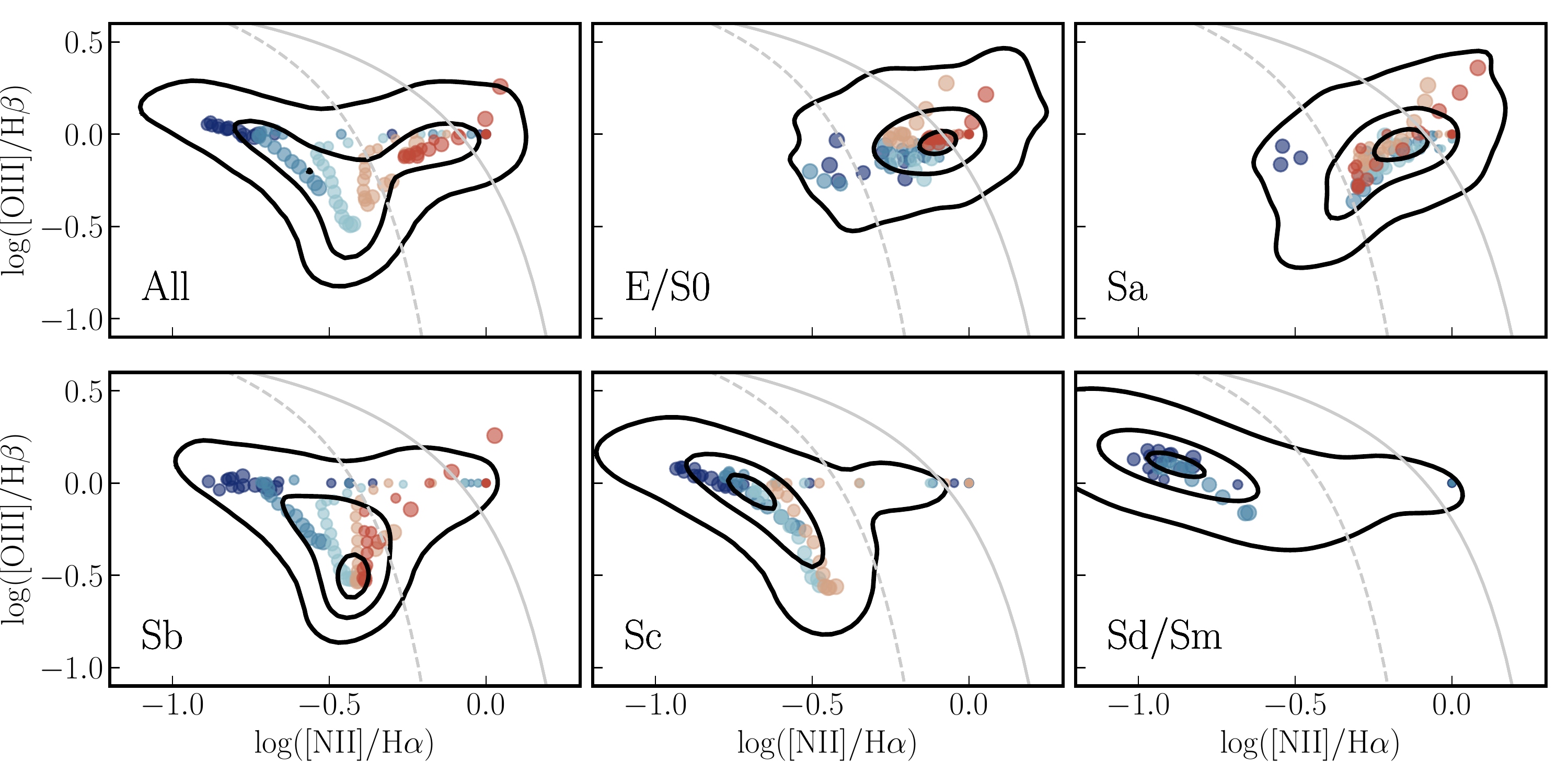}
\caption{The BPT diagnostic diagram using the radial averages for the Golden Sample. The contours enclose 90\%, 50\%, and 10\% of the line ratios for the radial bins. In each panel we segregate the sample according to their morphology (see the label in each panel). The circles represent the averaged values for different bins of stellar mass. The color of each circle represents the stellar mass, following the same color code as in Fig.~\ref{fig:Sstar_rad}. The sizes of the circles decrease with galactocentric distance. The solid and dashed gray lines represent the demarcation lines in this diagram by \citet{Kewley_2001} and \citet{Kauffmann_2003}, respectively.}
\label{fig:BPT_O3N2}    
\end{figure*}
Diagnostic diagrams using emission-line ratios are very useful tools to probe the ionization stage of the galaxies and kpc regions within them. Depending on the position of the galaxy/region in this diagram it can be associated with a different ionization process. The ionizing stage is segregated by the so-called demarcation lines. Prior to the large IFU surveys mapping a large number of targets, galaxies with line ratios below the Kauffmann demarcation line \citep{Kauffmann_2003} were considered as star-forming whereas galaxies above the Kewley demarcation line \citep{Kewley_2001} where considered as AGNs or LINERs. Galaxies in between were considered as 'composite', this is a mixture between both types of ionization. This picture has significantly changed thanks to angular-resolved observations \citep[for a review see][]{Sanchez_2020,Sanchez_2021}. In Fig.~\ref{fig:BPT_O3N2} we show the most well-known of those diagrams, the BPT diagram \citep{BPT} for the ratios derived for the radial bins of the MaNGA GS. We find that the distribution of flux ratios vary depending on the morphology, the stellar mass, and the galactocentric distance. For the entire GS (top left panel in Fig.~\ref{fig:BPT_O3N2}), we find a significant difference between low-mass and massive galaxies. On the one hand, central regions of massive galaxies are above the \citep{Kewley_2001} demarcation lines suggesting that the ionizing source for those regions is dominated by other process than star formation. Furthermore, as distance increases the flux ratios from massive galaxies move below the demarcation lines indicating that the ionization could due to star formation. These massive galaxies follow what has been identify as the composite/AGN branch using single-fiber spectroscopy\citep[e.g.,][ using SDSS dataset ]{Kauffmann_2003}. On the other hand, as stellar mass decreases,  most of the radial bins are below the Kauffmann demarcation line. This suggests that the ionization source for those galaxies, regardless the galactocentric distance, is due to star formation. Even more, the lowest mass bin exhibit the lowest values of the \nii/\ha\, ratio. These radial bins follow the star-formation branch observed using SDSS dataset \citep[e.g.,][]{Kauffmann_2003}.  

For the E/S0 morphological bin (top-middle panel of Fig.~\ref{fig:BPT_O3N2}), the radial values of those emission line ratios (black contours) are, mostly, above the \citet{Kauffmann_2003} demarcation line (gray-dashed line). Similar to the contours, these values are slightly below to the \citet{Kewley_2001} demarcation line. Although, this could indicate that the ionization source could be due to an AGN \citep[e.g.,][]{Husemann_2010,Husemann_2014}, this is only plausible in the central regions of these galaxies. It is also likely could also be the case that the ionization is due to hot-evolved stars \citep[also known as HOLMES, e.g.,][]{Binette_1994,Flores-Fajardo_2011}. To further constrain the source that ionizes the ISM it has been also require to measure the EW(\ha) \citep[e.g.,][]{Cid-Fernandes2010,Cid-Fernandes_2011, Barrera-Ballesteros_2016, Lacerda_2020}. Similarly to this study, \citet{Sanchez_2020} explored the distribution of the EW(\ha) for a large sample of IFS dataset at kpc scales within the BPT diagram for different stellar masses and morphologies. His results showed that indeed for massive E/S0 galaxies, the distribution of the emission line ratios in the BPT diagram is between the Kauffman and Kewley demarcation lines, with EW(\ha) values smaller than 6\AA\, \citep[below this threshold it is expected that the ionization source is mainly caused by HOLMES; e.g.,][]{Cid-Fernandes_2011}. We should also note that most of the averaged values for different stellar mass bins at the outskirts of E/S0 galaxies (small circles of different colors) are located close in the region where both ratios are close to one. As we mention above, we cannot rule out that these positions of the flux ratios in the BPT diagram is simply due to measure fluxes with low SNR.  
For Sa galaxies the distribution of the flux ratios measured in their radial bins spreads wider across the BPT diagram than for E/S0 galaxies. Nevertheless, as we mention for E/S0 galaxies, the outer radial bins regardless \Mstar are mostly concentrated where the ratios are close to one.  For the central radial bins of the most massive galaxies, their ratios are well above the Kewley demarcation line, whereas ratios from central regions in low-mass galaxies are well below the Kauffmann demarcation line.
The Sb galaxies show a similar distribution as the entire sample with low-mass galaxies lying below the Kauffmann demarcation line and massive galaxies spreading along the 'non star-forming' branch.
The flux ratios from most of the radial bins of Sc galaxies lies below the Kauffmann demarcation line, furthermore they lie in the so-called 'star-forming' branch. This suggest that for galaxies in this morphological type the ionization is due mostly to star formation across their optical extension, regardless the stellar mass. 
For the irregular galaxies (Sd/Sm), we find that the \oiii/\hb\, ratio is rather constant for the probed galaxies (\mbox{\oiii/\hb\,$\sim$ 1}). The  \nii/\ha\, ratio, on the other hand, covers a wide dynamical range than any other morphological type (\mbox{ 0.1 $<$ \nii/\ha\, $<$ 3}). For the irregular galaxies, we find flux ratios only for low-mass galaxies. In the BPT diagram, these ratios are located well below the Kauffmann demarcation line, suggesting that these ratios are the result of ionization due to star formation. Finally, we note that for those radial bins at the outskirts of galaxies both flux ratios tend to be close to one. We observe a similar behavior for the radial distribution of each line ratio (see Sec~\ref{sec:ORatios}). Rather than indicating a ionization stage other than star formation, this suggests that the low SNR from these emission lines at external radii do not allow us to derive their true ionization source. 


\subsection{Star-formation parameters}
\label{sec:SFRs}

The Balmer emission lines, are also a powerful tools to gauge the star formation rate (SFR) for galaxies/regions \citep[][and references therein]{Kennicutt_2012}. From the optical, following \citet{Kennicutt_1998}, we can use the extinction-corrected luminosity of the \ha\, emission line as proxy of the star formation rate. This calibration has been widely used in IFS studies \citep[e.g.,][]{Sanchez_2012,Cano-Diaz_2016,Cano-Diaz_2019}.  Other than the SNR threshold for the \ha\ emission line, we did not use any other selection criteria to derive the radial distribution of SFR, thus for some morphological types (e.g., E/S0) the radial profiles should be considered as upper limits of star formation. In this section we explore the radial distribution of the SFR density, \Ssfr, as well as its ratio with the different components of the baryonic mass: the specific SFR, \mbox{sSFR = \Ssfr/\Sstar}; and the star formation efficiency, \mbox{SFE = \Ssfr/\SmolAv}. The radial distributions of these parameters is fundamental to understand what drives or halts the star-formation activity in galaxies \citep[e.g.,][]{Colombo_2020, Ellison_2020}.   

\subsubsection{SFR surface density, \Ssfr}
\label{sec:Sigsfr}
\begin{figure}
\includegraphics[width=\linewidth]{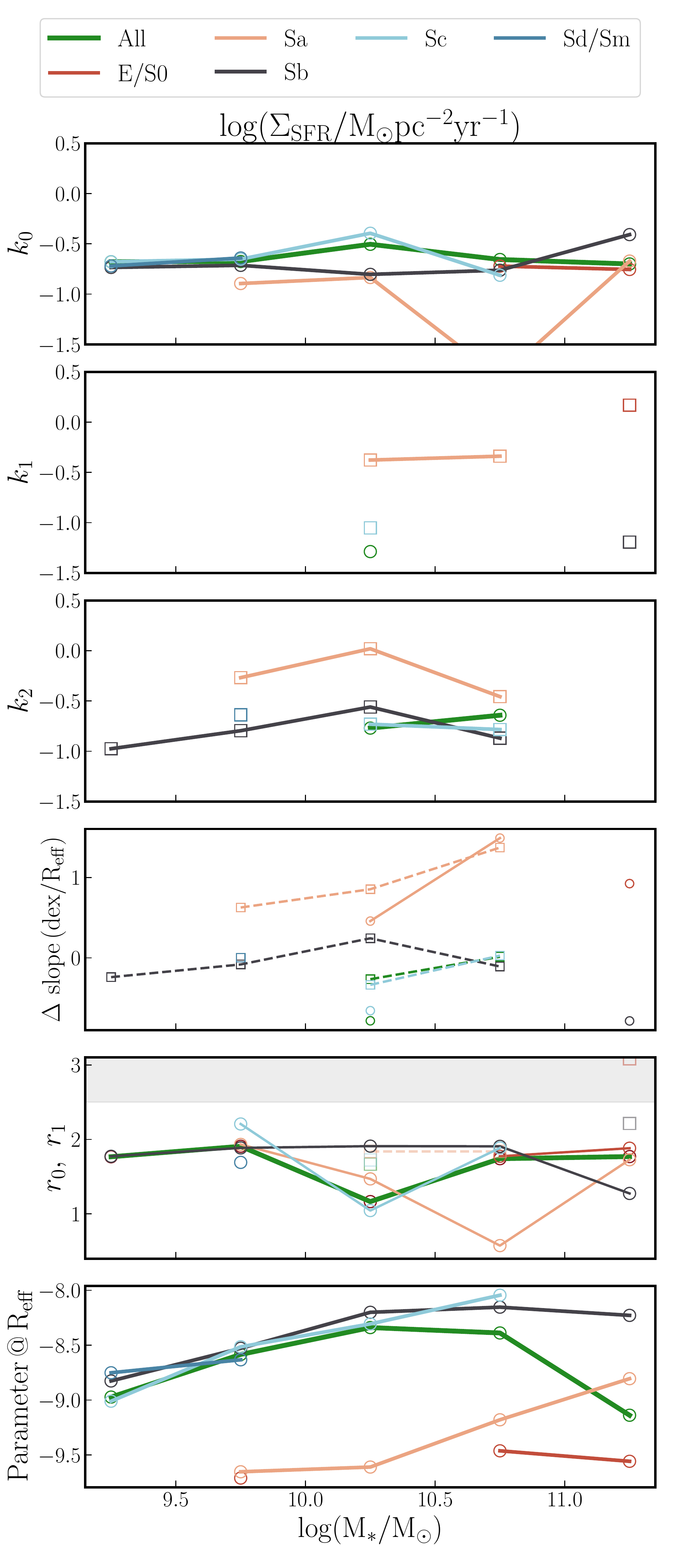}
\caption{Parameters derived from the piece-wise analysis for the radial distribution of the \Ssfr. The layout of the figure is similar as Fig.\ref{fig:ML}.}  
\label{fig:SigmaSFR}    
\end{figure}
In Fig.~\ref{fig:SigmaSFR} we plot the piece-wise analysis of the radial distribution of \Ssfr\, (see Fig.~\ref{fig:Ssfr_rad}). We find that the central gradient of \Ssfr, $k_0$, is negative regardless the stellar mass or morphology. Furthermore, $k_0$ is quite similar for the late-type galaxies regardless \Mstar\ (\mbox{$k_0 \sim$ -0.7 dex/\Reff}). For the Sa galaxies, the central slopes is slightly steeper than for late-type galaxies (except for intermediate stellar mass galaxies, where(\mbox{$k_0 \sim$ -2.0 dex/\Reff}). On the other hand, for E/S0 galaxies we estimate the central gradient only for the two most massive bins of \Mstar. These central gradient are similar to those derived for late-type galaxies with similar \Mstar. For those bins of \Mstar\ and morphology where the analysis detects an external gradient (i.e., $k_1$ and/or $k_2$), we find  a mix of flatter and steeper gradients, in any case, these are still negative. Regarding the characteristic value of \Ssfr (i.e., measured at \Reff), we find that for late-type galaxies it increases as \Mstar increases. For Sa-type galaxies, this \Ssfr\ is approximately one order of magnitude smaller for a given bin of \Mstar. The difference is larger for E/S0 galaxies, where the characteristic \Ssfr\ is $\sim$ \mbox{$10^{-9.5}$ \msunperpcsqperyr}. These results are in good agreement with previous measurements of the gradients and characteristic values of \Ssfr\ \citep{Sanchez_2020}. 

\subsubsection{specific SFR, sSFR}
\label{sec:ssfr}
\begin{figure}
\includegraphics[width=\linewidth]{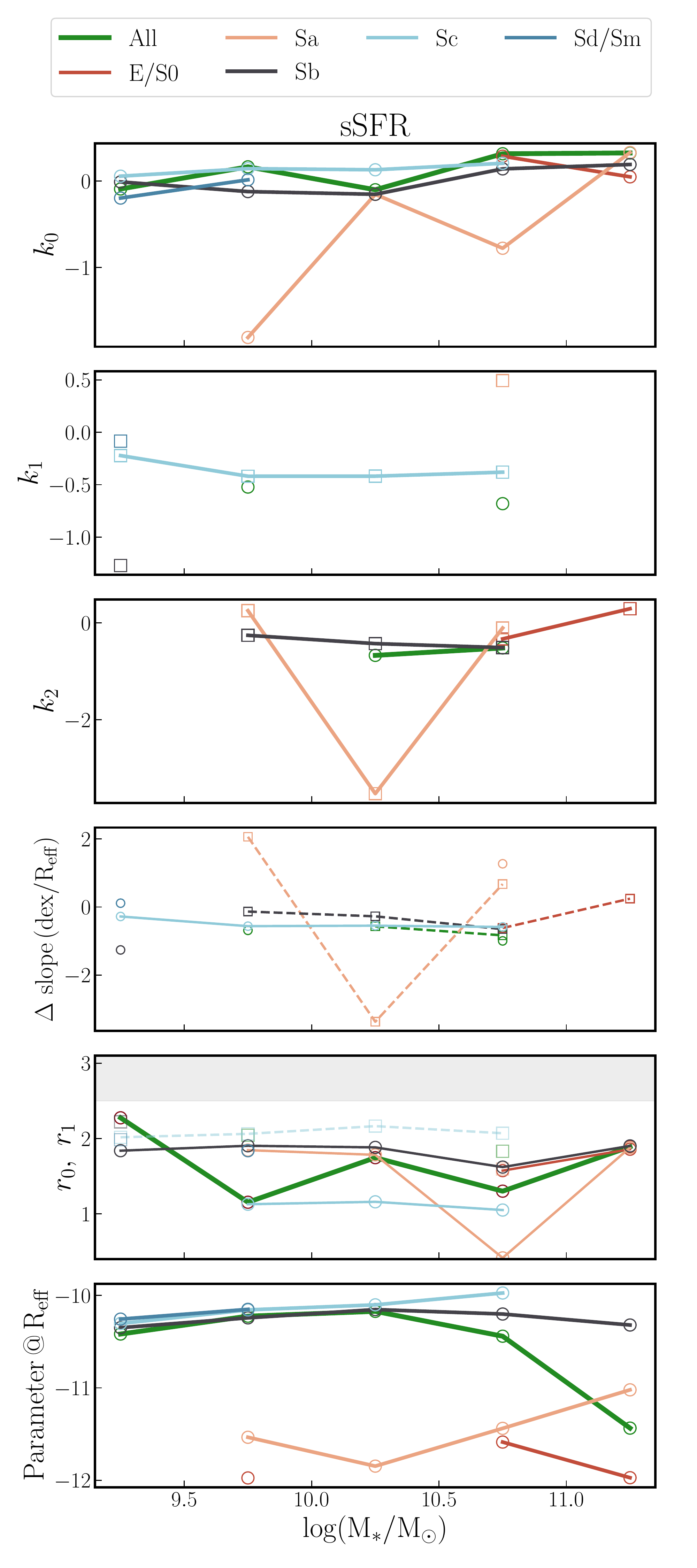}
\caption{Parameters derived from the piece-wise analysis for the radial distribution of the sSFR. The layout of the figure is similar as Fig.\ref{fig:ML}.}  
\label{fig:sSFR}    
\end{figure}

In Fig.~\ref{fig:sSFR}, we show the results of the piece-wise analysis of the radial distribution of the sSFR (see Fig~\ref{fig:sSFR_rad}). We find that the values of $k_0$ are relatively consistent for different bins of \Mstar within the late-type galaxies. If any, there is a mild increment of the slope (from negative to positive gradients) as \Mstar increases. As for the gradients of \Ssfr, Sa galaxies is the only morphological bin that shows significant variations in $k_0$. For those bins of \Mstar or morphology where we measure an external gradient ($k_1$ or $k_2$) their value is usually negative. Regarding the characteristic value of the radial distribution of sSFR (i.e., at \Reff), we find that it varies significantly depending on the morphology. The characteristic sSFR from early-type galaxies is at least one order of magnitude smaller in comparison to late-type galaxies. For each morphological type we do not see significant variations of the characteristic sSFR for different stellar mass bins. Qualitatively, these gradients are in agreement with previous results using a larger heterogeneous sample of galaxies \citep{Sanchez_2020}. 


\subsubsection{Star-forming Efficiency, SFE}
\label{sec:SFE}
\begin{figure}
\includegraphics[width=\linewidth]{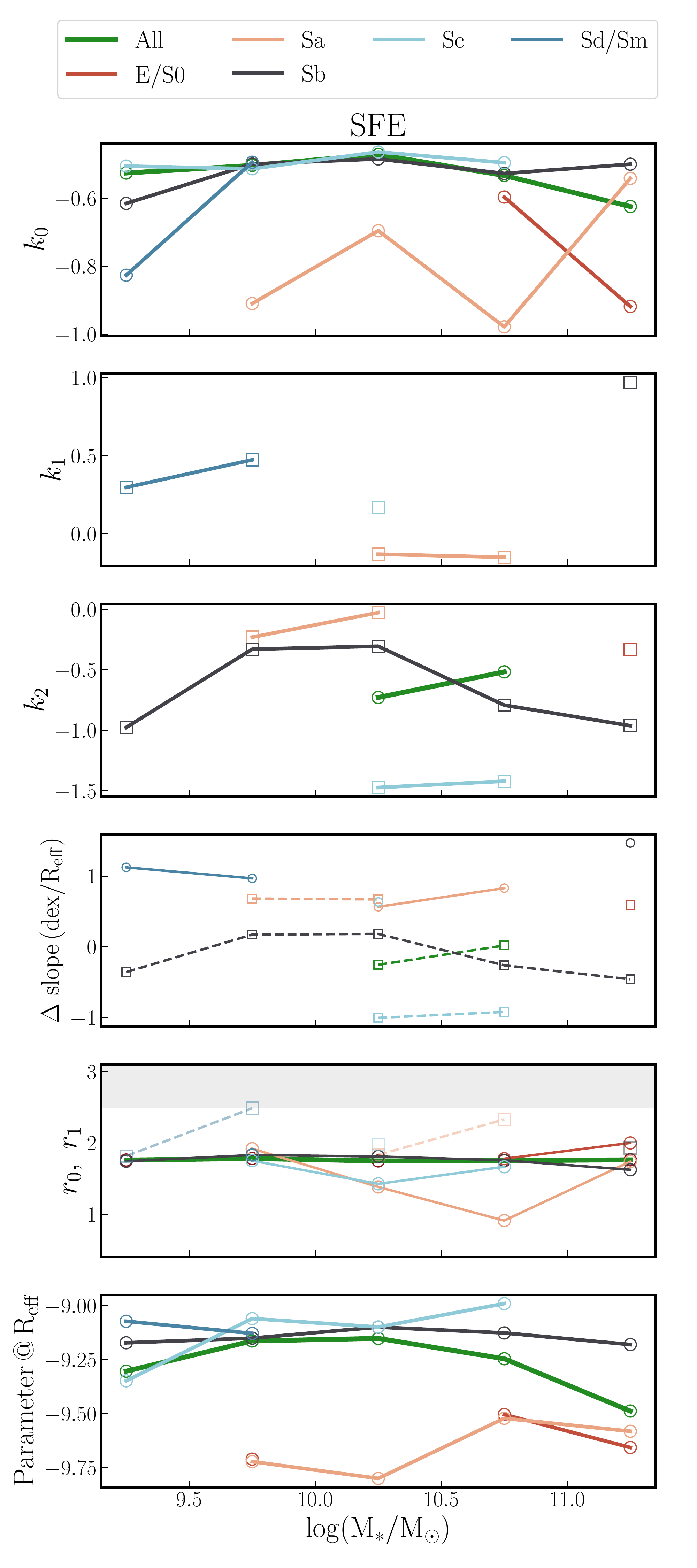}
\caption{Parameters derived from the piece-wise analysis for the radial distribution of the SFE. The layout of the figure is similar as Fig.\ref{fig:ML}.}  
\label{fig:SFE}    
\end{figure}

In Fig.~\ref{fig:SFE} we show the piece-wise analysis on the average of the radial distribution of the SFE segregated by morphology for different stellar mass bins (see Fig.~\ref{fig:SFE_rad}). Contrary to the gradients of sSFR, for the SFE the slopes for all the bins of \Mstar\ and morphology are negative (i.e., $k_0$< 0). Except for galaxies with the lowest mass, the value of this slope for late-type galaxies is relatively constant (\mbox{$k_0 \sim -0.5$ dex/\Reff}). On the other hand, for early-type galaxies the values of $k_0$ varies for different bins of \Mstar, nevertheless these slopes are significantly steeper than those derived from late-type galaxies. Although for most of the radial profiles a single gradient suffices to describe the radial trend of the SFE, for some morphological bins the piece-wise analysis detected a gradient at their outskirt ($k_2$, e.g., Sb galaxies). The value of this gradient depends on \Mstar\ with low and high masses having steeper gradients than the derived central values for the same mass bin. However, we note that this values may not be representative of the radial trend at this galactocentric distances as they measure the very outer part of these galaxies. Similar to the characteristic values of sSFR, we find that for the SFE these values are similar for late-type galaxies regardless the probed \Mstar\ (SFE \mbox{$\sim$ 10$^{-9.1}$~yr$^{-1}$}). For the early-type galaxies this characteristic value is \mbox{$\sim$ 0.6 dex} smaller in comparison to late-type galaxies for a similar mass-bin. The trends presented here are in agreement with previous studies exploring the radial distribution of the SFE using spatially resolved observations of the molecular gas \citep[e.g., ][]{Leroy_2008, Villanueva_2021}.

\subsection{Chemical Abundances}
\label{sec:Chem}

Thanks to the emission of different chemical species in the optical it has been possible to have an estimation of the amount of heavier elements than hydrogen or helium in the ISM. There is a plethora of chemical abundance calibrators in the literature, in particular of the oxygen abundance \citep[][and references therein]{Maiolino_2019A&ARv}. In this analysis we present the radial distribution of the oxygen abundances using a fiducial calibrator for this chemical abundance. However,  in Appendix ~\ref{app:OH_calib} we present the impact on the radial profiles using different calibrators. 



\subsubsection{Oxygen Abundance}
\label{sec:OH}
\begin{figure}
\includegraphics[width=\linewidth]{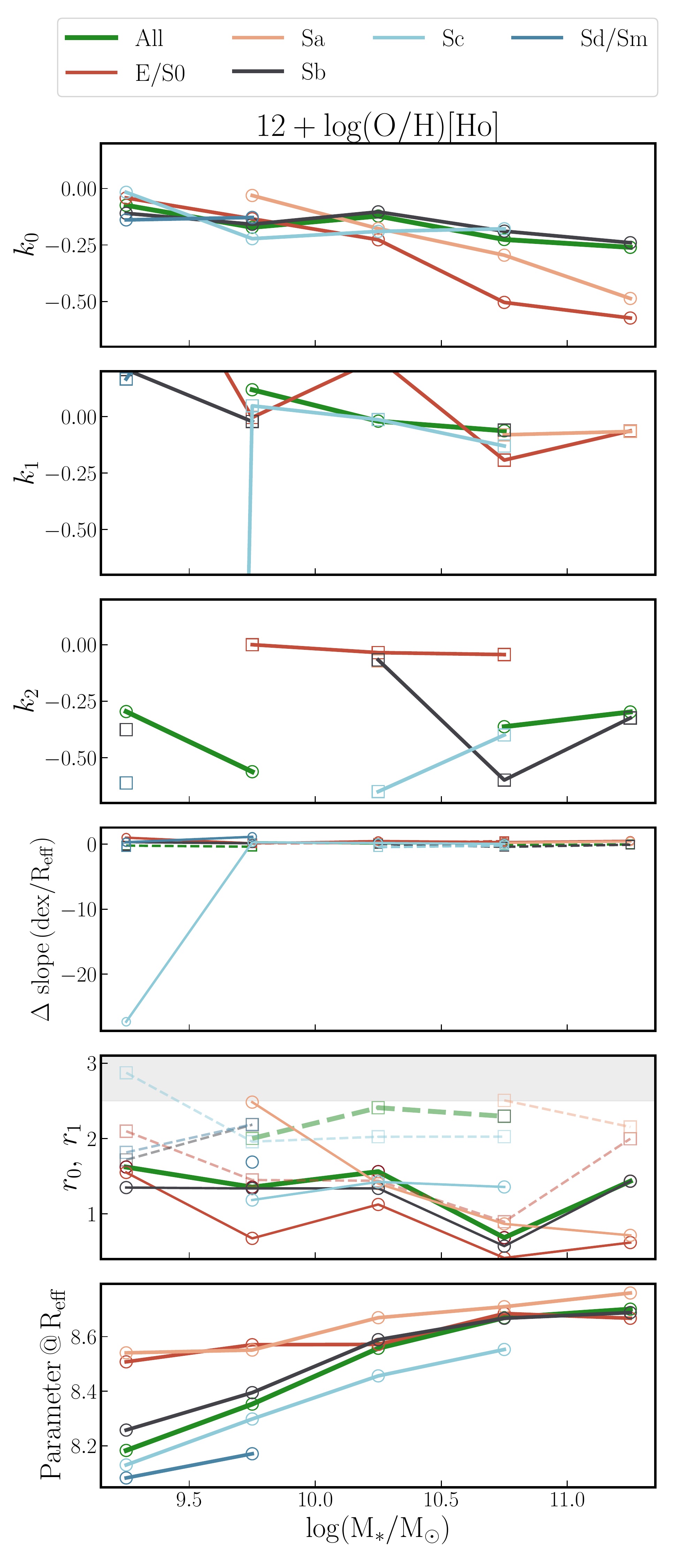}
\caption{The radial distribution of the oxygen abundance using the Ho calibrator. The layout of the figure is similar as Fig.\ref{fig:ML}.}  
\label{fig:OH_Ho}    
\end{figure}

The oxygen abundance has been widely used to gauge the chemical stage of galaxies as well as its evolution  \citep[e.g.,][and reference therein]{Maiolino_2019A&ARv}. As we mention above, there is a large amount of abundance calibrators in the literature using different methodologies (e.g., direct estimations, photoionization models, or hybrid ones). In this section we present the analysis of the radial distribution of the oxygen abundance using an empirical calibrator derived using a neuronal network analysis from \citet{Ho_2019}. Nevertheless, we present the same analysis of the radial distribution of the oxygen abundance using different calibrators in Appendix \ref{app:OH_calib}  \footnote{Oxygen and nitrogen abundances presented in this study are derived using the script \texttt{pyOxy} (\url{https://github.com/cespinosa/pyOxy}).}. When necessary, we briefly discuss the differences in the results using different calibrators. We note that the radial distributions of the oxygen and nitrogen abundances presented in this study have been derived using the entire radial distribution of the emission lines presented in Sec.~\ref{sec:Flux}, which may include contributions from ionization sources other than star formation.

In Fig.~\ref{fig:OH_Ho} we show the result of the piece-wise analysis of the radial distribution of the oxygen abundance for the MaNGA golden sample (see Fig.~\ref{fig:OH_Ho_rad}). We find that the central gradient of, $k_0$, is negative regardless the morphology or \Mstar. For late-type galaxies (from Sb to Sd/Sm) we find little changes of $k_0$ with respect to \Mstar\ with an average value of \mbox{$k_0 \sim -0.13$ dex/\Reff}. If any, we find a slightly flatter gradient for the lowest mass bin in comparison to massive ones. Interestingly, we find similar central gradients from early-type low-mass galaxies (\mbox{$\log(\Mstar/{\rm M}_\odot) \lesssim 10.5$}). For larger masses, early-type galaxies tend to have steeper gradients in comparison to late-type ones. For those radial profiles where we are able to measure an external gradient ($k_1$ and $k_2$) we find that these gradients range from positive to negative values. For the characteristic  oxygen abundance (measure at \Reff), we find that for late-type galaxies it increases with \Mstar. On the other hand, it is relatively constant for early-type galaxies \mbox{12+$\log({\rm O/H}) \sim$ -8.6}. Furthermore, massive early-type galaxies tend to have similar characteristic abundance than their late-type counterparts. 

These results are in partial agreement with previous results using the MaNGA dataset and others IFS data. On the one hand, we find more subtle variations of the slope ($k_0$) with \Mstar\ than those reported by \citet{Belfiore_2017}. Differences between that study and the work presented here are expected due to the differences in the samples (selection, sizes, etc) as well as the different oxygen calibrator adopted for each work. It could also be the case that the methodology for deriving the slopes of the gradients has an impact for these different works. In this work we employ a piece-wise analysis to take into account that the radial distribution could have different gradients at different galactocentric distances whereas \citet{Belfiore_2017} employed a linear fit with a single slope to describe the radial gradient of the oxygen abundance. 
On the other hand, regarding the morphology we find differences in the slope of the oxygen abundance for early-type galaxies only for massive ones. Similar results have been reported recently using a larger sample of MaNGA galaxies \citep{Boardman_2021}. Our results suggest, at least for this sample of galaxies and adopted calibrator, that the central gradient for late-type galaxies is rather constant for a wide range of \Mstar. 

The above results could vary significantly depending on the adopted calibrator. In Appendix~\ref{app:OH_calib} we present a similar analysis as the one derived above using other two different calibrators (the empirical calibrator O3N2 from \citealt{Marino_2013} and the theoretical one from \citealt{KK_2004}). In this appendix we describe the differences in the gradients depending on the used abundance calibrator. In general, we find that the adopted calibrator could have a significant impact in the derived gradients for different morphologies and stellar masses.  


\subsubsection{Nitrogen/Oxygen Abundances ratio}
\label{sec:NO}
\begin{figure}
\includegraphics[width=\linewidth]{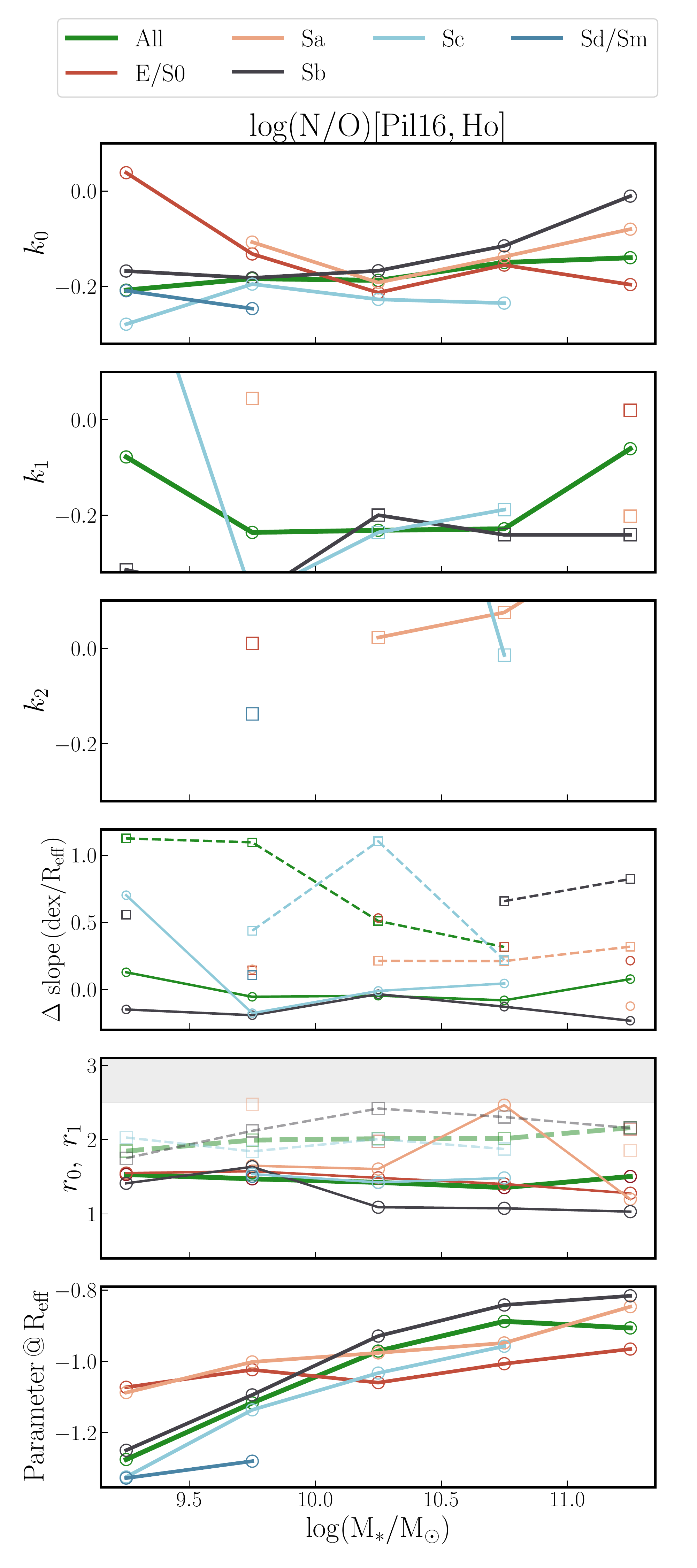}
\caption{The radial distribution of the N/O ratio. The layout of the figure is similar as Fig.\ref{fig:ML}.}  
\label{fig:NO}    
\end{figure}
Having a measurement of chemical abundances with different nucleosynthesis origins, or their ratios, allow us in principle to quantify the scales of star formation across the evolution of galaxies. On the one hand, the oxygen abundance measures the chemical enrichment from short-lived massive stars. On the other hand, the nitrogen abundance partially measures the enrichment to the ISM from stars with a lower mass and longer lifetimes. In this section we explore the radial distribution of their ratio in our sample of MaNGA golden galaxies. For the oxygen abundance we use the same calibrator described in the previous section whereas for the nitrogen abundance we use the calibrator derived by \citet{Pilyugin_2016}. 

In Fig.~\ref{fig:NO} we present the piece-wise analysis of the radial distribution of the N/O ratio for our sample of galaxies. Contrary to the oxygen abundance analysis in the previous section, we find a significant impact of both the stellar mass and morphology in setting the gradient of the N/O ratio in the central portion of our sample, $k_0$. For late-type Sb galaxies, $k_0$ increases with \Mstar. In other words, for this morphological type the central slope changes from negative to almost flat gradients as \Mstar\ increases. The flatness in the central region for the most massive Sb galaxies could be a consequence of the line ratios tracing diffuse ionized gas in regions dominated by a bulge, instead of tracing star formation. Thus, the measurements of the N/O ratio for the central region may not be reliable -- a similar scenario is likely occurring for E/S0 low-mass galaxies. However, we note that \citet{Espinosa-Ponce_2022} found similar results using a large sample of \textsc{Hii} regions with spectroscopic information. For Sc galaxies, $k_0$ is negative and similar regardless the stellar mass except for the lowest mass bin where the slope is steeper. For Sd/Sm galaxies the slope is similar as those derived from the Sc-type. For most of the bins of morphology and \Mstar\ where the piece-wise fit is able to measure $k_1$, we find that it is negative, slightly steeper than $k_0$. However, for those bins with three gradients we find that the outer one (i.e., $k_2$) is positive and significantly steeper. We consider that this may be an spurious artifact due to the lax cut in SNR. This affects the radial values of the N/O ratio at the outskirts of galaxies. According to our analysis, the change in slopes for the radial distributions occurs at $\sim$ 1.5 \Reff, regardless the stellar mass or morphology. The characteristic values of the N/O ratio at \Reff\, increases monotonically with \Mstar, regardless the morphology. However, for a given stellar mass bin, the average value for the Sb galaxies is the largest in comparison to other morphological types, among late-type galaxies. For early-type galaxies this characteristic value is relatively similar regardless \Mstar\ (\mbox{$\log({\rm N/O}) \sim -1.0$}). 

Considering only our late-type sample, our central gradients from the N/O ratio differ from those derived by \citet{Belfiore_2017} using a smaller sample of the MaNGA survey. Contrary to our results, they found that the gradient of the radial distribution decreases with \Mstar. As for the radial distribution of the oxygen abundance, these differences could be caused by the difference in the sample or in the adopted calibrators. Our results thus suggest that for late-type massive galaxies there is a flattening in the central gradient of N/O. On the other hand, using a large sample of \textsc{Hii} regions with spectroscopic information drawn from the CALIFA survey and the same abundances calibrators, \citep{Espinosa-Ponce_2022} found similar results as those derive in this study. In particular they also found flat gradients in the central part of massive Sb galaxies. This may suggest that even when using spectral information from \textsc{Hii} the dominant ionization mechanism detected could be do to diffuse ionized gas.

\subsection{ISM properties}
\label{sec:ism}
The ratio of the flux of different emission lines allow us to provide estimations of physical properties of the ISM. In this section we explore the radial distribution of two parameters, that are fundamental to understand the energetic of the ISM, the electron density, $n_e$, and the ionization parameter, $U$.   
\subsubsection{Electronic Density, $n_e$}
\label{sec:ne}
\begin{figure}
\includegraphics[width=\linewidth]{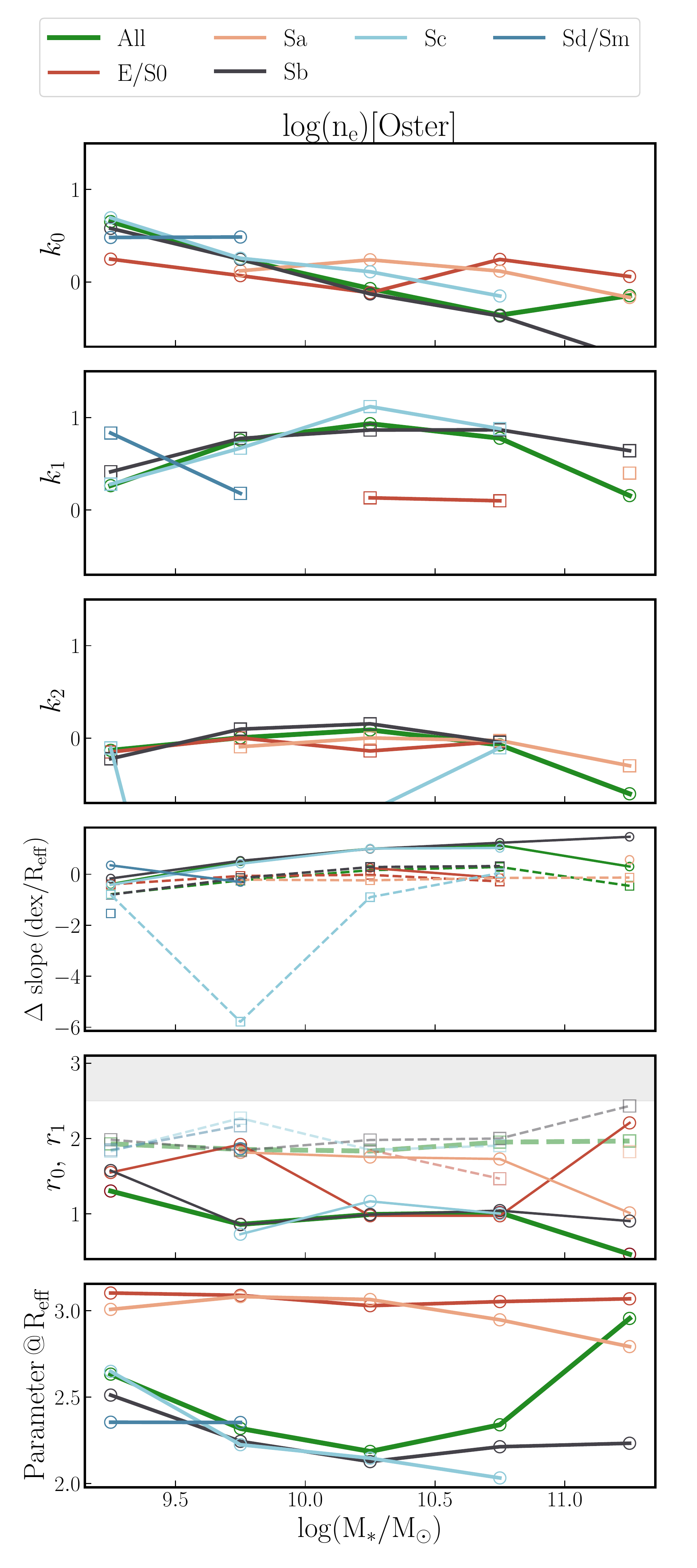}
\caption{The radial distribution of the electronic density derived from the Oster et al. calibrator. The layout of the figure is similar as Fig.\ref{fig:ML}.} \label{fig:ne}    
\end{figure}

Usually the electron density is estimated using the emission line ratio from a single ion. To gauge $n_e$, we follow \citet{Espinosa-Ponce_2022}. They derived this density using the [\ion{S}{ii}] doublet solving the equation: 
\begin{equation}
  \frac{[\textsc{Sii}]\lambda 6717}{[\textsc{Sii}]\lambda 6731} = 1.49 \frac{1+3.77x}{1+12.8x}
\end{equation}

\noindent where $x=10^{-4}n_{e}t^{-1/2}$ and $t$ is the electron temperature in units of $10^{4}$ K \citep{McCall_1985}. They assume a fiducial electron temperature expected for the usual conditions of an \ion{H}{ii} region ($t=10^{4}$ K). Although this doublet is sensitive only to a narrow range of densities \citep[$\sim$50 to $\sim$7000 cm$^{-3}$,][]{Osterbrock_2006}, it is still useful to gauge the radial distribution of $n_e$ in our sample of MaNGA galaxies. 

In Fig.~\ref{fig:ne} we plot the results of the piece-wise analysis of the radial distribution of $n_e$ for our golden sample (see Fig~\ref{fig:ne_rad}). We find that the central gradient of the radial distribution of $n_e$, $k_0$, changes significantly with \Mstar, going from positive to negative as the stellar mass increases. We find rather flat gradients for early-type galaxies. For those galaxies where we are able to estimate an external gradient, $k_1$, we find steeper positive values in almost all the bins of morphology and stellar mass. We find that  the characteristic electron density measure at \Reff\ for late-type galaxies at different stellar mass bins is relatively constant (\mbox{$n_e~\sim 10^{2.1}- 10^{2.5}$~cm$^{-3}$}). On the other hand, we find larger densities for massive early-type galaxies (\mbox{$n_e~\sim 10^{3.0}$~cm$^{-3}$}). Similar radial trends as those derived in this study have been reported using large spectroscopic data sets of \ion{H}{ii} regions \citep{Espinosa-Ponce_2022}. These trends have been attributed to the fact that denser material is located in regions of high pressure. Usually regions of high pressure are located at the central part of galaxies with little dependence on stellar mass or morphology \citep{Barrera-Ballesteros_2021a}. 

\subsubsection{Ionization Parameter}
\label{sec:U}
\begin{figure}
\includegraphics[width=\linewidth]{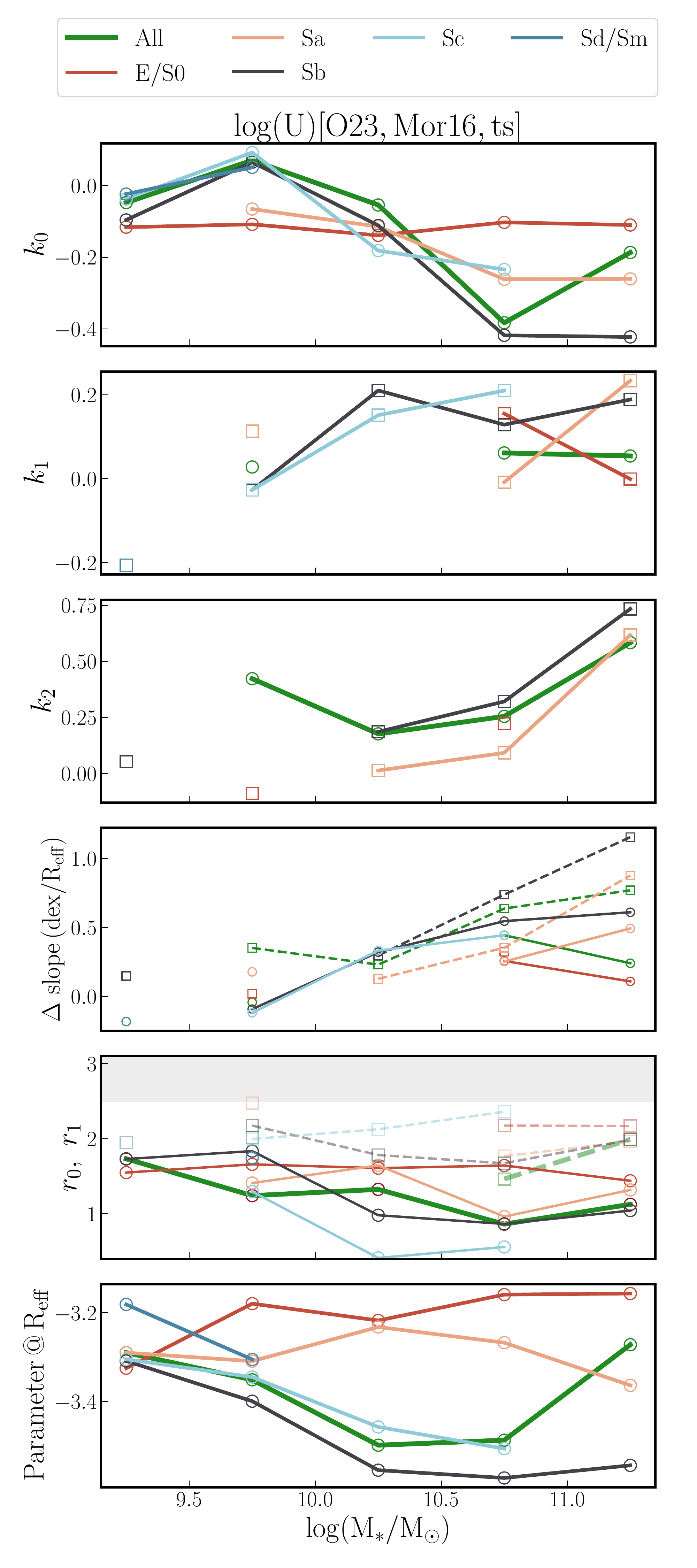}
\caption{The radial distribution of the ionization parameter derived from the calibrator presented by \citep{Morisset_2016}. The layout of the figure is similar as Fig.\ref{fig:ML}.}  
\label{fig:U_Mor16}    
\end{figure}

The ionization parameter ($U$) measures the ratio between the number of ionizing photons and the  number of atoms of hydrogen.  
Despite its importance, it is rather difficult to estimate $U$ observationally. Although it is usually gauged using emission line ratios from a given element (e.g., [\ion{O}{iii}]/ [\ion{O}{ii}]), this parameter depends on the geometry of the explored regions, the hardness of the ionizing spectra, among other properties. For this study, we estimate the ionization parameter and the above line ratio. We follow the relation derived from recent photoionization models from \citet{Morisset_2016}. 

In Fig.~\ref{fig:U_Mor16} we show the results from the piece-wise analysis from the radial distribution of the ionization parameter in our sample (Fig.~\ref{fig:U_rad}). In contrast to the early-type galaxies where the central gradient has small variations ($k_0 \sim -0.1 dex/\Reff$), late-type galaxies have significant variations of their central gradients for different bins of stellar mass. Late-type galaxies in the lowest mass bin tend to have a similar negative gradient as those derived for early-type galaxies whereas for the next mass bin (\mbox{$\log(\Mstar/{\rm M}_\odot) \lesssim 9.7$}) we find that galaxies have a positive gradient. For the next two mass bins, late-type galaxies have negative gradients with $k_0$ decreasing as \Mstar increases. For the most massive bin (where it is only possible to measure the central gradient from Sb galaxies) $k_0$ remains with a similar negative slope as in the previous mass bin. Regardless the stellar mass or the morphology, we find positive steep gradients in the outskirts of the galaxies in our sample. We find that the characteristic value of $U$ measured at \Reff\ decreases with \Mstar\ for late-type galaxies whereas for early-type objects this value of $U$ is relatively constant for different bins of \Mstar\ (\mbox{$U \sim 10^{-3.2}$ }). Qualitatively our results are in partial agreement to those derived using a sample of \ion{H}{ii} regions from the CALIFA survey \citep{Espinosa-Ponce_2022}.

\section{Kinematic Properties}
\label{sec:kin_prop}

%
\begin{figure*}
\includegraphics[width=\linewidth]{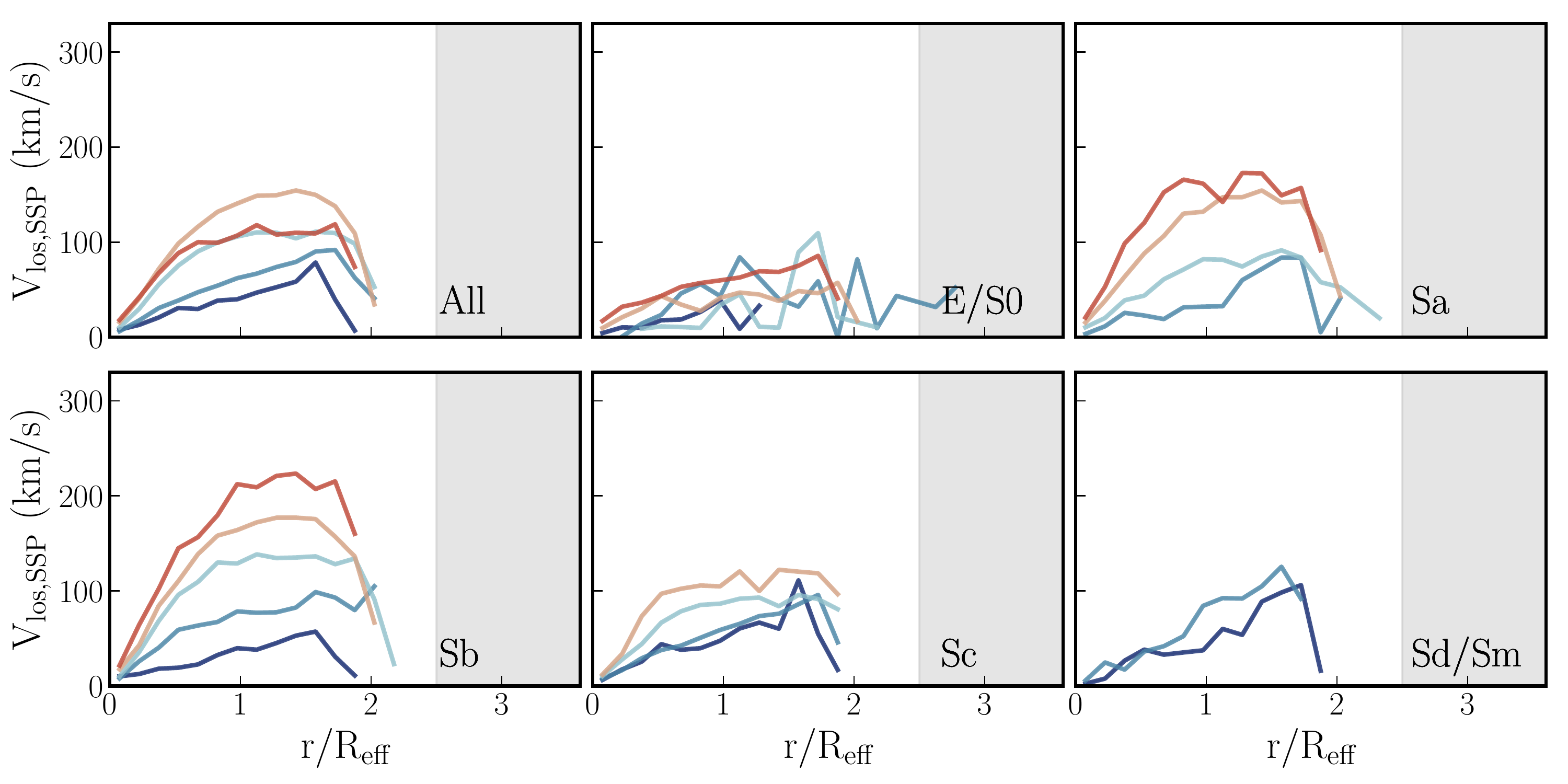}
\caption{The average radial distribution of the line-of-sight stellar velocity, \mbox{$\mathrm{V_{los, SSP}}$}. Similar to the distribution of Fig.~\ref{fig:Sstar_rad}, the gradients are average by stellar mass and morphology. In each panel, each solid lines represents the average profile per total stellar mass bin.}  
\label{fig:velssp}    
\end{figure*}
\begin{figure*}
\includegraphics[width=\linewidth]{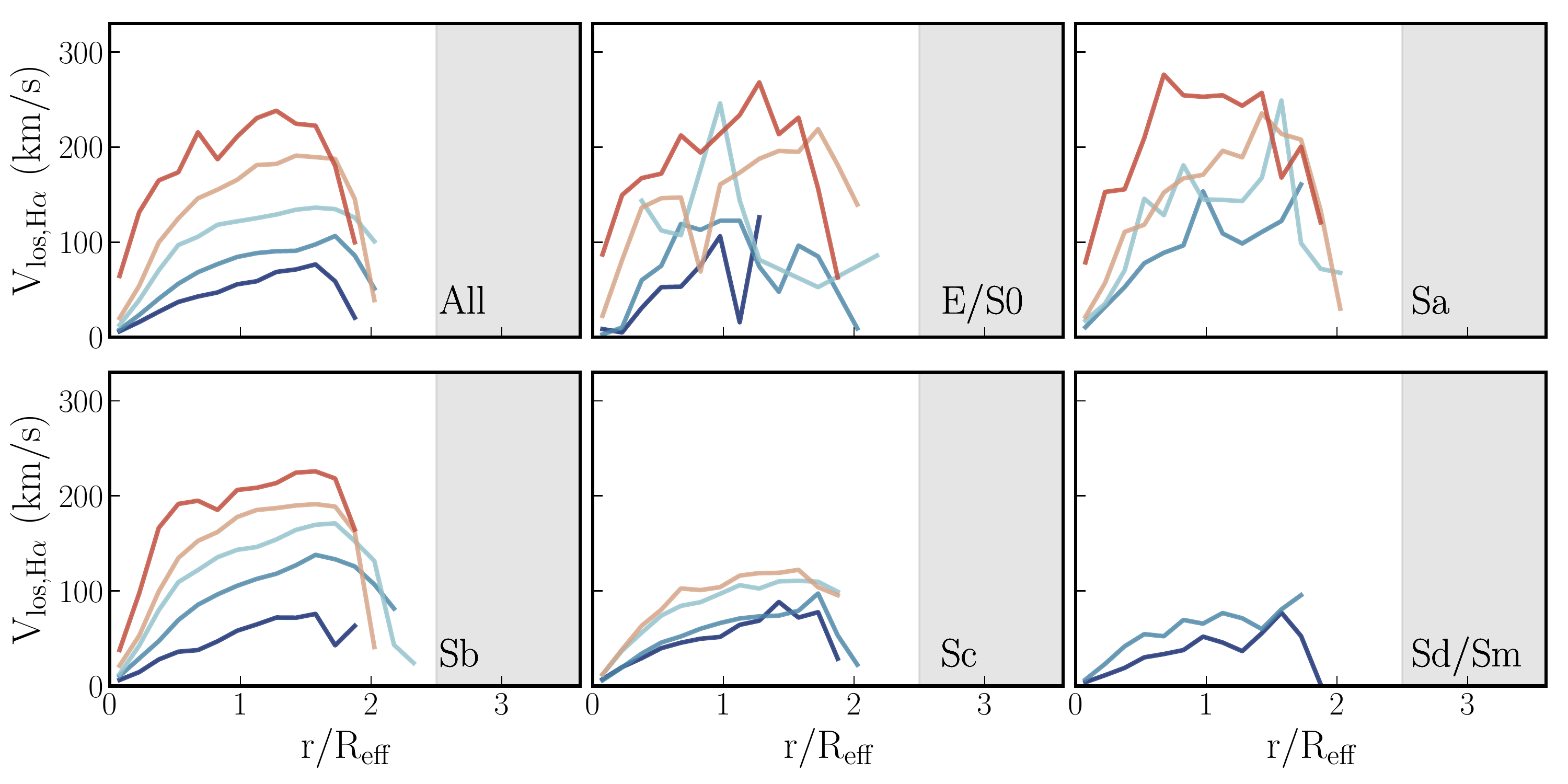}
\caption{The radial distribution of the line-of sight ionized gas velocity, \mbox{$\mathrm{V_{los, H\alpha}}$}. The layout of the figure is similar as Fig.\ref{fig:velssp}.}  
\label{fig:velgas}    
\end{figure*}
\begin{figure*}
\includegraphics[width=\linewidth]{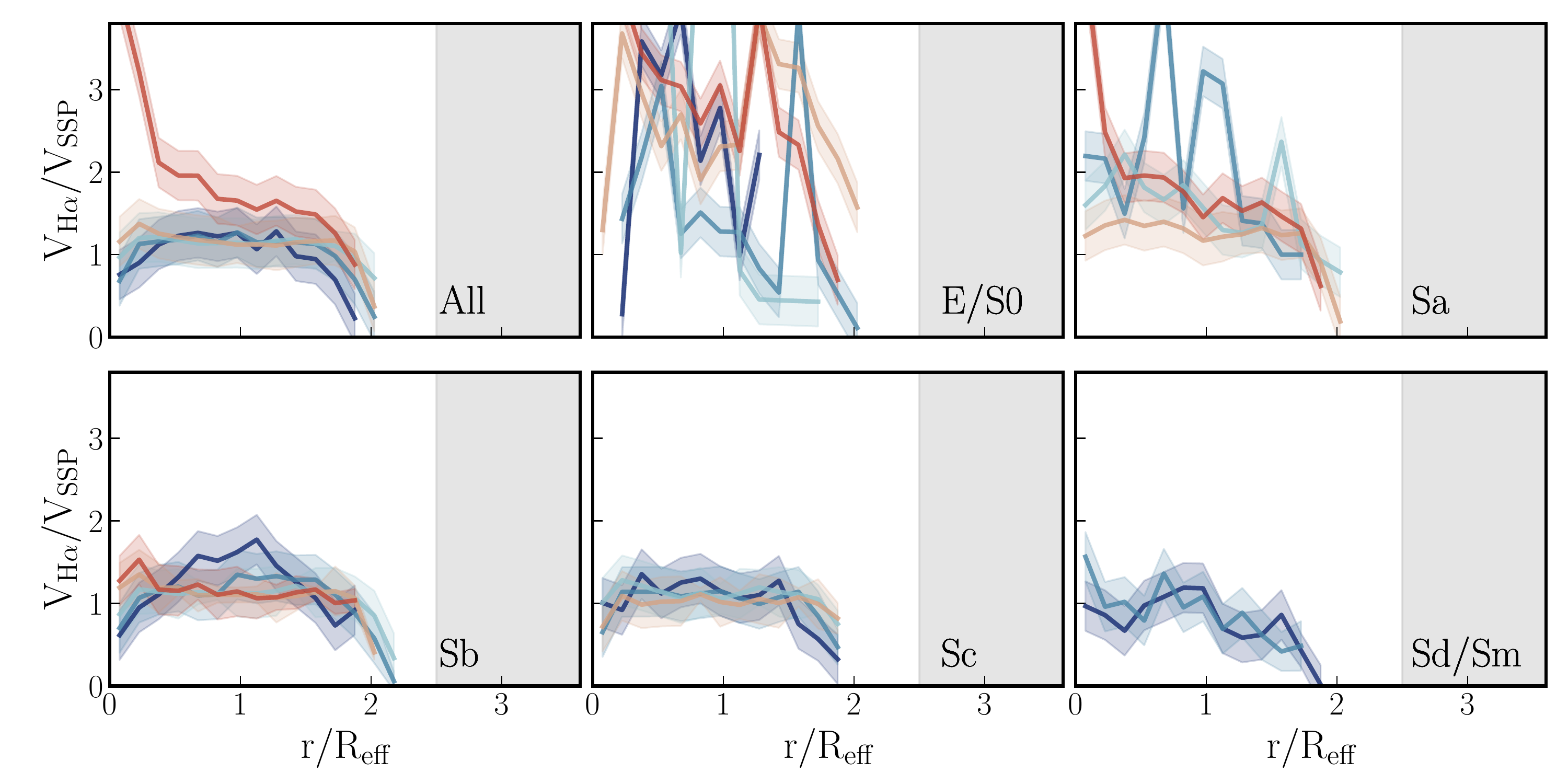}
\caption{The ratio of the radial profiles presented in Figs.\ref{fig:velgas} and \ref{fig:velssp}. The shaded area in the gradients represent the standard deviation of the radial distribution in each radial beam.  The layout of the figure is similar as Fig.\ref{fig:velssp}.}  
\label{fig:velrat}    
\end{figure*}

To provide a reliable estimation of the properties of the stellar component from the continuum, the SSP fitting technique must be able to also provide an estimation of at least the first two moments of the line-of-sight velocity distribution (LOSVD) from the stellar component. In a similar fashion, the emission-line analysis should also provide an estimation of at least the two moments of the LOSVD of the ionized gas. Given the fact that the kinematic radial profiles follow different trends an their shape is usually fitted by a non-linear function \citep[e.g., ][]{Lopez-Coba_2017,Barrera-Ballesteros_2018}, in this section we will provide qualitatively description of those kinematic features.  

In Fig.~\ref{fig:velssp} we plot the radial distribution of the first moment of the LOSVD, \Vlosspp, averaged for different stellar mass and morphologies. To create these radial profiles of \Vlosspp\, for each galaxy and each radial bin, we averaged the absolute values of velocity from the receding and the approaching sides of the stellar velocity field. 
The shape of the \Vlosspp\, profiles segregated only by stellar mass depends strongly on it (top left panel of Fig.~\ref{fig:velssp}): massive galaxies exhibit a rising profile with a flattening at $\sim$ 1.5 \Reff, whereas low-mass galaxies shows a monotonically increasing of \Vlosspp. Furthermore, the absolute values of each radial profile increases with the stellar mass (e.g., measuring \Vlosspp\ at \Reff), except for the most massive bin where the radial profile \Vlosspp\ is below the profile from the second most massive bin. We also find that, regardless \Mstar, at large galactocentric distance \Vlosspp has a significant drop. These drops could be indicating the maximum radii at which it is possible to have a reliable measurement of the rotational curve for the galaxies, beyond that radii the measurements the SNR of the continuum may not be enough to have a proper estimation of  \Vlosspp. For consistency with the radial analysis of the other stellar properties we do not attempt any further selection on the radial profiles of \Vlosspp.  
Segregated by morphology and stellar mass, the radial profiles of \Vlosspp\, reveal interesting features. For E/S0 galaxies (middle top panel of Fig.~\ref{fig:velssp}) we find strong radial variations of \Vlosspp\ for different  bins of \Mstar\ -- except for the most massive bin, where we observe a monotonic increment of \Vlosspp\ with radius. These strong variations could suggest that early-type galaxies are supported by random motions instead of ordered ones, however when we plot the radial distribution of the \lamssp\ parameter, we find that these galaxies appear to be supported by these two components. We will come to this point below. For Sa galaxies we find a similar trend for the radial distributions of \Vlosspp\  as those observed using the entire GS for different bins of \Mstar; the gradient of \Vlosspp\ becomes steeper as \Mstar increases, only the lowest probed mass bin shows a monotonically increment of \Vlosspp, while the rest of profiles show a plateau at large galactocentric distances. 
For late-type galaxies (Sb, left bottom panel of Fig.~\ref{fig:velssp}) we find that the shape of the radial profile of \Vlosspp\, varies for different stellar masses. Massive galaxies show a steeper gradient, and a flattening at large radii in comparison to low-mass galaxies that exhibit a monotonically increase of \Vlosspp. For Sc galaxies (middle bottom panel of Fig.~\ref{fig:velssp}) we observe a monotonically increment of \Vlosspp\ with radius with little dependence on the stellar mass. As for the other stellar properties, for Sd/Sm galaxies we are able to measure the radial gradients of \Vlosspp\, only for low-mass galaxies. Thus the radial profiles of \Vlosspp\ show a monotonically increment with no signatures of a plateau.  

In comparison to \Vlosspp, we find similar trends in the radial distribution of the velocity derived for the ionized gas, \VlosHa\, (Fig.~\ref{fig:velgas}). Averaging the gradients only by stellar mass, the increments for massive galaxies are stepper in comparison to low-mass galaxies. Similar trends are also observed when we average \VlosHa\, for morphology and \Mstar bins in comparison to those derived for \Vlosspp. Although the  radial average are similar, we note that \VlosHa\, shows larger velocities in comparison to \Vlosspp.

To further quantify these differences, we plot in Fig.~\ref{fig:velrat} the radial distribution of the \VlosHa/\Vlosspp\ ratio using the radial profiles of \VlosHa and \Vlosspp presented in Figs~\ref{fig:velgas} and \ref{fig:velssp}, respectively. We find interesting features for the radial profiles of this ratio. When segregating the entire GS only by \Mstar we find that the most massive \Mstar\ bin shows a significant increment of \VlosHa\ in comparison to \Vlosspp across the optical extension of the galaxies, even more the radial profile of the \VlosHa/\Vlosspp\ ratio spikes at the central region of these galaxies with \mbox{\VlosHa/\Vlosspp\, $\sim$ 4} . For the rest of the stellar mass bins, the shape and values of this ratio is relatively similar. For the very center of galaxies the ratio is slightly below or close to 1, as the galactocentric distance increases this ratio remains relatively constant with \mbox{\VlosHa/\Vlosspp\ $\sim$ 1.2}, for radii larger than r/\Reff\ $\sim$ 1.5, this ratio drops to values close to zero. For early-type galaxies, this ratio strongly varies across the radial extension of the galaxies, therefore it is not possible to provide a reliable description of ratio. For massive Sa galaxies (\mbox{$\log(\Mstar/{\rm M}_\odot) $<$ 10.5$}) we find that the radial profiles of the \VlosHa/\Vlosspp\ ratio have similar shapes and values as those derived for the entire sample. Sa galaxies with smaller stellar mass exhibit strong variations in their radial profiles of their velocity ratio. Except for the lowest-mass bin of Sb galaxies, these galaxies show a rather constant \VlosHa/\Vlosspp\ ratio across their optical extension \mbox{\VlosHa/\Vlosspp\ $\sim$ 1.4}. Low-mass galaxies exhibit a rising profile reaching a peak of \mbox{\VlosHa/\Vlosspp\ $\sim$ 2} at r/\Reff\ $\sim$ 1.2, after this radius the ratio decreases to values close to zero. For Sc galaxies the radial distribution of this ratio is relatively constant for the different mass bins (\mbox{\VlosHa/\Vlosspp\ $\sim$ 1}). Finally, for the Sd/Sm galaxies the \VlosHa/\Vlosspp\ ratio decreases as radius increases -- we note that we are only able to estimate this ratio for low-mass galaxies. Although the ratio at the center of these galaxies is larger than 1, it decreases reaching values close to zero at their outskirts. This analysis highlights the well-known interplay between the morphology and kinematic structure of galaxies \citep[e.g., ][]{Cappellari_2016ARAA}.  For Sa galaxies, where the \VlosHa/\Vlosspp\, ratio is the largest,  \VlosHa\, traces ionized gas probably located in the mid-plane of the galaxy whereas \Vlosspp\, is coupled with non-ordered motion in the line-of sight. On the other hand, for late-type galaxies both stars and ionized gas traced similar velocities, as both trace the ordered motions from the disk. 

\begin{figure*}
\includegraphics[width=\linewidth]{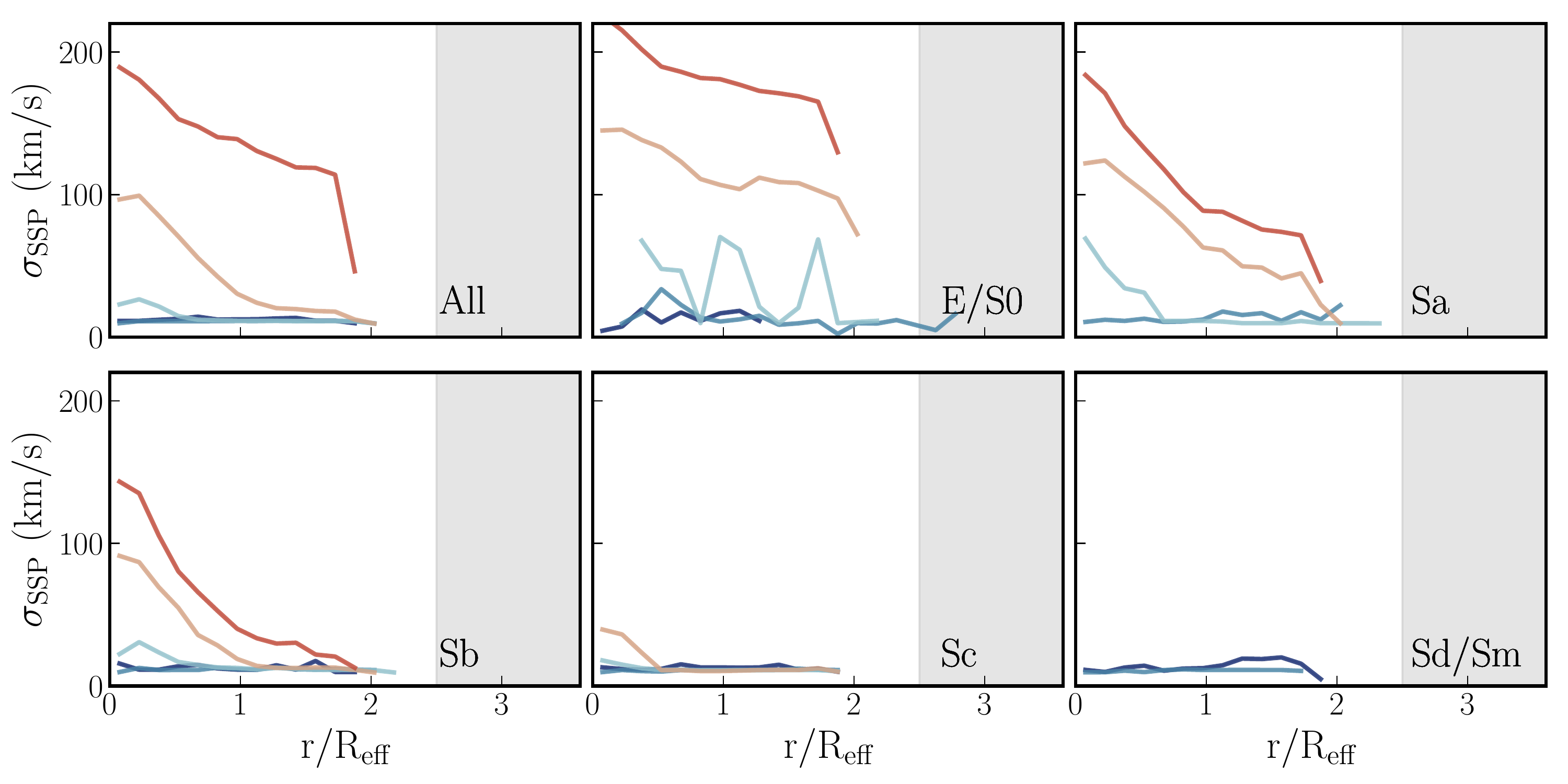}
\caption{The radial distribution of the stellar velocity dispersion, \mbox{$\mathrm{\sigma_{SSP}}$}. The layout of the figure is similar as Fig.\ref{fig:velssp}.}
\label{fig:dispssp}    
\end{figure*}
\begin{figure*}
\includegraphics[width=\linewidth]{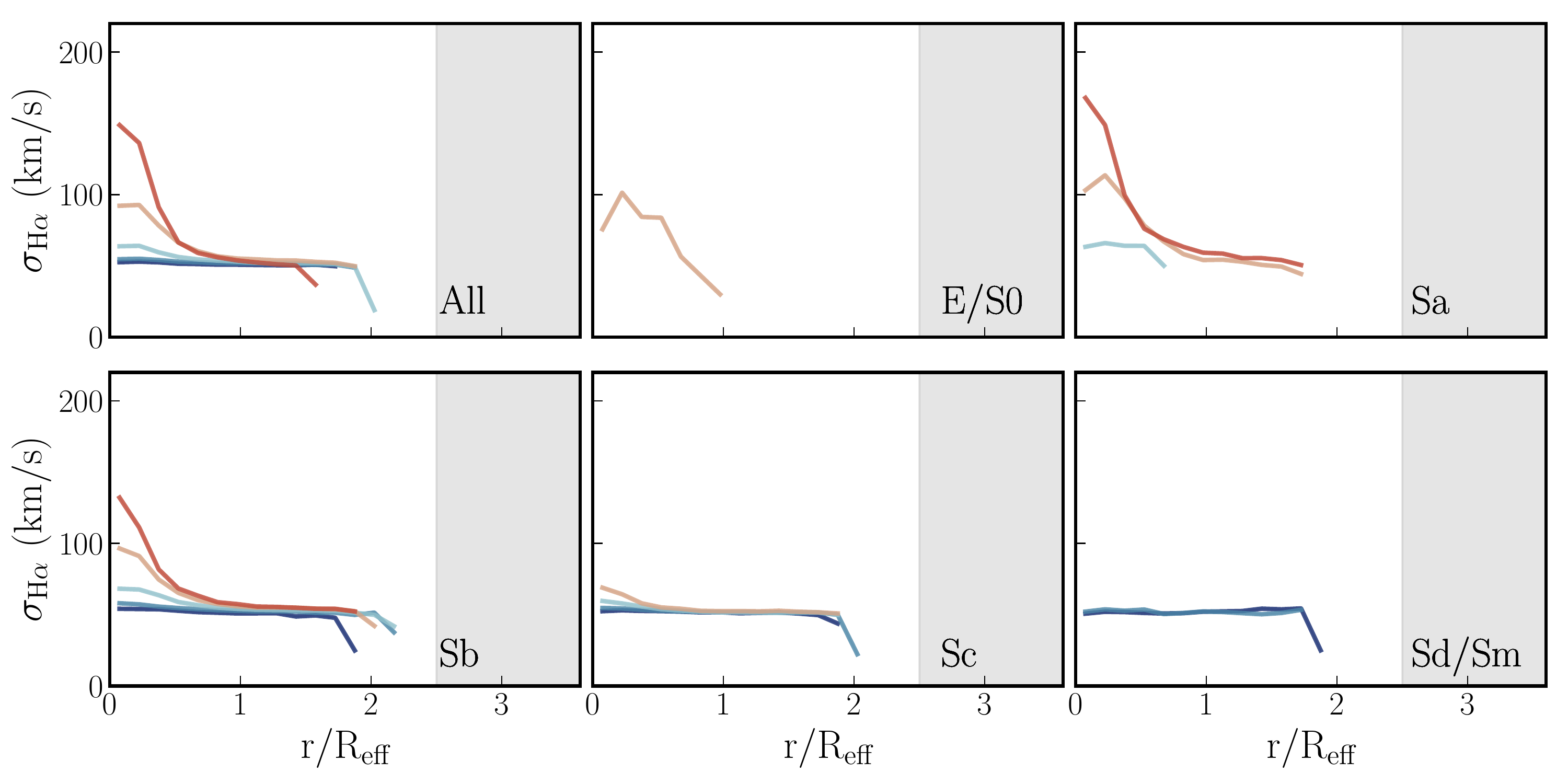}
\caption{The radial distribution of the stellar velocity dispersion, \mbox{$\mathrm{\sigma_{H\alpha}}$}. The layout of the figure is similar as Fig.\ref{fig:velssp}.}  
\label{fig:dispgas}    
\end{figure*}
\begin{figure*}
\includegraphics[width=\linewidth]{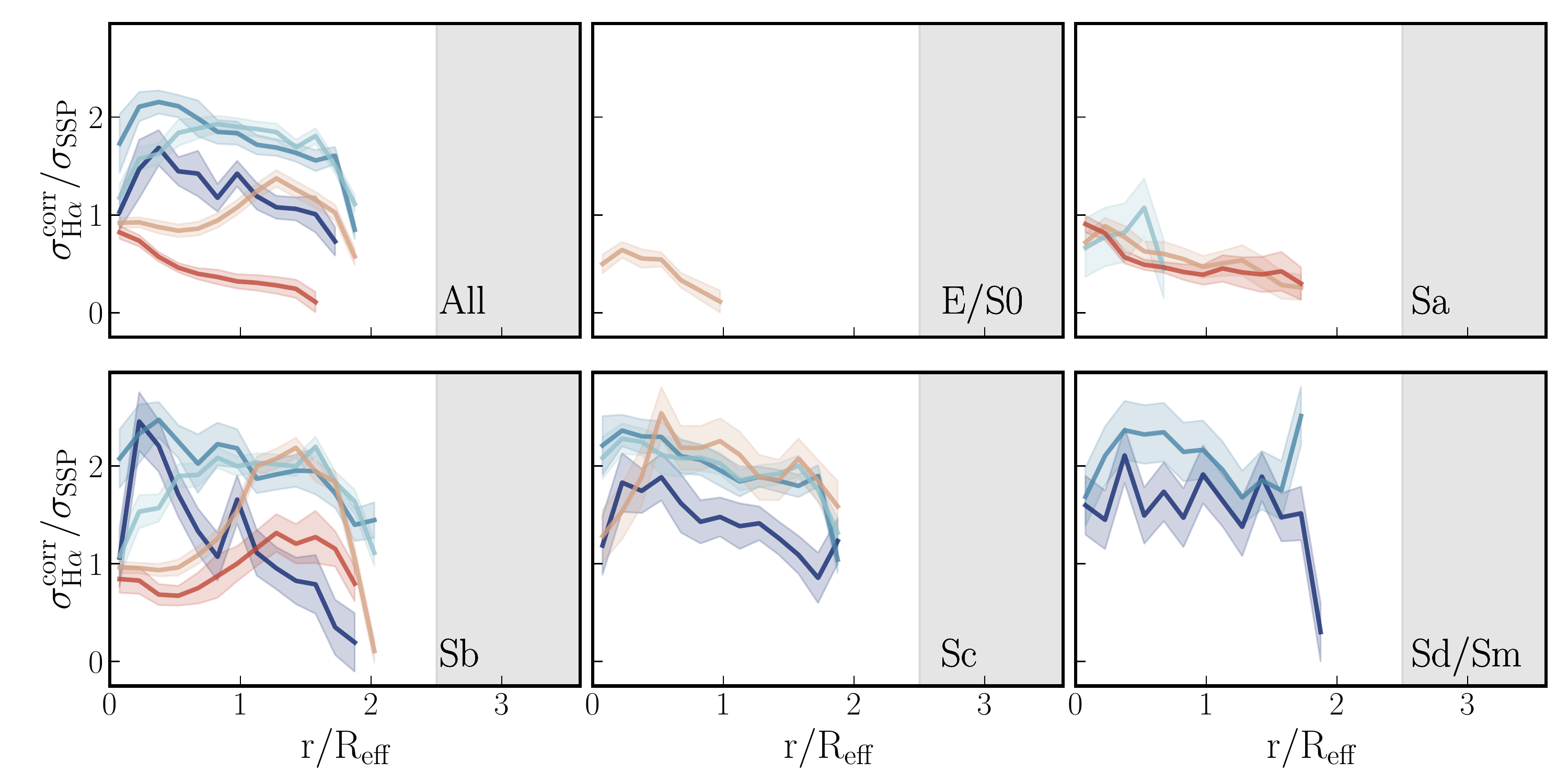}
\caption{The radial distribution of the \slosGASc/\slosSSP\, ratio. The layout of the figure is similar as Fig.\ref{fig:velssp}.}  
\label{fig:disprat}    
\end{figure*}
\begin{figure*}
\includegraphics[width=\linewidth]{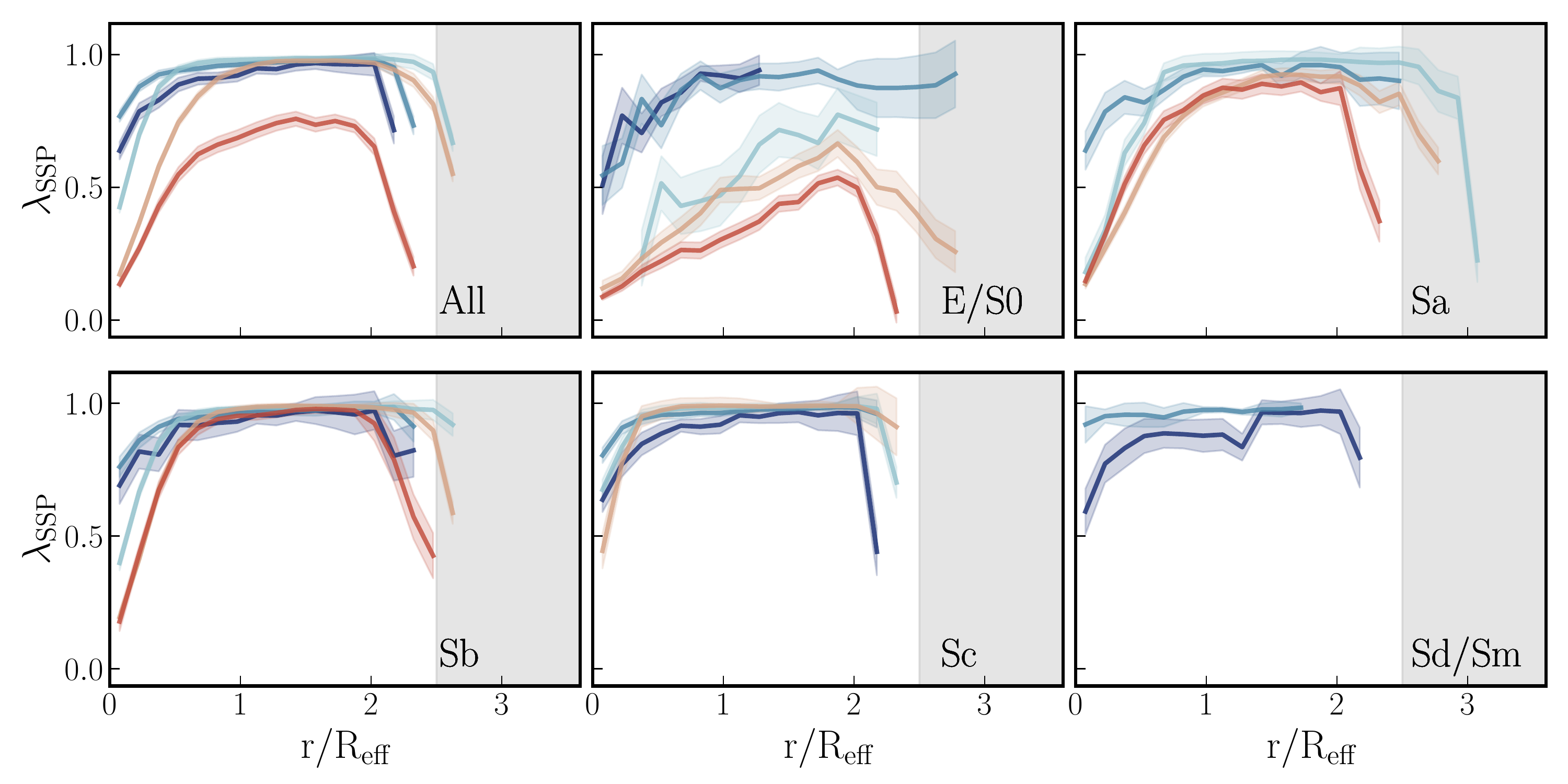}
\caption{The radial distribution of the $\lambda$ parameter for the stellar component. The layout of the figure is similar as Fig.\ref{fig:Sstar}.}  
\label{fig:lambda}    
\end{figure*}
The SSP as well as the emission line analysis provides an estimation of the second momentum of the LOSVD, also known as the stellar and ionized gas velocity dispersions, (\slosSSP, \slosGAS, respectively). In Fig.~\ref{fig:dispssp} we plot the radial distribution of \slosSSP\, for different stellar masses and morphologies. The top left panel of this figure shows the large impact that stellar mass has in the radial profiles of \slosSSP. Massive galaxies ($\MstarMsun > 11.0$, dark red profile) have the largest values of \slosSSP. The central velocity dispersion for these massive galaxies reach the largest value (\mbox{$\sim$ 200 km s$^{-1}$}) and their average radial profile decreases with galactocentric distance. For the lower stellar mass bin (\mbox{$10.5 < \MstarMsun < 11.0$}, yellow profile), the shape of \slosSSP\, radial profile is similar although its central value is significantly smaller than the one derive for the most massive galaxies, and for large radius it reach at flat gradient close to 10 km s$^{-1}$.  For the remaining stellar mass bins the radial profiles are basically flat with a constant velocity dispersion of \mbox{$\sim$ 10 km s$^{-1}$}. For massive E/S0 galaxies \slosSSP\ decreases with radius, the central dispersion for the most massive galaxies is the largest of the entire sample (\mbox{$\sim$ 250 km s$^{-1}$}). For low-mass E/S0 the radial profiles are flat with values close to 0 km s$^{-1}$. Sa galaxies show a similar trend as those described for E/S0 galaxies, although the drop of \slosSSP\ with radius is sharper for massive Sa galaxies than their E/S0 counterparts. For Sb galaxies the trends are similar than for Sa galaxies, however the drops for massive galaxies is even sharper than for Sa galaxies. For Sc and Sd/Sm galaxies we  report flat gradients of \slosSSP with dispersions close to 0 km s$^{-1}$. 


Similar to \slosSSP, in Fig.~\ref{fig:dispgas} we plot the radial distribution of \slosGAS. We find similar trends for different stellar masses and morphologies in comparison to those trends reported above for \slosSSP. However we note two significant differences. On the one hand, the flattening of the \slosGAS\, radial profiles occurs at higher velocity dispersion than for \slosSSP. This is expected since the estimation of \slosGAS\, is significantly limited by the spectral resolution of the instrument \citep[\mbox{$\sim$ 45 km s$^{-1}$},][]{Law_2021}. On the other hand, we find smaller values of \slosGAS\, in comparison of \slosSSP as well as sharper drops of the velocity dispersion for massive galaxies. To account for these difference we follow a similar analysis as for the line-of-sight velocities, namely, we derive the ratio \slosGASc/\slosSSP, where \slosGASc, considers a correction in quadrature to \slosGAS\, in order to account for the instrumental resolution (\mbox{$\slosGASc^2 = \slosGAS^2 - \sigma^2_{ins}$}), with $\sigma_{ins}$ = 45 km s$^{-1}$ (Sanchez et al., submitted). 

In Fig.~\ref{fig:disprat} we show the radial distribution of the above ratio. When the sample is segregated only by stellar mass (top left panel), we observe different trends for the different stellar mass bins. For low and intermediate-mass galaxies we find that the radial profiles \slosGASc/\slosSSP $>$ 1, suggesting that \slosGAS\ across these galaxies is larger than \slosSSP. Furthermore the radial profiles of this ratio increase and peak up to certain galactocentric distance (r/\Reff\ $\sim$ 0.5), beyond this radius this ratio decreases distance. On the other hand, for the most massive galaxies \slosGASc/\slosSSP\ $<$ 1 for all the radial bins. Although the central value is close to one, this ratio decreases with radius reaching values close to zero. For the intermediate stellar mass bin (\mbox{$10.5 < \MstarMsun < 11.0$}, yellow profile) the central portion of the ratio is rather flat with \slosGASc/\slosSSP $\sim$ 1, however at r $\sim$ \Reff this ratio increases to $\sim$ 1.5, suggesting that for galaxies of that \Mstar, the dispersions are similar in their central portion with an increment of \slosGAS\, at larger radii in comparison to \slosSSP. 

For early-type galaxies (top middle panel of Fig.~\ref{fig:disprat}), we are able to make this comparison only for one stellar mass bin (\mbox{$10.5 < \MstarMsun < 11.0$}, yellow profile). This profile shows only \slosGASc/\slosSSP\ $<$ 1 with values decreasing with radius. As expected, \slosSSP\ is larger for E/S0 galaxies in comparison to \slosGASc. For Sa galaxies the radial distributions of \slosGASc/\slosSSP\ decreases with radius, however all the values are below 1 -- except for the lowest mass bin sampled, although the values are highly variant. For these galaxies as for the E/S0 sample, \slosSSP\ dominates over \slosGASc. For late-type Sb galaxies we find different trends for different mass bins (bottom left panel of Fig.~\ref{fig:disprat}). The most massive Sb galaxies show central values of \slosGASc/\slosSSP\ slightly smaller than 1, however as the galactocentric distance increases \slosGASc/\slosSSP\ has values above one for radii beyond 1.5 \Reff. This change in tendency suggest that although for the central portion of this massive Sb galaxies \slosSSP\ is larger than \slosGASc, at larger radii the situation is the other way around, turbulent motions of the ionized gas component dominate of the ones from the stellar component. The lower mass bin of Sb galaxies (yellow profile) have a similar yet more evident radial trend of the \slosGASc/\slosSSP\ ratio in comparison to red profile. For distances smaller than the \Reff\ this ratio is 1, however it increases reaching values close to 2 around 1.5 \Reff. Regardless the galactocentric distance, we find for intermediate mass bins that the \slosGASc/\slosSSP\ ratio is larger than 1. For Sb galaxies with \mbox{$10.0 < \MstarMsun < 10.5$} (cyan profile), we find that this ratio increases with radius up to \Reff, beyond that point the ratio is relatively flat reaching a constant value of \slosGASc/\slosSSP\ $\sim$ 2.  For Sb galaxies with \mbox{$9.5 < \MstarMsun < 10.0$}, the dispersion ratio decreases with distance from $\sim$2 to 1. For Sb galaxies, the radial profile with the sharpest variation of this ratio is the one corresponding for the lowest mass bin (\mbox{$9.5 < \MstarMsun$}, dark-blue profile). For Sb galaxies, these objects have the largest \slosGASc/\slosSSP\ ratio in their center, however as the galactocentric distance increases this ratio decreases reaching values close to zero at their outskirts. For Sc galaxies, the \slosGASc/\slosSSP\ ratio is larger than 1 regardless the probed stellar mass and radii. In general, for the probed bins of \Mstar\ we find that this ratio decreases with distance; with values close to 2 in the center to values close to 1 in their outskirts. Finally, for the Sd/Sm galaxies we also find that this ratio is larger than 1 for the probed bins of \Mstar (low-mass galaxies) and radii. The values of this ratio are relatively constant for different radii the \slosGASc/\slosSSP\ $\sim$  1.5 - 2. Thanks to the spatially resolved information of the MaNGA survey, we are able to probe the kinematic stage of the demographics in the nearby universe by comparing the kinematic properties of the stellar and the ionized gas. Furthermore, these radial profiles are also useful to explore dynamical differences in these components. For instance, by measuring the differences between \Vlosspp\ and \VlosHa\ as well as \slosSSP\ and \slosGASc\ we are able to explore how each of these components trace the potential well of the galaxies.

In Fig.~\ref{fig:lambda} we plot the radial distribution of the so-called spin parameter for the stellar component \lamssp\, for different stellar masses and morphologies. This parameter, which varies between 0 and 1, gauges what sets the kinematic structure of a galaxy \citep{Cappellari_2004}. Values of \lamssp\, close to 1 suggest that the region/galaxy is rotational supported whereas values close to zero suggest that the region/galaxy is supported by non-ordered motions. In general \lamssp\, increases with distance, although the central shape of this increment varies depending on the stellar mass and morphology. Segregated only by stellar mass, the central value of \lamssp\, decreases with \Mstar. This parameter increases steeply with distance, reaching the value of 1 at small galactocentric distances for most of the stellar mass bins -- except for the most massive bin, where the radial profiles of \lamssp\ reach a plateau at \mbox{\lamssp\, $\sim$ 0.7}. For E/S0 galaxies the radial distributions of \lamssp\ vary significantly depending on \Mstar. For low-mass galaxies the radial distribution rise very steeply becoming almost flat with values close to 1. For intermediate-mass and massive galaxies the radial profiles of  \lamssp\, monotonically increases. We note that the two more massive bins shows a strong drop of \lamssp\ at their outskirts. As we mention above, kinematic properties such as \Vlosspp\ may not be reliable at large radii due to our conservative cuts in SNR. These results suggest that for galaxies with this morphology the stellar mass plays a major role in setting the kinematic stage of the stellar component; low-mass E/S0 galaxies appear to be kinematically supported by rotation while massive E/S0 galaxies are supported by random motions or rotation, depending on the location. For the rest of morphological bins the trends are similar for different bins of \Mstar. Most of the profiles exhibit an almost flat radial distribution close to 1 -- except for the most massive galaxies. Other than the E/S0 morphological bin, we are able to measure the radial profile of \lamssp\ for the most massive stellar bin only for Sa and Sb galaxies. In these morphological bins the radial profile of the most massive bin shows a sharp drop for their central region, suggesting that for these galaxies their centers are supported by random motions while the rest of the galaxy is supported by ordered motions. These results highlights the fact that, in comparison to the morphology, \Mstar\ appears to play a major role in setting the radial distribution of \lamssp.


\section{Radial profiles from the entire MaNGA sample}
\label{sec:ALL}

As we mention in Sec.~\ref{sec:sample}, in order to provide a reliable estimation of the radial distribution of the physical properties, we select from the entire MaNGA sample ($\sim$ 10000 targets) those galaxies that satisfy several criteria including good radial coverage from the fiber bundle as well as reliable spatial sampling and resolution. This selection criteria yields what we refer in this study as the {\it Golden Sample, GS} ($\sim$ 13 \% of the entire MaNGA sample). In this section we derive the radial distribution of the physical properties derived above for a much larger sample of MaNGA galaxies. The only selection criteria for this analysis is low-inclined galaxies (i.e., the major/minor axis ratio has to be smaller than 0.45). This criteria yields a sample of $\sim$ 7500 galaxies. Evidently, the entire sample provides a much better coverage on \Mstar\, and morphology. Since the GS provides those galaxies with the best spatial coverage and resolution, here we only provide a qualitatively comparison between the radials profile from the GS and those derive from the entire sample. We refer the reader to visit the following web page where we include the same analysis performed above for the entire sample: \url{http://ifs.astroscu.unam.mx/MaNGA/Pipe3D_v3_1_1/rad_profiles/}

{\bf M/L ratio, and \Sstar:} In general, the radial profiles of the $M/L$ ratio for the entire sample show similar negative trends as those derived for the GS (see Figs.~\ref{fig:ML} and \ref{fig:ML_rad}). As for the GS, the central gradient of the radial profiles of $M/L$ ratio depends mainly on \Mstar. Furthermore, the characteristic value of this ratio strongly depends on the morphology. Similarly, for the entire sample the radial profiles of \Sstar\ decrease with radius as for the GS. However, we note sharp drops at large radii for low-mass galaxies regardless the morphology. On the other hand, contrary to the GS, the characteristic value of \Sstar\ for the entire sample appears to depends strongly on both \Mstar\ and morphology. 

{\bf Stellar age, metallicity, and extinction:} The central gradients of the stellar age are slightly flatter for the entire sample in comparison to the GS. As for \Sstar, we note that for some bins of morphology and \Mstar\, the age shows a sharp drop at the outskirts of galaxies. For the gradients of stellar metallicity, although we find similar negative central slopes in both the complete and the golden sample, for the entire sample we find step positive external gradients in different bins of morphologies and stellar mass. As for other measurements such as those derived from the emission lines, we are finding that for those galaxies with those positive gradients at the outskirts the continuum flux is dominated by the noise, therefore those values of metallicity are note reliable. Finally, for the optical extinction, we find that the gradients of the entire sample are similar to the GS, however they are slightly smoother, this is that the gradients at the outskirts tend to be similar as those derived in the central part of galaxies.  

{\bf Emission lines fluxes, EW(\ha), and line ratios:} Although the general behavior of the slopes of the radial gradients for the different emission lines (\ha,\hb, \nii, and \oiii) for the entire sample is similar to the one measured from the GS (i.e., negative gradients), we find that regardless the emission line, the central gradients for early-type galaxies are similar in comparison to those derived from late-type galaxies. We recall that in Sec.~\ref{sec:Flux} we find that the early-type galaxies show sharper negative gradients in comparison to late-type for the probed bins of \Mstar. In contrast to the GS, for the entire sample it is more difficult to observe the difference between the characteristic flux measured at the \Reff\ for early-type galaxies than for late-type ones. 
For the radial distribution of the EW(\ha) using the entire sample, we find similar trends as those measured by the GS both in slopes and characteristic radius. As we discuss in the next section, we consider that those variations in the gradients are caused mainly by the impact of the bulge. 
Regarding the radial distribution of the emission line ratios (\ha/\hb, \nii/\ha, and \oiii/\hb), we find similar trends with respect to the GS as well as similar characteristic values. Other properties derived from the emission line ratios for the entire sample such as \avgas, the \avgas/\avssp\ ratio, \SmolAv, \fgas, show similar radial distributions as those derived from the GS.


{\bf \Ssfr, sSFR, and SFE:}  The radial distribution of the \Ssfr\ for the entire sample is similar to the one derived for the GS. We find a negative gradient across the entire extension of the galaxies. Although we note some deviations from this gradient depending on the stellar mass and morphology (e.g., almost flat gradient for the lowest-mass bin of early-type galaxies). The characteristic values of \Ssfr\ measured at \Reff\ varies depending on both, \Mstar\, and the morphology: it usually increases from early to late type galaxies for a given bin of \Mstar.
Similar to the GS, the entire sample shows an almost flat gradient for the sSFR (although at large radii some profiles show drops), regardless the stellar mass and morphology. As for the GS, there is a clear segregation between the sSFR measure at \Reff\ for early and late-type galaxies: early galaxies have significant smaller sSFR in comparison to late-types for the same stellar mass bin.
For the radial distribution of the SFE we find a characteristic negative gradient regardless the stellar mass and morphology. However, we find significant variations of the SFE measured at \Reff for both stellar bins of \Mstar\ and morphology. 

{\bf Oxygen abundance, N/O ratio, electron density, and ionization parameter:} The radial distribution of the oxygen abundance using the entire sample, and the Ho calibrator \citep{Ho_2019}, is very similar as the one derived from the GS adopting the same calibrator. However, the piece-wise analysis provides a better description of the radial profiles when using the entire sample in comparison to the GS. For the entire sample, the slope of the central gradient decreases, $k_0$ with \Mstar. It is important to note, that the extension of the central gradient also varies depending on \Mstar. The extension where the piece-wise analysis detects $k_0$ is significantly smaller for massive galaxies in comparison to galaxies with lower mass. When comparing the large extension of the galaxies, the slope of the radial distribution is similar for different bins of \Mstar -- except for the lowest-mass bin, where the slope is flatter in comparison to other mass bins. The characteristic oxygen abundances are the same regardless the sample. These results are in agreement with those presented previously in the literature using the same sample of galaxies \citep[e.g., ][]{Barrera-Ballesteros_2016, Boardman_2021}. We also find similar distributions for the N/O ratio, the electron density, and the ionization parameter  for both samples. These results suggest that the GS is a representative sample of the entire MaNGA sample regarding the estimation of radial properties of the ionized gas. 

{\bf Stellar and ionized gas kinematics:} For the entire sample we find that the radial distribution of \Vlosspp\ is similar to the one derived for the GS for most of the bins of morphology and \Mstar. We only find significant differences for E/S0 low-mass galaxies, instead of strong radial variations in \Vlosspp, we find that their radial profile is flat an close to zero km~s~$^{-1}$. Similar to \Vlosspp, we find common trends between the entire sample and the GS for the radial distribution of \VlosHa. As above, the entire sample provides a smoother radial profiles of \VlosHa\, for E/S0 galaxies in comparison to those derived for the GS. From the comparisons above it is clear that the radial distribution of the \VlosHa/\Vlosspp\ ratio is similar when using the entire sample than the GS. Nevertheless we note,from the entire sample, that the most massive E/S0 galaxies are those responsible for the large ratio observed for the entire population of massive galaxies (e.g., top left panel of Fig.\ref{fig:velrat}). Although, the radial distribution of both \slosSSP, and \slosGAS\ are similar for both the entire and the GS, we find significant difference when we compare the radial distribution of the \slosGASc/\slosSSP\ ratio between the two samples. Contrary to the GS, the radial distribution of this ratio when using the entire sample segregated only by \Mstar is relatively flat and close to 1. Similar trends are observed for the Sb galaxies. Finally, we find similar radial trends for \lamssp between the entire sample and the GS. The above results show that the most significant difference, kinematically speaking, when using a sample that provides the best conditions to derive radial profiles  and the entire MaNGA sample is observed in their velocity dispersion of both components. This could indicate that rather than affecting the systemic velocity of the galaxies using a large sample of galaxies average a non-linear property such as the velocity dispersion.  
\section{Discussion}
\label{sec:Disc}

Along this study we explore the radial distribution of the different physical properties that can be derived from the optical spectra (i.e., the stellar and ionized gas properties).  We use a piece-wise analysis to account for possible variations on the slopes of the gradients from those radial distributions. In general, we find that for those parameters that related to a  to an absolute property, a gradient with a single slope usually suffices to describe them (e.g., \Sstar, \avssp, emission-line fluxes). On the other hand, we find that some relative properties a gradient with different slopes is needed to provide a good representation of the radial distribution of those parameters (e.g., M/L, EW(\ha), emission-line ratios). We also find that the election of a gradient with a single or several slopes for the radial distribution of a given property also depends on both the morphology and the stellar mass: for a significant fraction of parameters the slopes from early-type galaxies differ from those measured from late-type ones. This is also valid for the absolute values of the radial distributions (measured by their characteristic value at \Reff).  

One of the physical properties that best exemplifies the need of using gradients with different slopes to describe its radial distribution is the EW(\ha) -- see the piece-wise analysis and the radial distribution in Figs. \ref{fig:EW} and \ref{fig:EW_rad}, respectively (see also the line ratios \nii/\ha\ and \oiii/\hb). As we describe in Sec.~\ref{sec:Flux}, the sign of the slopes vary from positive to negative for the central and external part of the radial distribution of the EW(\ha), respectively. This transition of gradients is evident for Sb galaxies, regardless the total stellar mass. On the other hand, early-type galaxies (E/S0, Sa) show a flat gradient regardless \Mstar. As we mention in Sec.~\ref{sec:Flux} the EW(\ha) measures the star-formation activity. Thus, regions with an EW(\ha) $\lesssim$ 6\AA\ correspond to a ionization source different than star formation (e.g., diffuse ionized gas, DIGs, from  HOLMES), whereas larger values of EW(\ha) are associated to star-formation; the larger the value of EW(\ha) the more star-formation activity is occurring. 

The change in the slopes for Sb galaxies reflects what we consider is the impact of the galaxy's structure in particular the presence of a bulge in the center of galaxies. For instance, the central value of EW(\ha) from the most massive Sa galaxies is similar to those derived from E/S0 and Sa galaxies (i.e., below 6\AA) suggesting that even though the galaxy has a late-type morphology in its center it has similar properties as an early-type galaxy. As the galactocentric distance increases the EW(\ha) increases reaching a peak around $\sim$ 1.2\Reff. This could indicate a composite stage where DIGs and star-forming regions cohabit and as the galactocentric distance increases the star-formation increases overcoming the contribution from DIGs reaching a maximum contribution at $\sim$ 1.2\Reff. For galactocentric distances larger than \mbox{$\sim$ 1.2\Reff}, EW(\ha) decreases with radius reaching again values below 6\AA. The slope of this decrease is similar to the negative slope derived from other late-type galaxies (Sc and Sd/Sm). This indicates that for further distances the properties of an Sb galaxy resemble those expected from late-type galaxies. Our results thus suggest that particular type of galaxies -- such as the Sb galaxies -- can be considered as a composition between an early-type galaxy in their center and a disk one in their outskirts.  

Using IFS data, different works have suggested a similar scenario for galaxies with bulges. Using a photo-spectral decomposition of galaxies \citet{Mendez-Abreu_2019} obtain the spectra for individual structural components of S0 galaxies in particular their bulges and disks \citep[see also,][]{Johnston_2017, Mendez-Abreu_2021}. They found that indeed the physical properties of these galaxies are different between their center and their outskirts. For instance, the properties from the ionized gas in the central region resembles those from early-type galaxies whereas the properties from the outskirts are consistent with those derived from late-type galaxy. In other words, our results suggest that for those galaxies where we find significant differences in the slopes of their gradient is due to the fact that the physical properties are different across their optical extension. Furthermore, for bulge galaxies our results also agree with the scenario in which bulges (or central parts of the galaxies) were formed at early ages of the universe either by monolithic collapse or major mergers whereas the outskirts of galaxies were likely formed after the formation of the bulge via different evolutionary channels (e.g., gas accretion or wet minor-mergers). Moreover, for the entire sample of galaxies our results also support the scenario in which galaxies form in an inside-out fashion.

\section{Summary and Conclusions}
\label{sec:DisConc}

Using the MaNGA sample (the largest IFU sample up to date, with $\sim$ 10000 galaxies), we present one of the most comprehensive explorations of the radial distribution from physical properties derived from both, the stellar continuum and the ionized gas emission lines in the optical (including their main kinematic properties). From the entire sample we select a so-called {\it Golden Sample}, in other words, we select the closest targets with the best spatial coverage ($\sim$1400 galaxies). Given the size of the sample we are able to disentangle the impact of two fundamental global properties: the total stellar mass, \Mstar, and the morphology. To quantify the gradients of those radial distributions, we make use of a piece-wise analysis allowing us to measure changes in the slope of those radial profiles as well as its characteristic value (i.e., measured at \Reff). The second allows to quantify how the absolute values of a given property changes depending on either the stellar mass or the morphology. We also explore how these radial distributions vary when considering a larger samples of galaxies at different distances (or physical spatial resolution) and with different spatial coverage ($\sim$ 7000 galaxies).  

In general, we find that most of the physical properties from both components decreases with distance (e.g., \Sstar, and \ha\ flux) with \Mstar\ and the  morphology modulating their gradient as well as their characteristic values for some observables. Here we summarize the main results from this study for the different properties derived from the stellar continuum and ionized gas emission lines using the  {\it Golden Sample}:

\begin{itemize}
    \item The stellar mass surface density, \Sstar, as well as the stellar mass-to-light ratio, $M/L$, decrease with radius. Their slopes and characteristic  values at \Reff\ become steeper and larger as \Mstar\ increases. Although morphology does not seem to significantly affect these slopes, the characteristic values for early-type galaxies are larger in comparison to late-type ones for a given bin of \Mstar. These results are in agreement with those derived for a heterogeneous sample of galaxies \citep{Sanchez_2021}.
    
    \item Both the luminosity-weighted stellar age and metallicity show in general negative central slopes regardless \Mstar\ and morphology. Although these gradients are close to flat, we find a mild trend with \Mstar, with the slopes becoming stepper as \Mstar increases. The characteristic values from both properties increases with \Mstar. Early-type galaxies are older and more metal rich in comparison to late-type galaxies of similar \Mstar. The central slope of the radial distribution of the stellar optical extinction, \avssp, is close to zero  for late-type galaxies and positive for early-type ones; external slopes are positive. The characteristic stellar extinction is significantly affected by morphology; late-type galaxies have larger values of \avssp in comparison to early-type object of similar \Mstar . 
    %
    
    \item The radial distributions of the flux from the brightest emission lines have a negative slope, with similar values for the probed lines. We do not find significant differences in their slope for different stellar mass or morphology. However, depending on the emission line, their fluxes at \Reff\ depends on \Mstar\ and morphology. 
    
    \item From these emission lines we derive the radial distribution of their ratios. Depending on the ratio, the slope of the gradients can depend on both \Mstar\ and morphology. The \ha/\hb\, line ratio allows us to estimate the optical extinction, \avgas\ which in turn allows us to estimate the radial distribution of the molecular gas mass density, \SmolAv. We also present the radial distribution of properties derived from \avgas: \avgas/\avssp, and \fgas. As explored by previous spatially resolved studies \citep{Li_2021}, we find that the radial distribution of the  \avgas/\avssp ratio is not constant, but decreases with radius. Furthermore, the slope becomes steeper as \Mstar increases. On the other hand, \fgas\ has positive gradients regardless \Mstar\ or the morphology. 
    
    \item The slopes of the \Ssfr\ gradients are negative with a similar value regardless \Mstar\ or the morphology. On the other hand, the central slopes of the sSFR radial distribution have a mild variation from negative to positive whereas the outer gradient have negative slopes. The slope of the gradient of SFE have negative values. For these three parameters, late-type galaxies have large characteristic values in comparison to early-type galaxies at similar \Mstar.  
    
    \item We find that in general the radial gradients from the central portion of the oxygen abundance derive from emission-line calibrators have negative slopes. The exact value of those gradients depend on the calibrator. Contrary to previous studies, we do not find a strong impact of \Mstar\ on those gradients for late-type galaxies. The values of the oxygen abundance at \Reff\ strongly depend on both \Mstar\ and the adopted calibrator. 
    
    \item The MaNGA dataset allow us to measure the radial distribution of the line-of-sight velocity from the stellar and ionized gas component. We find that, for both components, massive galaxies have a steeper increase in \Vlos; these galaxies reach the expected plateau at smaller galactocentric distances in \Vlos\ than low-mass galaxies. Morphology also have a significant role in shaping those radial profiles: late-type galaxies have steeper gradients than those derived from early-type galaxies. On the other hand, the velocity dispersion from both components decrease with radius for the most massive galaxies, whereas for the low-mass galaxies the profiles is flat and close to 0 km/s.
    
    \item Using the kinematic information from each component we explore the radial distribution of the \VlosHa/\Vlosspp\ and \slosGASc/\slosSSP\ ratios. Although the \VlosHa/\Vlosspp\ is relatively constant -- close to unity -- for different bins of \Mstar\ and morphology, we find that in the center of massive galaxies this ratio reaches values larger than 3. This is due to the large central value of \VlosHa\ in comparison to \Vlosspp\ in early-type galaxies. On the other hand, the radial distribution of the \slosGASc/\slosSSP\ ratio varies significantly depending on both \Mstar\ and the morphology. 
    
    \item In general, we find that the radial properties derived from the \emph{Golden Sample} for the stellar and ionized gas components are similar to those derive from a larger low-inclined sample ($\sim$7500 galaxies). 
    
\end{itemize}

The radial distributions observed from the different parameters presented in this study are a significant evidence that the main processes responsible to shape the formation and evolution of galaxies appears to occur at kpc scales. These radial profiles are the result of the scaling relations derived at kpc scales, thus at the same galactocentric distance the physics that regulates the properties of both the stellar component, the ionized gas, and the interplay between them is similar. Furthermore, these radial distributions suggest an inside-out formation scenario. Global properties such as the stellar mass or morphology play a secondary role in setting the local properties of galaxies in the local universe.    

\section*{Acknowledgments}

J.B-B acknowledges support from the grant IA-101522 (DGAPA-PAPIIT, UNAM) and funding from the CONACYT grant CF19-39578. L.C. thanks the support from the grant IN103820 (DGAPA-PAPIIT, UNAM). 
This research made use of Astropy,\footnote{http://www.astropy.org} a community-developed core Python package for Astronomy \citep{astropy:2013, astropy:2018}. 

Funding for the Sloan Digital Sky Survey IV has been provided by the Alfred P. Sloan Foundation, the U.S. Department of Energy Office of  Science, and the Participating  Institutions. 

SDSS-IV acknowledges support and  resources from the Center for High Performance Computing  at the University of Utah. The SDSS 
website is www.sdss.org.

SDSS-IV is managed by the  Astrophysical Research Consortium 
for the Participating Institutions 
of the SDSS Collaboration including 
the Brazilian Participation Group, 
the Carnegie Institution for Science, 
Carnegie Mellon University, Center for 
Astrophysics | Harvard \& 
Smithsonian, the Chilean Participation 
Group, the French Participation Group, 
Instituto de Astrof\'isica de 
Canarias, The Johns Hopkins 
University, Kavli Institute for the 
Physics and Mathematics of the 
Universe (IPMU) / University of 
Tokyo, the Korean Participation Group, 
Lawrence Berkeley National Laboratory, 
Leibniz Institut f\"ur Astrophysik 
Potsdam (AIP),  Max-Planck-Institut 
f\"ur Astronomie (MPIA Heidelberg), 
Max-Planck-Institut f\"ur 
Astrophysik (MPA Garching), 
Max-Planck-Institut f\"ur 
Extraterrestrische Physik (MPE), 
National Astronomical Observatories of 
China, New Mexico State University, 
New York University, University of 
Notre Dame, Observat\'ario 
Nacional / MCTI, The Ohio State 
University, Pennsylvania State 
University, Shanghai 
Astronomical Observatory, United 
Kingdom Participation Group, 
Universidad Nacional Aut\'onoma 
de M\'exico, University of Arizona, 
University of Colorado Boulder, 
University of Oxford, University of 
Portsmouth, University of Utah, 
University of Virginia, University 
of Washington, University of 
Wisconsin, Vanderbilt University, 
and Yale University.

\bibliography{sample631}{}
\bibliographystyle{rmaa}
\appendix

\section{Radial Profiles and Gradients}
\label{app:profiles_grads}

In this Appendix we include the averaged radial profiles for the properties explored in this study. As we mention in Sec.~\ref{sec:grad_profiles}, we present these radial profiles for different bins of morphology and stellar mass. 
\begin{figure*}
\includegraphics[width=\linewidth]{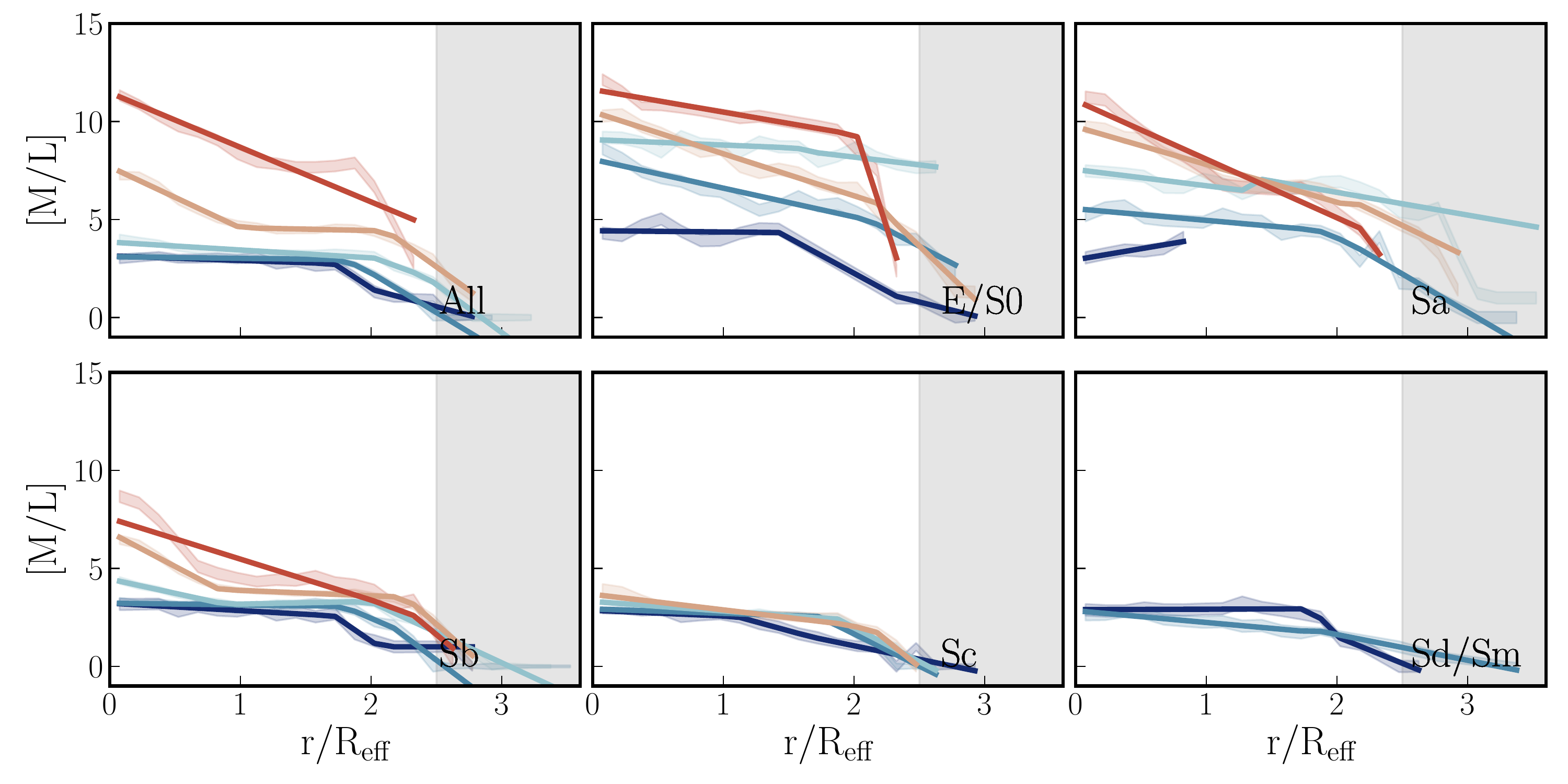}
\caption{Similar to Fig.~\ref{fig:Sstar_rad} for the M/L ratio.}  
\label{fig:ML_rad}    
\end{figure*}

\begin{figure*}
\includegraphics[width=\linewidth]{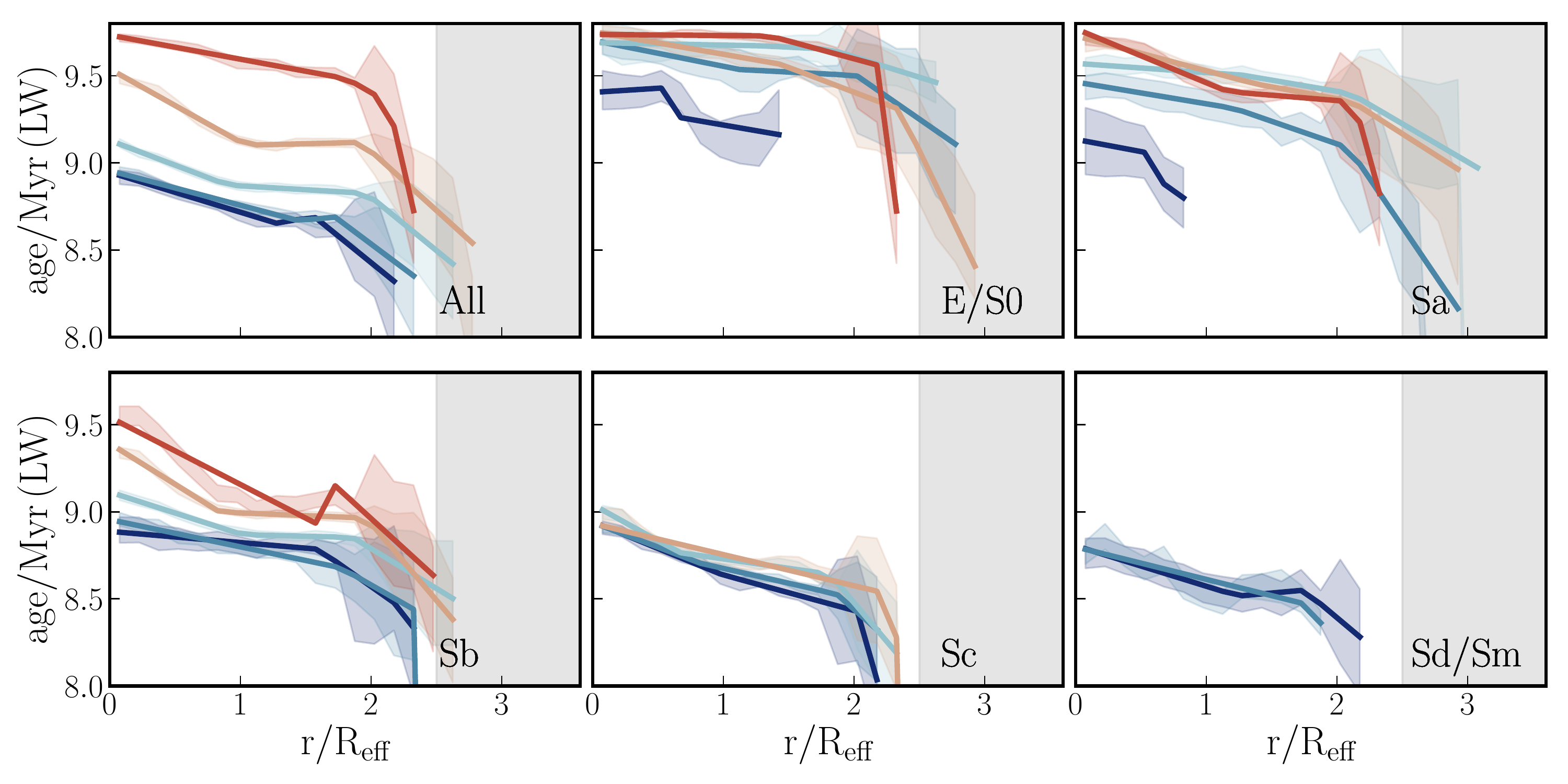}
\caption{Similar to Fig.~\ref{fig:Sstar_rad} for the luminosity-weighted age of the stellar population.}  
\label{fig:age_rad}    
\end{figure*}

\begin{figure*}
\includegraphics[width=\linewidth]{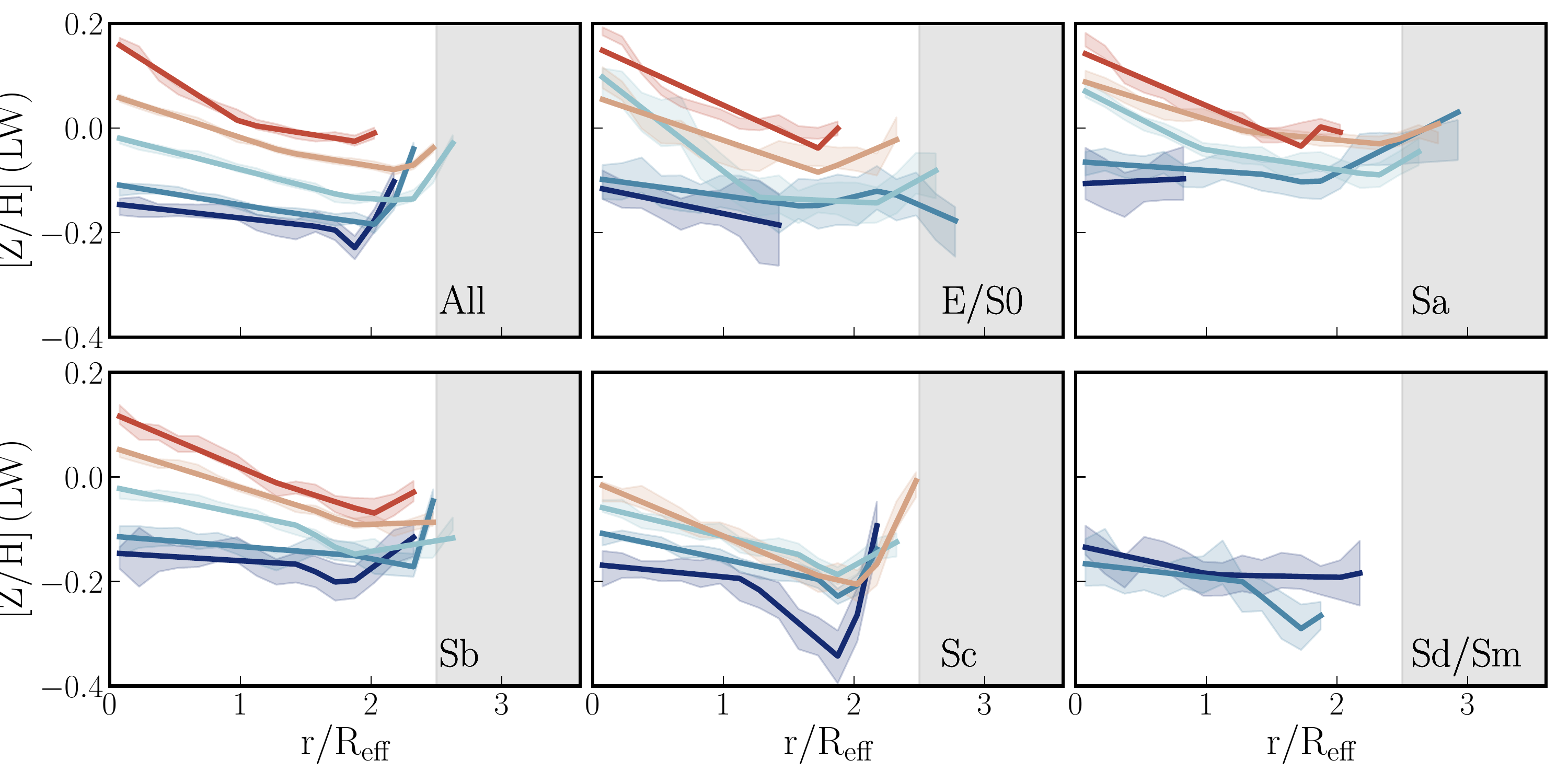}
\caption{Similar to Fig.~\ref{fig:Sstar_rad} for the luminosity-weighted stellar age.}  
\label{fig:ZH_rad}    
\end{figure*}

\begin{figure*}
\includegraphics[width=\linewidth]{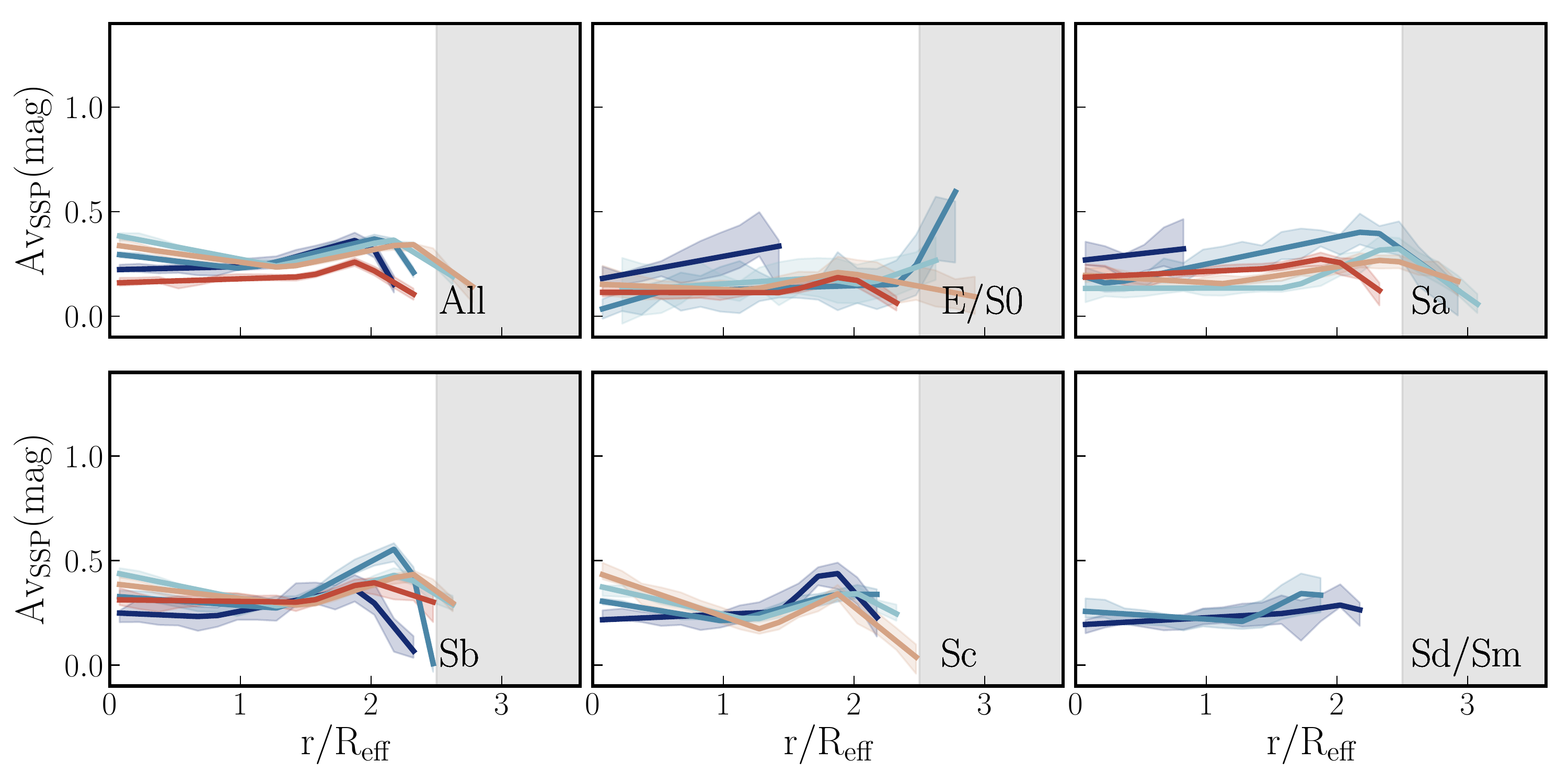}
\caption{Similar to Fig.~\ref{fig:Sstar_rad} for the luminosity-weighted stellar age.}  
\label{fig:av_ssp_rad}    
\end{figure*}

\begin{figure*}
\includegraphics[width=\linewidth]{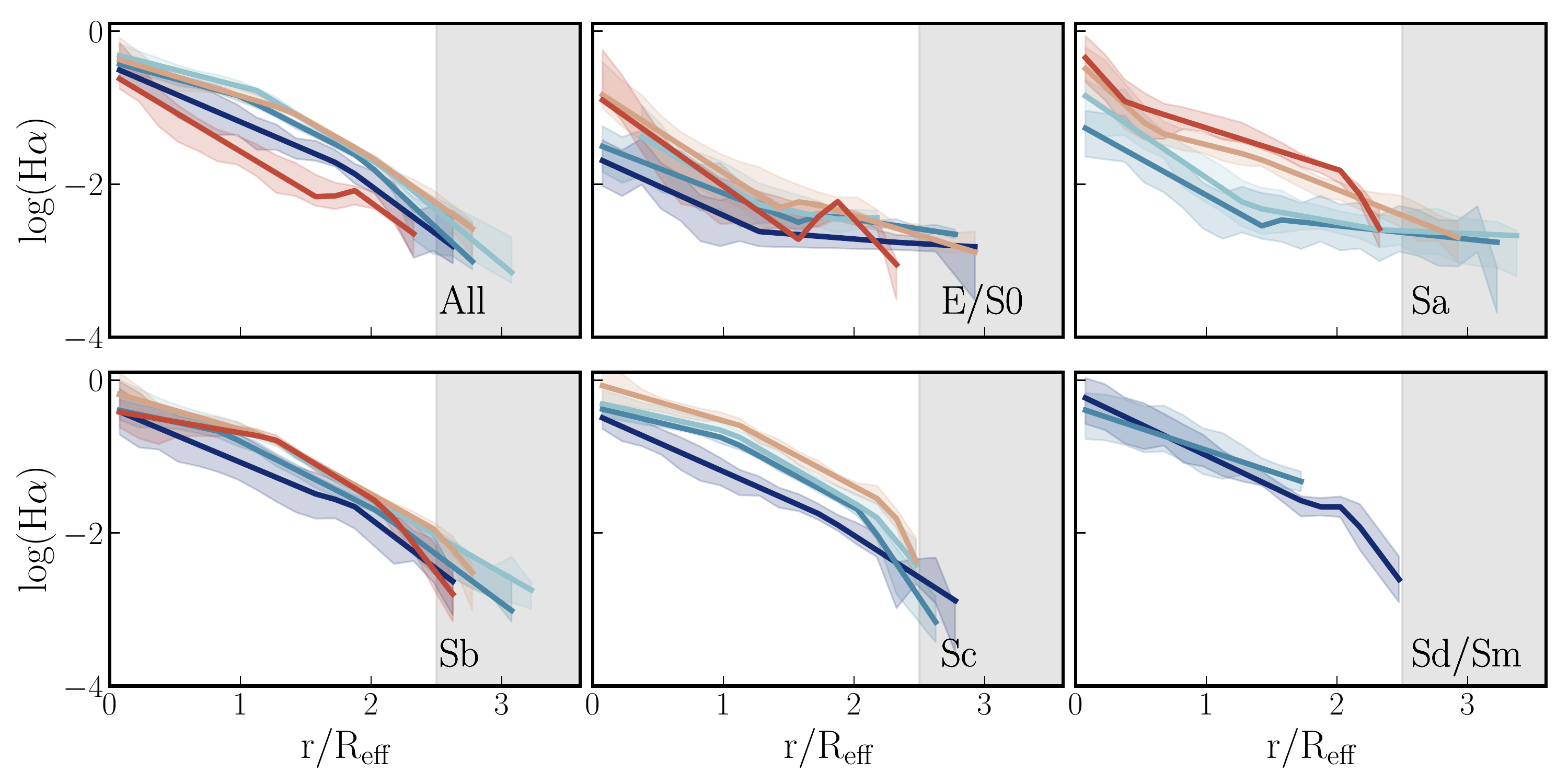}
\caption{Similar to Fig.~\ref{fig:Sstar_rad} for the \ha\ emission line.}  
\label{fig:ha_rad}    
\end{figure*}

\begin{figure*}
\includegraphics[width=\linewidth]{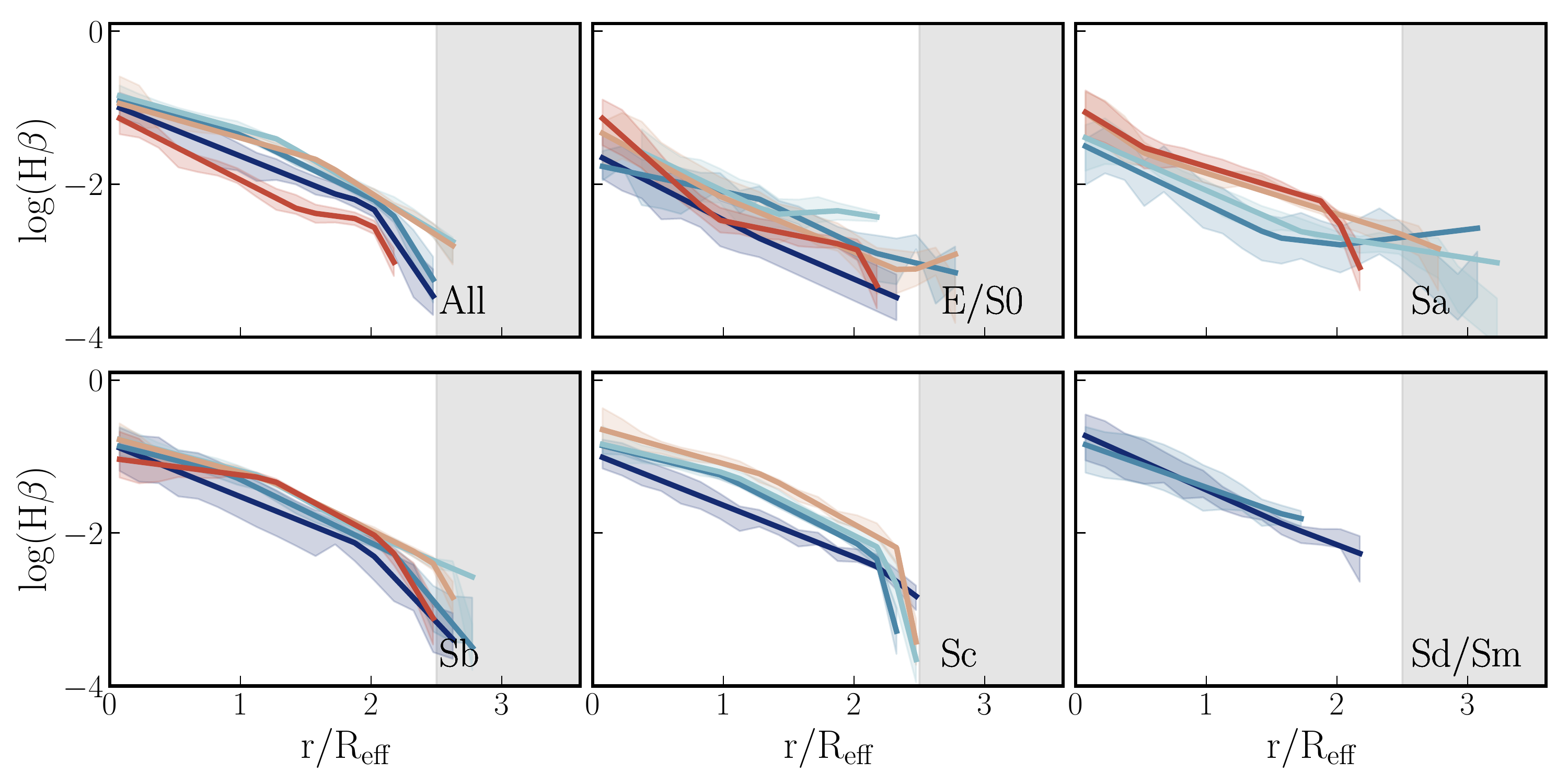}
\caption{Similar to Fig.~\ref{fig:Sstar_rad} for the \hb\ emission line.}  
\label{fig:hb_rad}    
\end{figure*}

\begin{figure*}
\includegraphics[width=\linewidth]{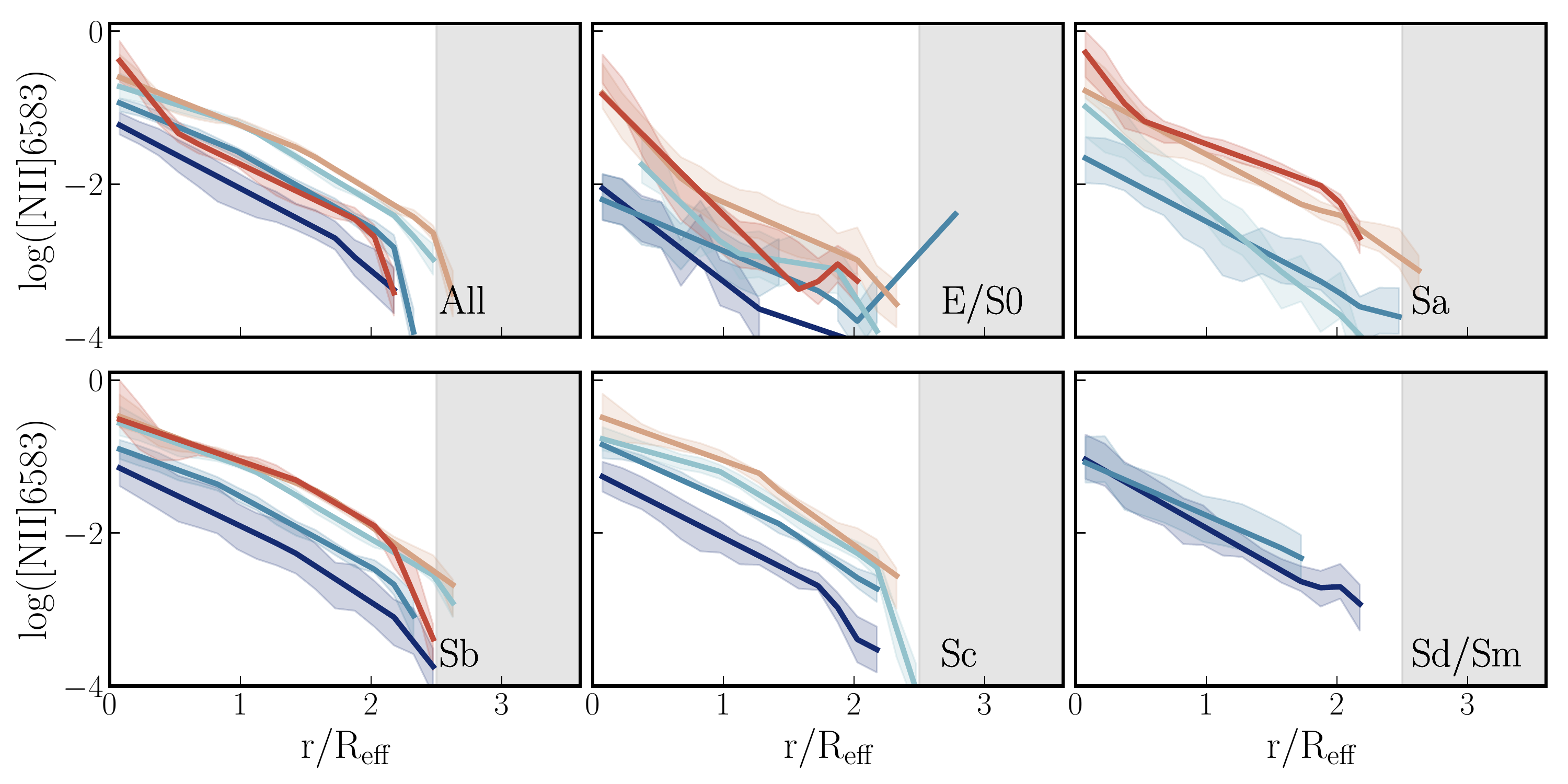}
\caption{Similar to Fig.~\ref{fig:Sstar_rad} for the [NII] emission line.}  
\label{fig:n2_rad}    
\end{figure*}

\begin{figure*}
\includegraphics[width=\linewidth]{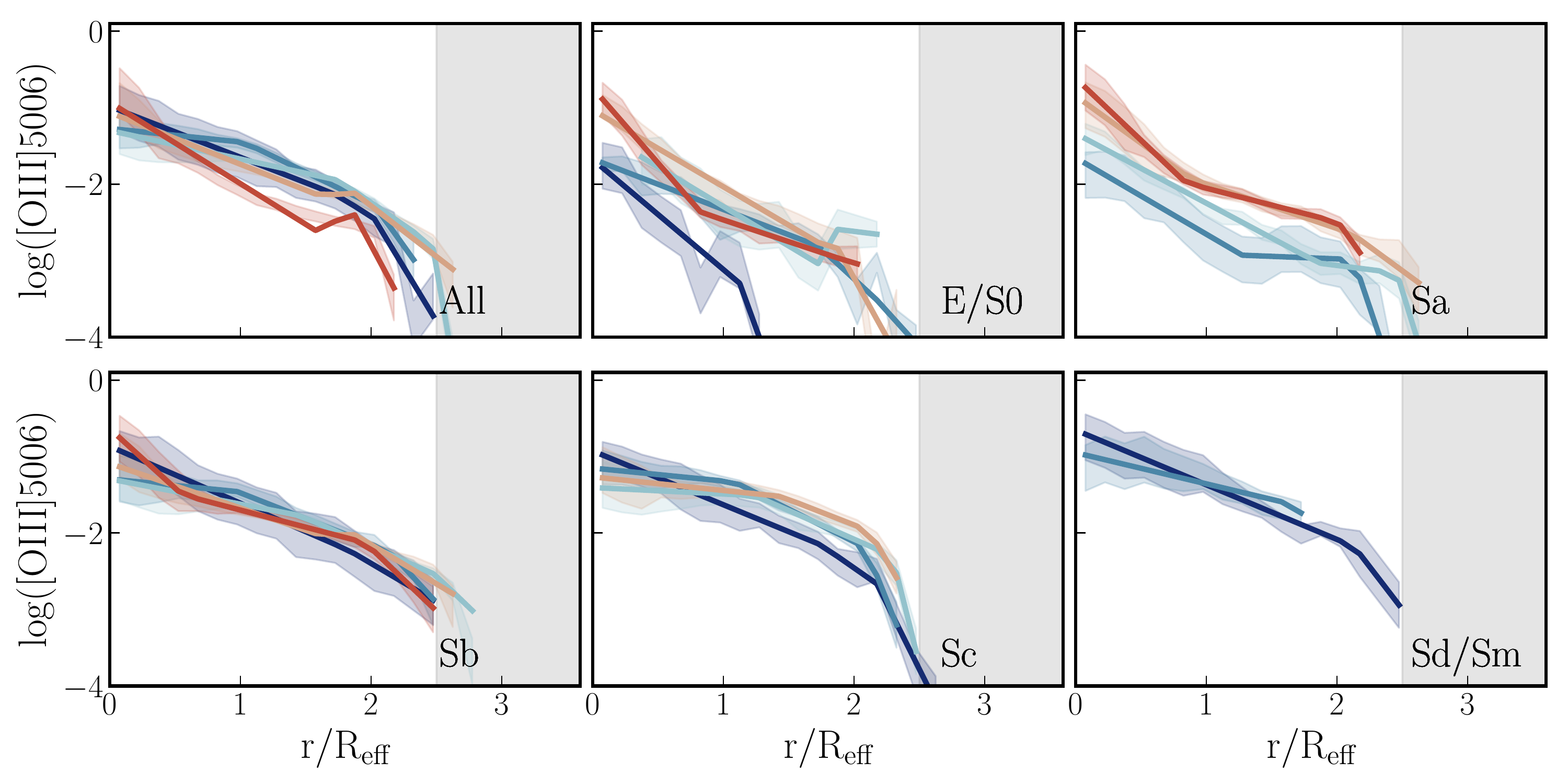}
\caption{Similar to Fig.~\ref{fig:Sstar_rad} for the [OIII] emission line.}  
\label{fig:o3_rad}    
\end{figure*}

\begin{figure*}
\includegraphics[width=\linewidth]{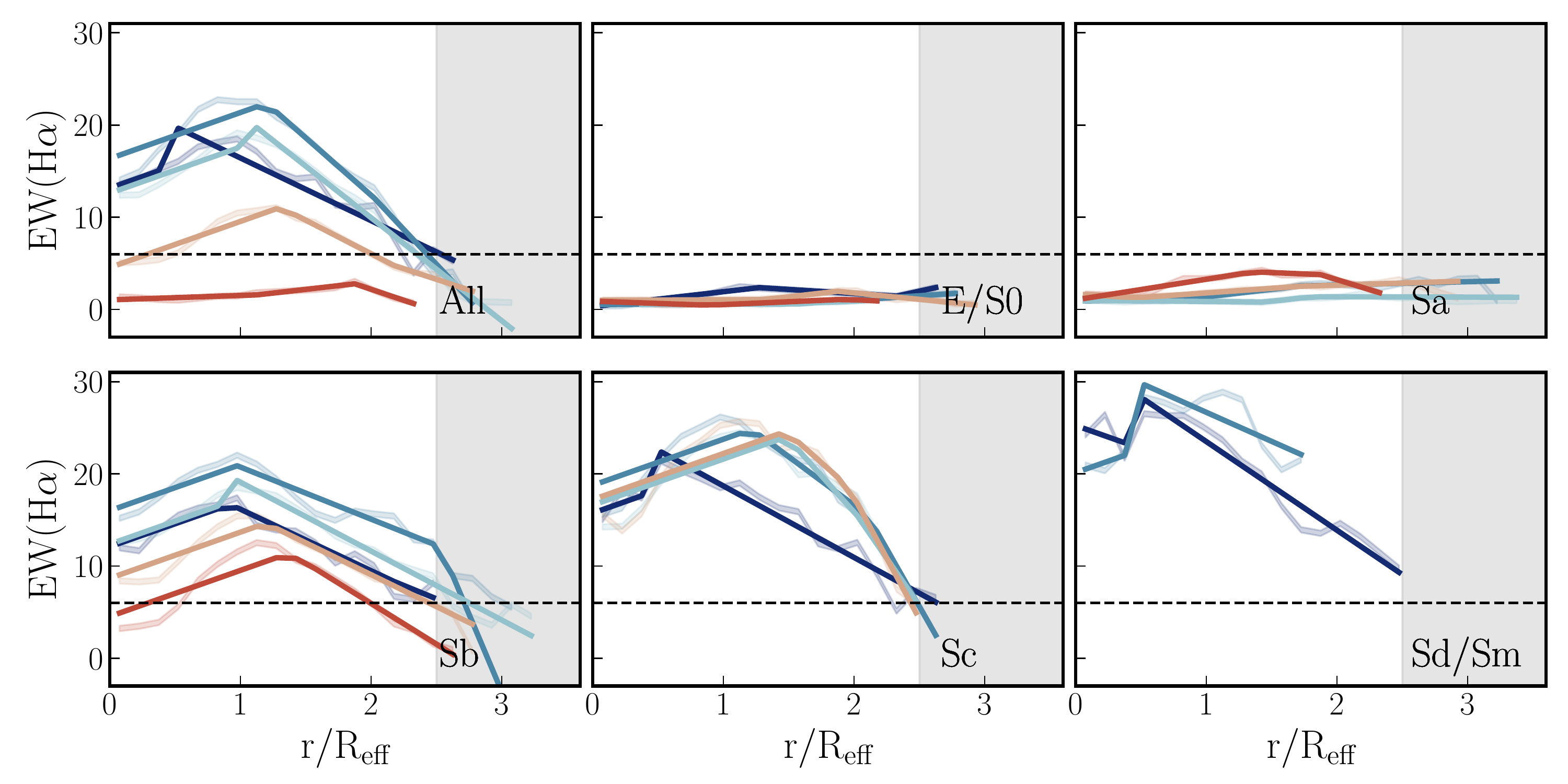}
\caption{Similar to Fig.~\ref{fig:Sstar_rad} for EW(\ha). The dashed horizontal lines in all the panels represent an EW(\ha) of 6\AA.}  
\label{fig:EW_rad}    
\end{figure*}

\begin{figure*}
\includegraphics[width=\linewidth]{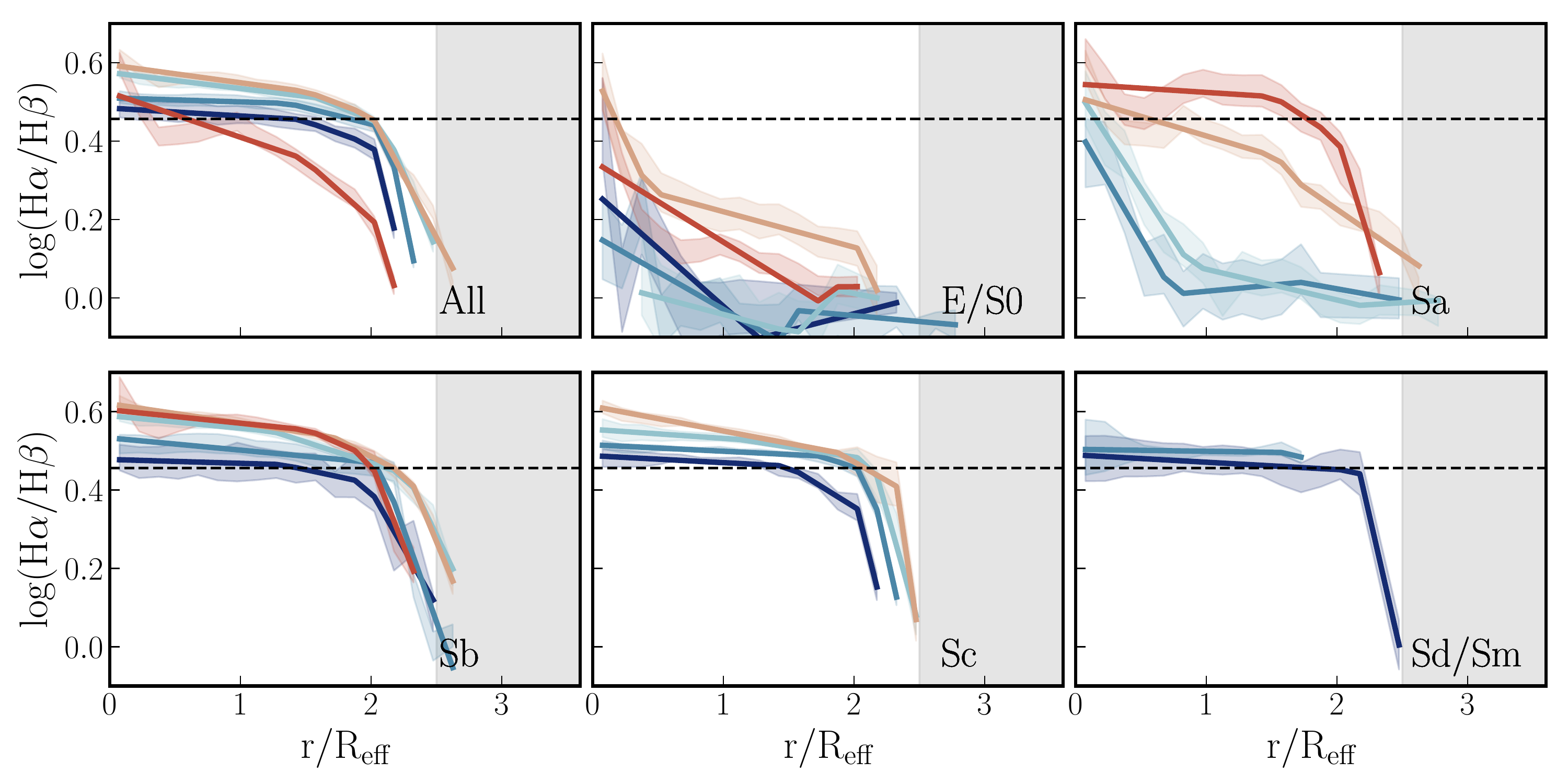}
\caption{Similar to Fig.~\ref{fig:Sstar_rad} for the \ha/\hb\, line ratio. The dashed horizontal lines in all the panels represent the canonical value of this ratio }  
\label{fig:BD_rad}    
\end{figure*}

\begin{figure*}
\includegraphics[width=\linewidth]{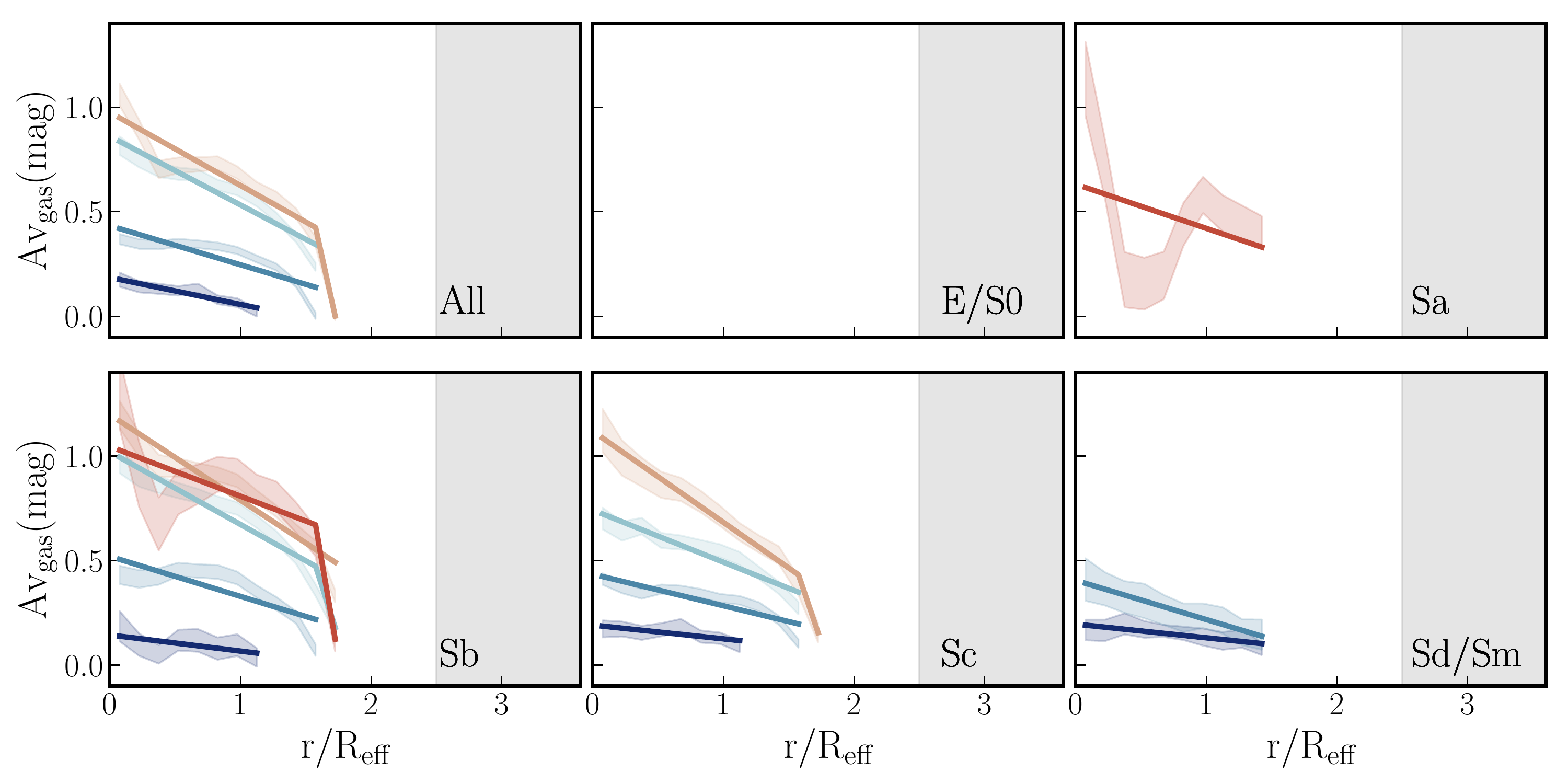}
\caption{Similar to Fig.~\ref{fig:Sstar_rad} for the optical extinction derived from the Balmer decrement. }  
\label{fig:avgas_rad}    
\end{figure*}

\begin{figure*}
\includegraphics[width=\linewidth]{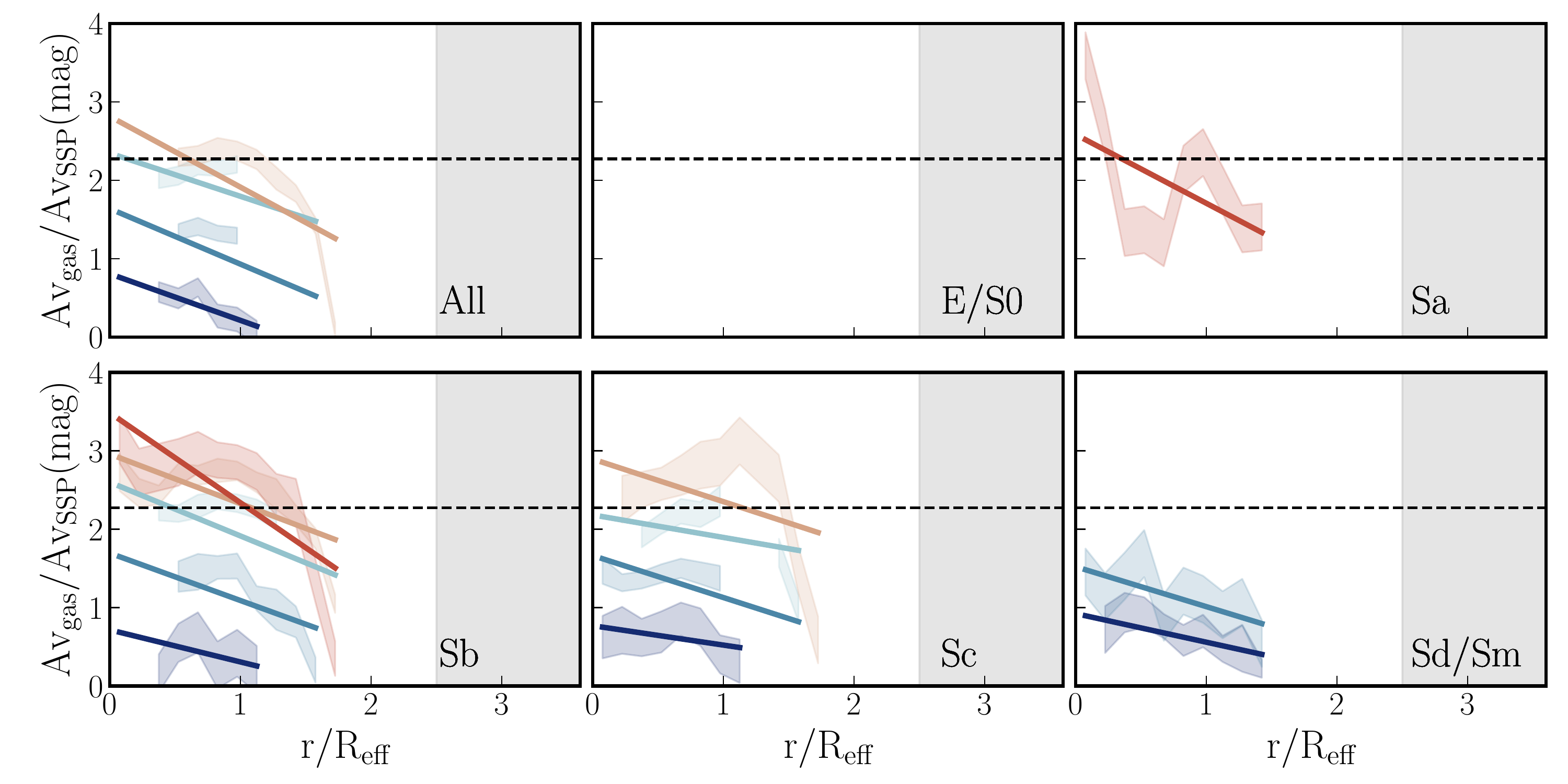}
\caption{Similar to Fig.~\ref{fig:Sstar_rad} for the \avgas/\avssp ratio. The dashed line in all the panels represents the value derived from \citet{Calzetti_1997}:\avgas/\avssp $\sim$ 2.27.}  
\label{fig:avgasssp_rad}    
\end{figure*}

\begin{figure*}
\includegraphics[width=\linewidth]{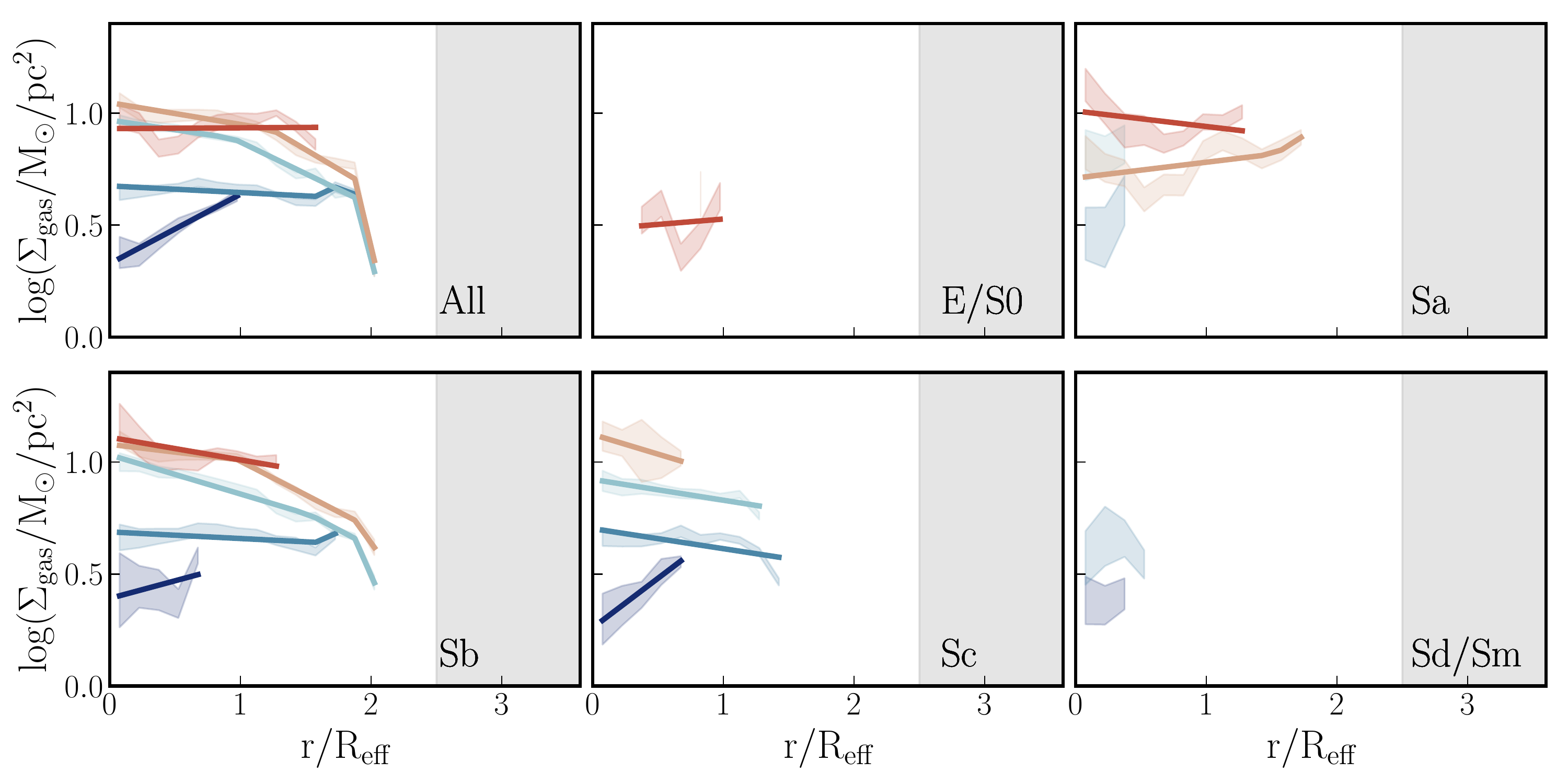}
\caption{Similar to Fig.~\ref{fig:Sstar_rad} for the \avgas/\avssp ratio.}  
\label{fig:mol_rad}    
\end{figure*}

\begin{figure*}
\includegraphics[width=\linewidth]{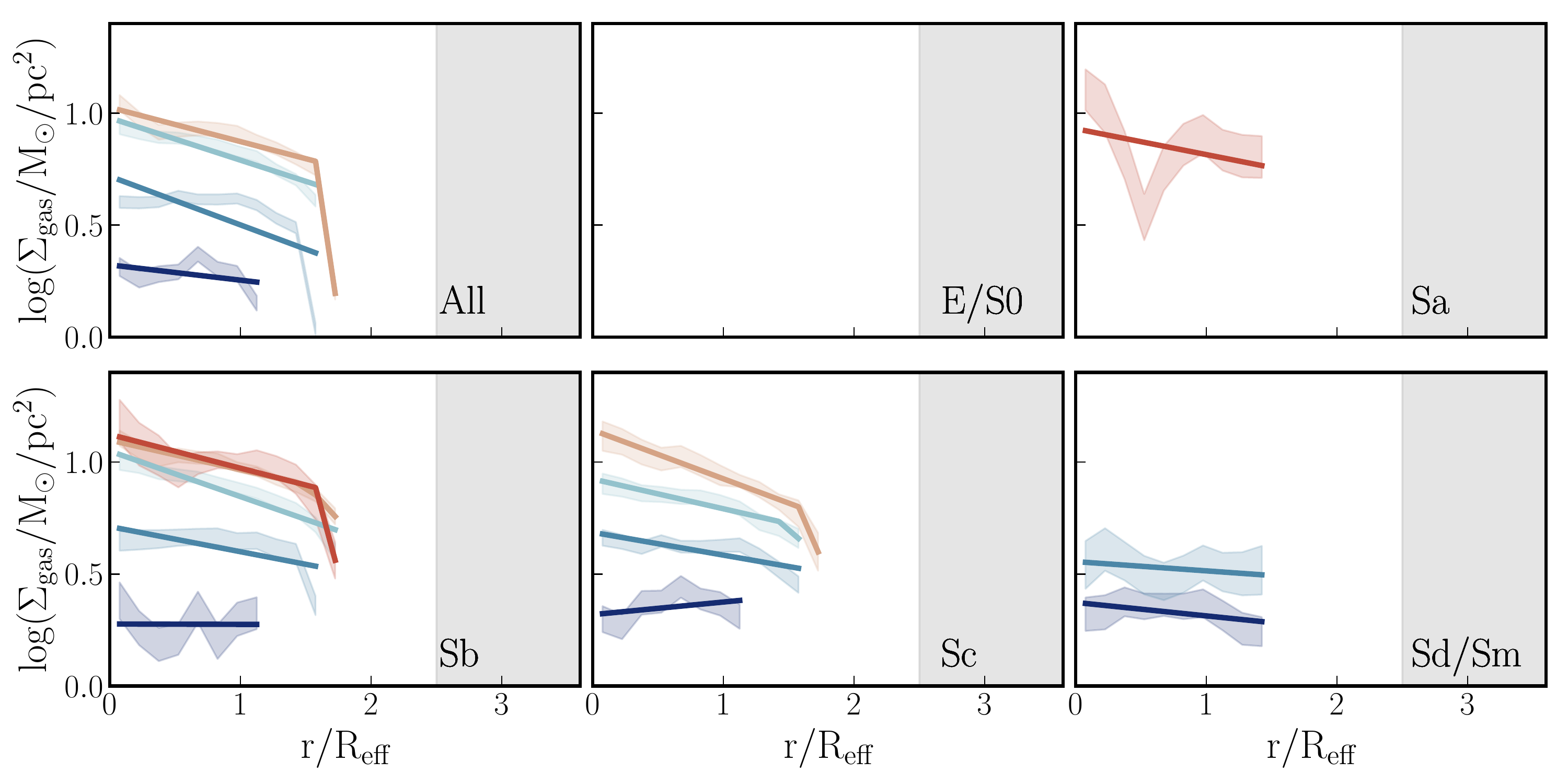}
\caption{Similar to Fig.~\ref{fig:Sstar_rad} for gas fraction, \fgas.}  
\label{fig:fgas_rad}    
\end{figure*}

\begin{figure*}
\includegraphics[width=\linewidth]{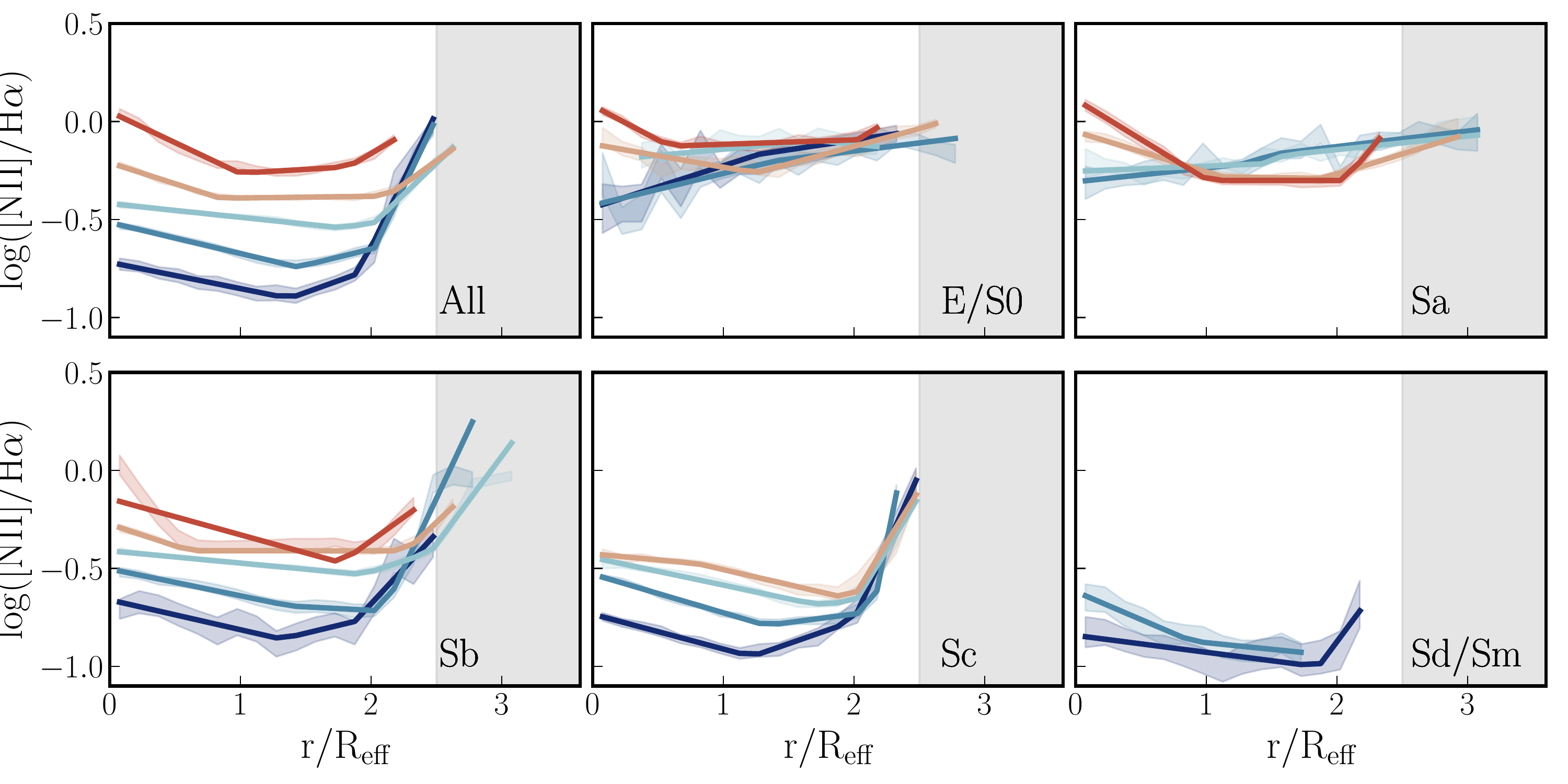}
\caption{Similar to Fig.~\ref{fig:Sstar_rad} for emission lines ratio [NII]/\ha.}  
\label{fig:N2_rad}    
\end{figure*}

\begin{figure*}
\includegraphics[width=\linewidth]{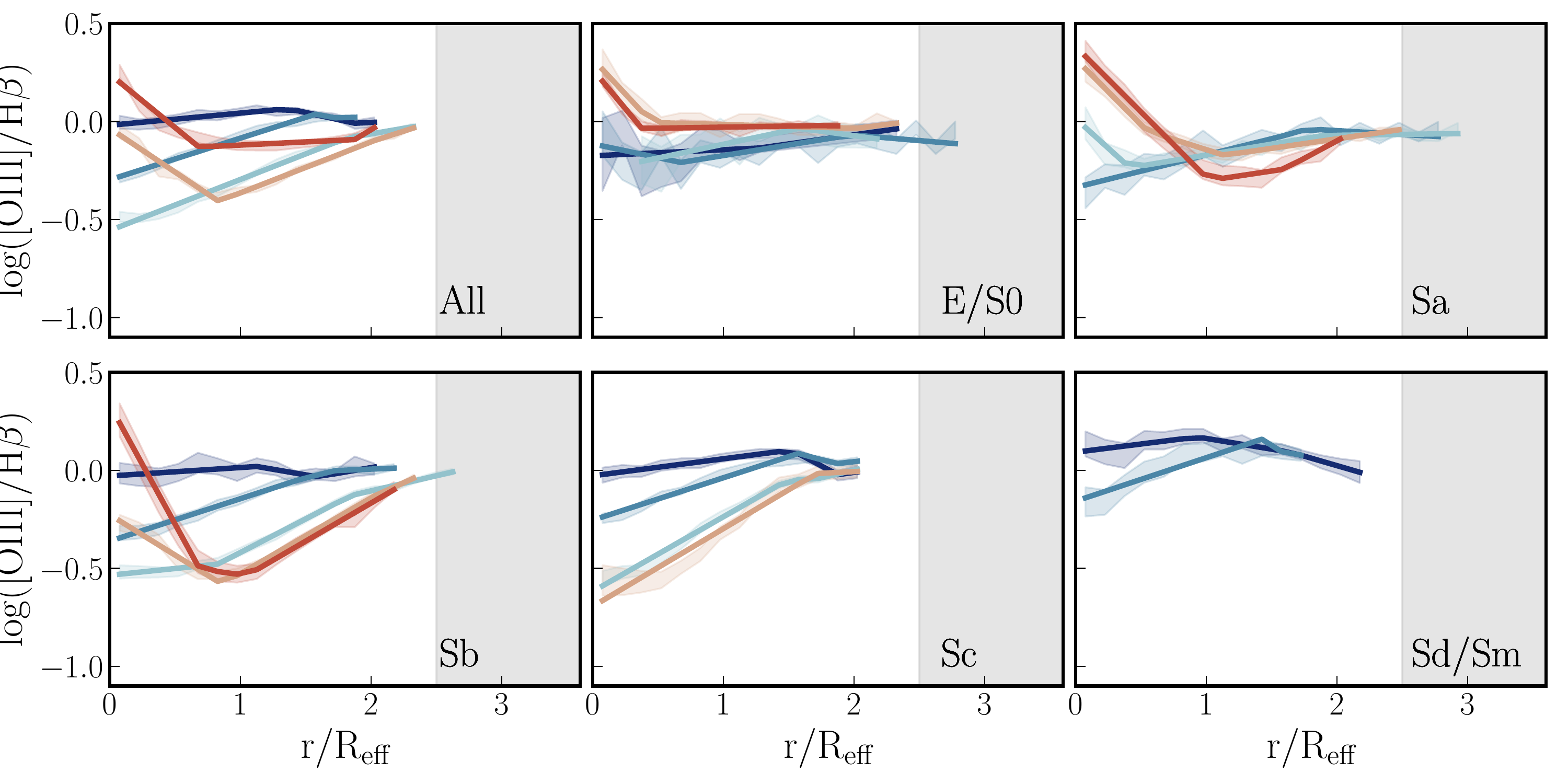}
\caption{Similar to Fig.~\ref{fig:Sstar_rad} for emission lines ratio [OIII]i/\hb.}
\label{fig:O3_rad}    
\end{figure*}

\begin{figure*}
\includegraphics[width=\linewidth]{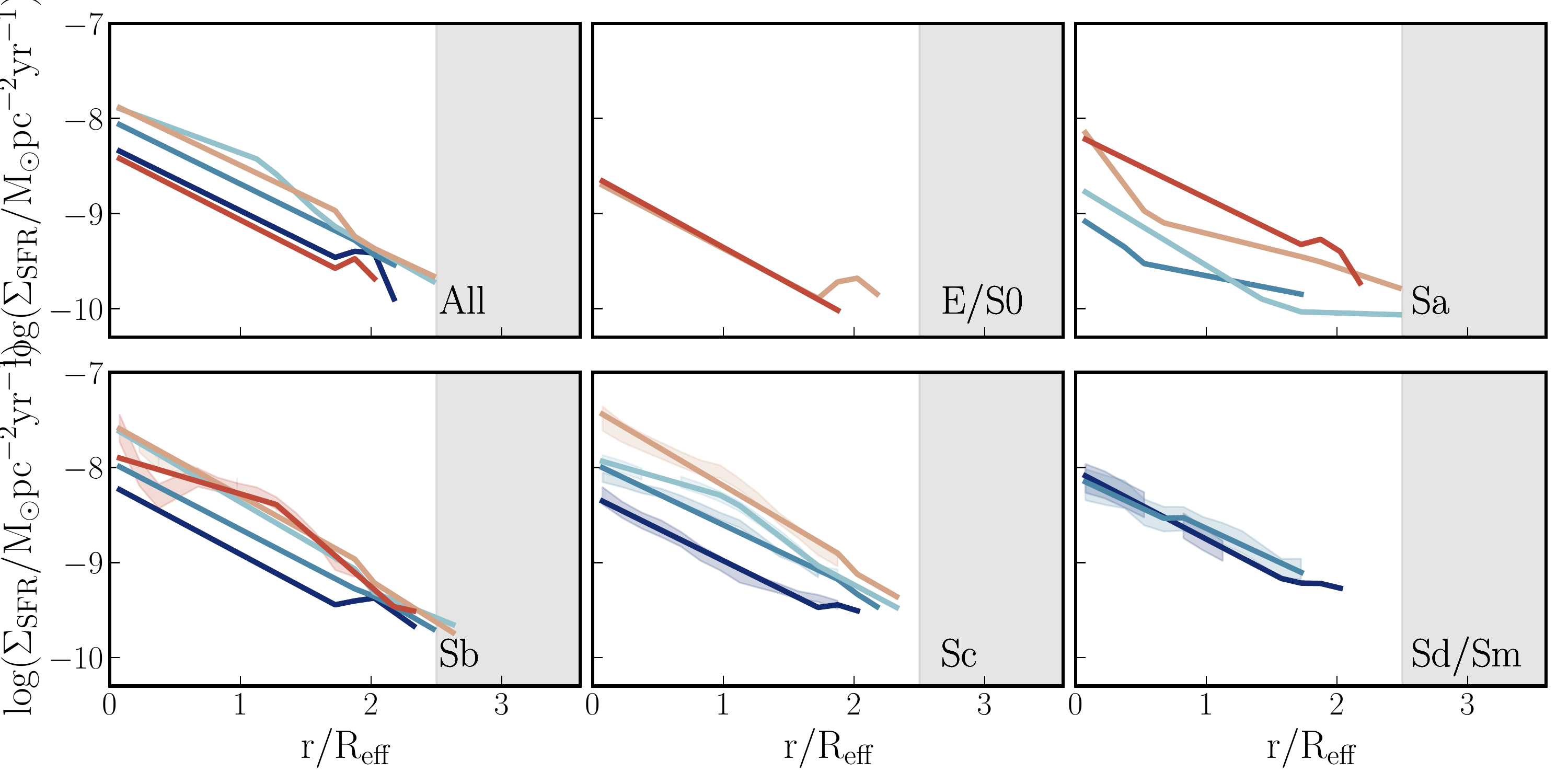}
\caption{Similar to Fig.~\ref{fig:Sstar_rad} for \Ssfr.}
\label{fig:Ssfr_rad}    
\end{figure*}

\begin{figure*}
\includegraphics[width=\linewidth]{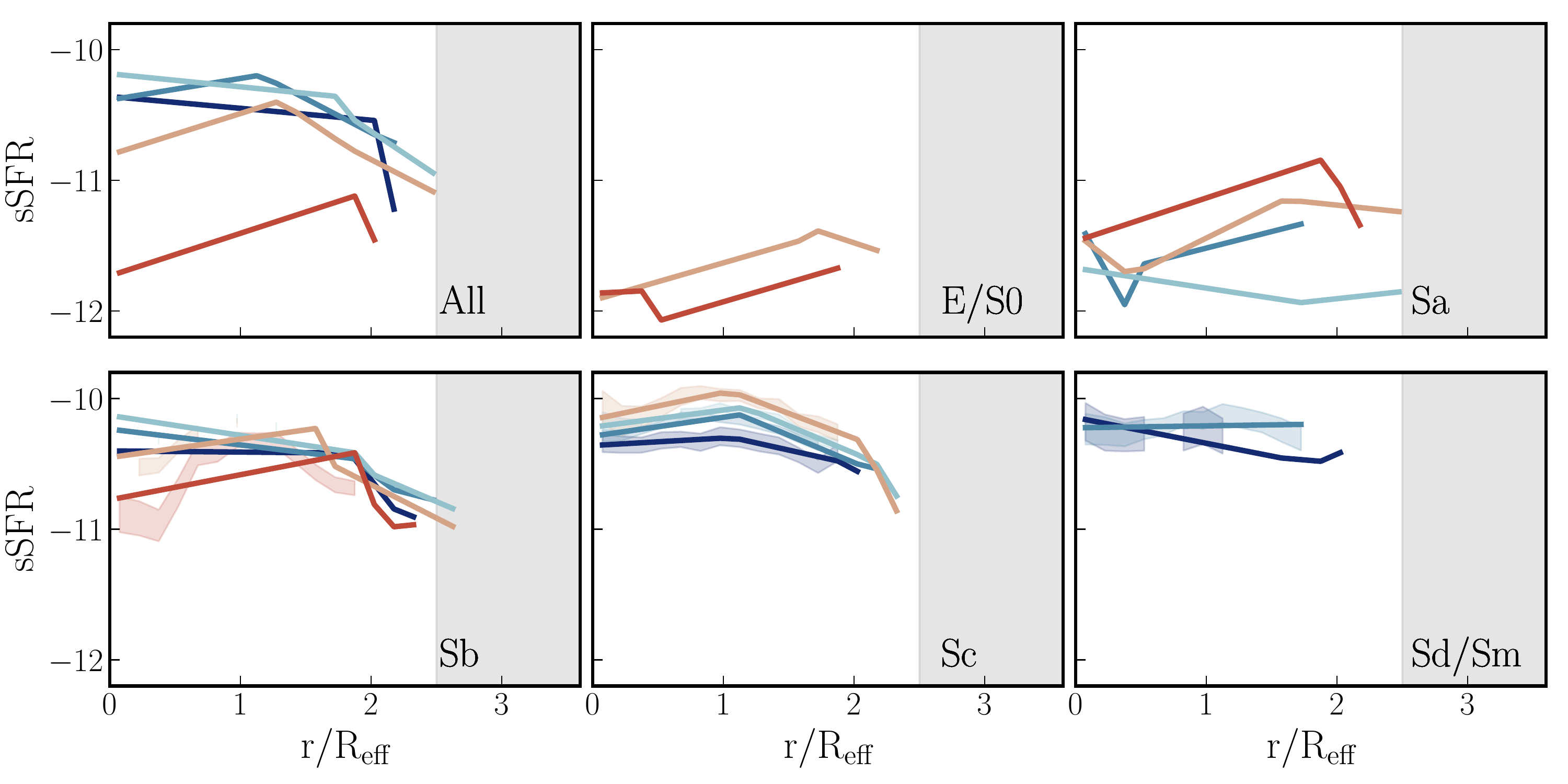}
\caption{Similar to Fig.~\ref{fig:Sstar_rad} for sSFR.}
\label{fig:sSFR_rad}    
\end{figure*}

\begin{figure*}
\includegraphics[width=\linewidth]{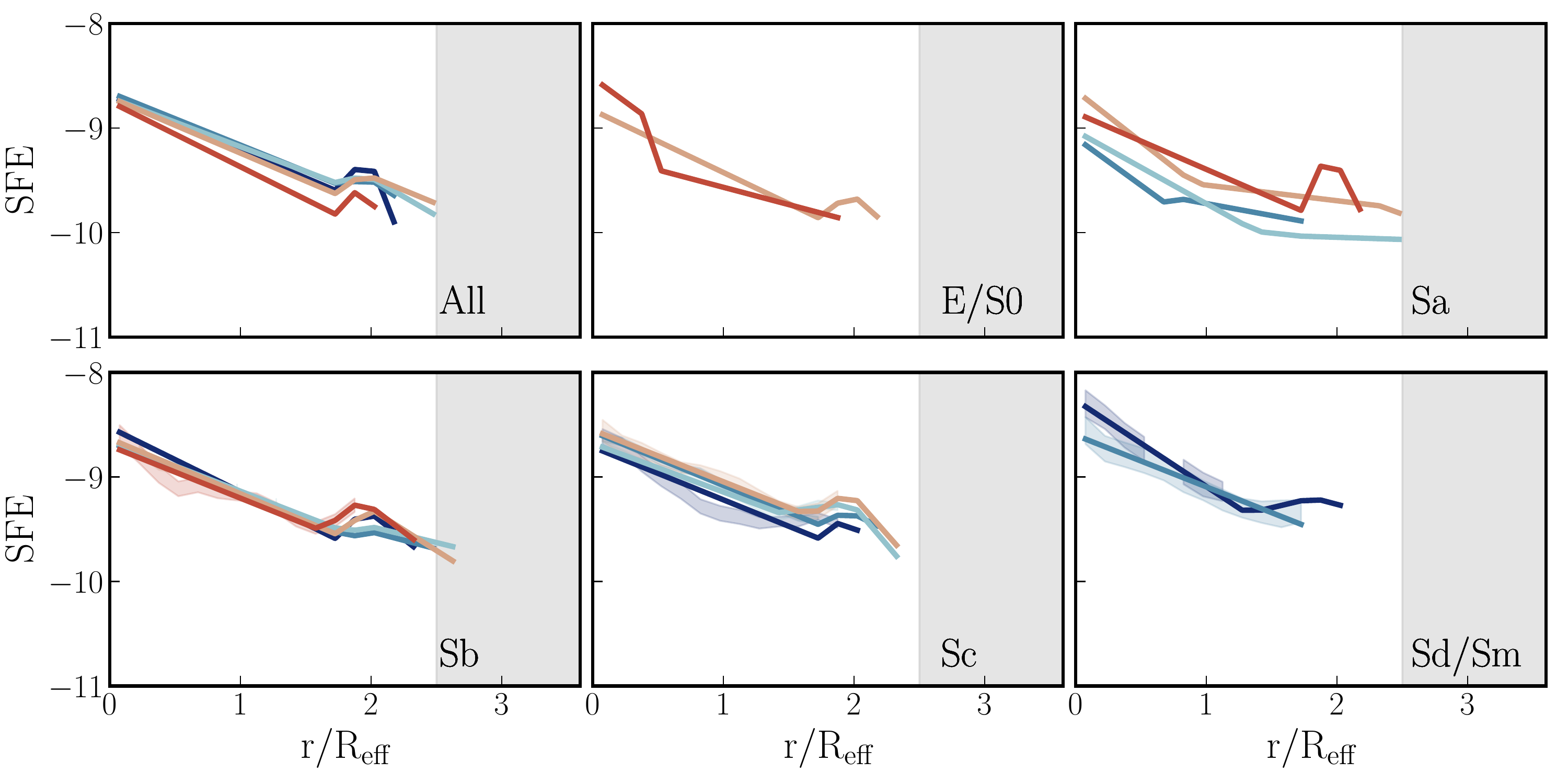}
\caption{Similar to Fig.~\ref{fig:Sstar_rad} for SFE.}
\label{fig:SFE_rad}    
\end{figure*}

\begin{figure*}
\includegraphics[width=\linewidth]{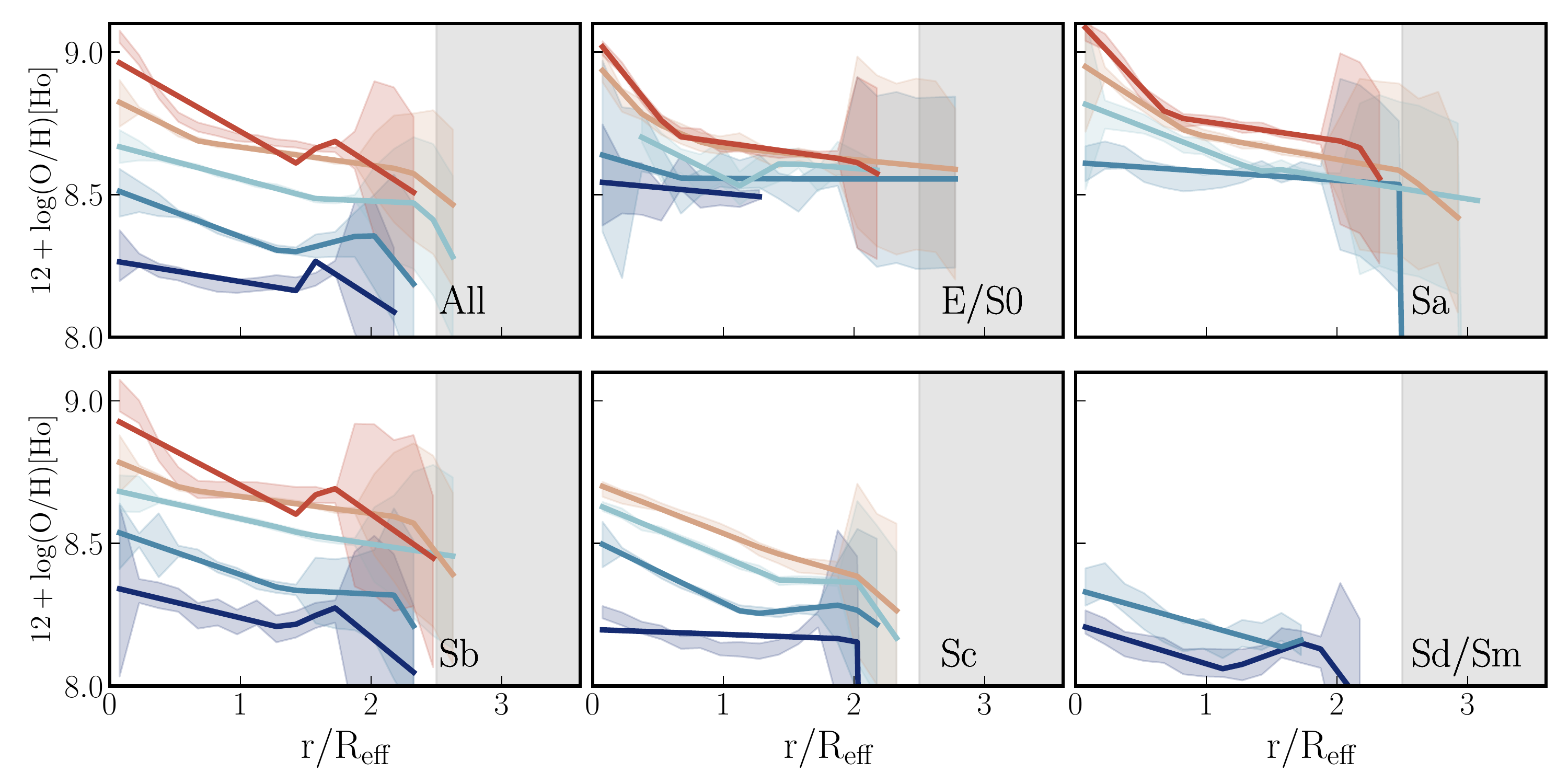}
\caption{Similar to Fig.~\ref{fig:Sstar_rad} for the oxygen abundance derived using the calibrator presented in \citet{Ho_2019}}
\label{fig:OH_Ho_rad}    
\end{figure*}

\begin{figure*}
\includegraphics[width=\linewidth]{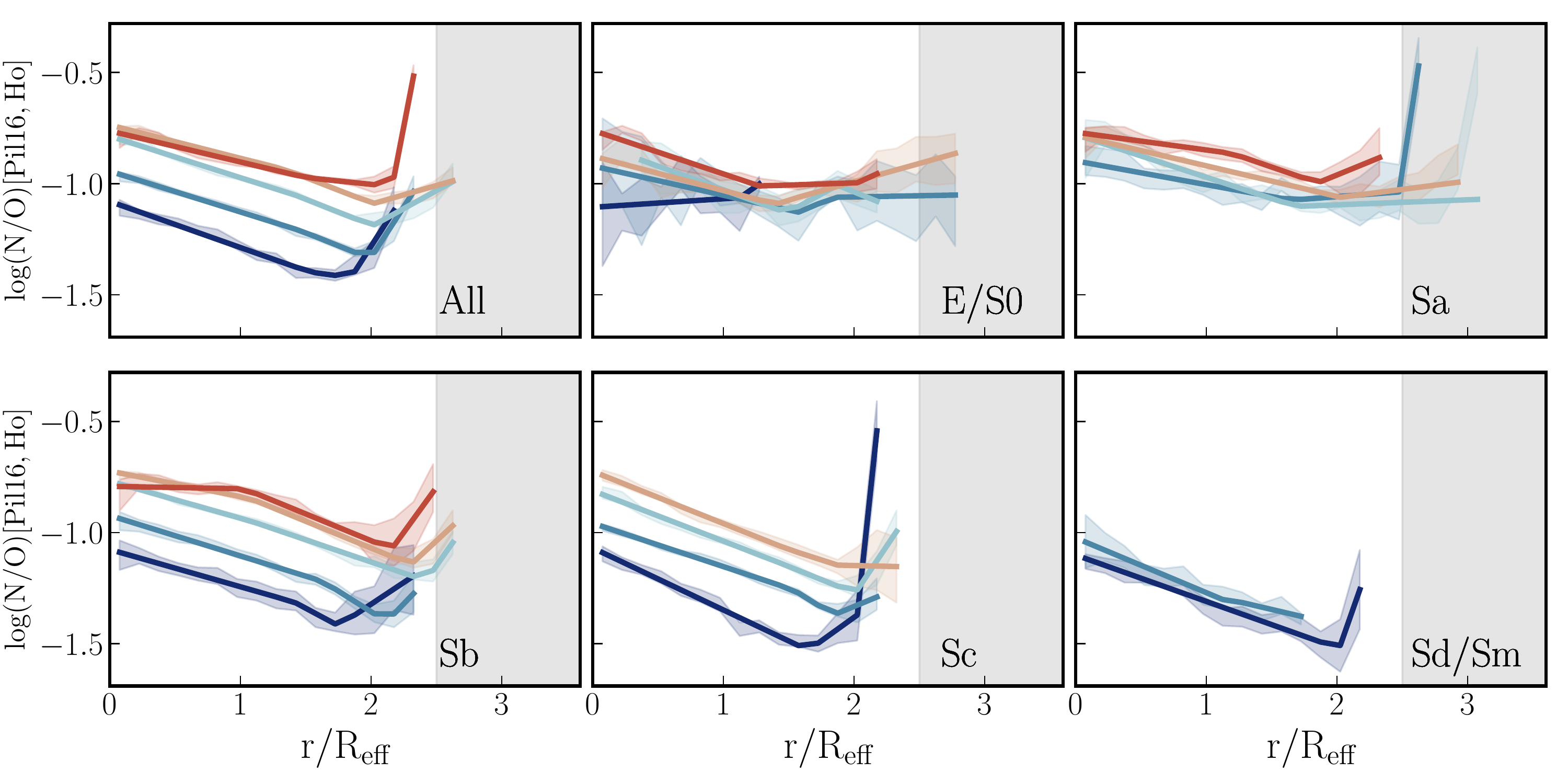}
\caption{Similar to Fig.~\ref{fig:Sstar_rad} for the N/O ratio.}
\label{fig:ON_rad}    
\end{figure*}

\begin{figure*}
\includegraphics[width=\linewidth]{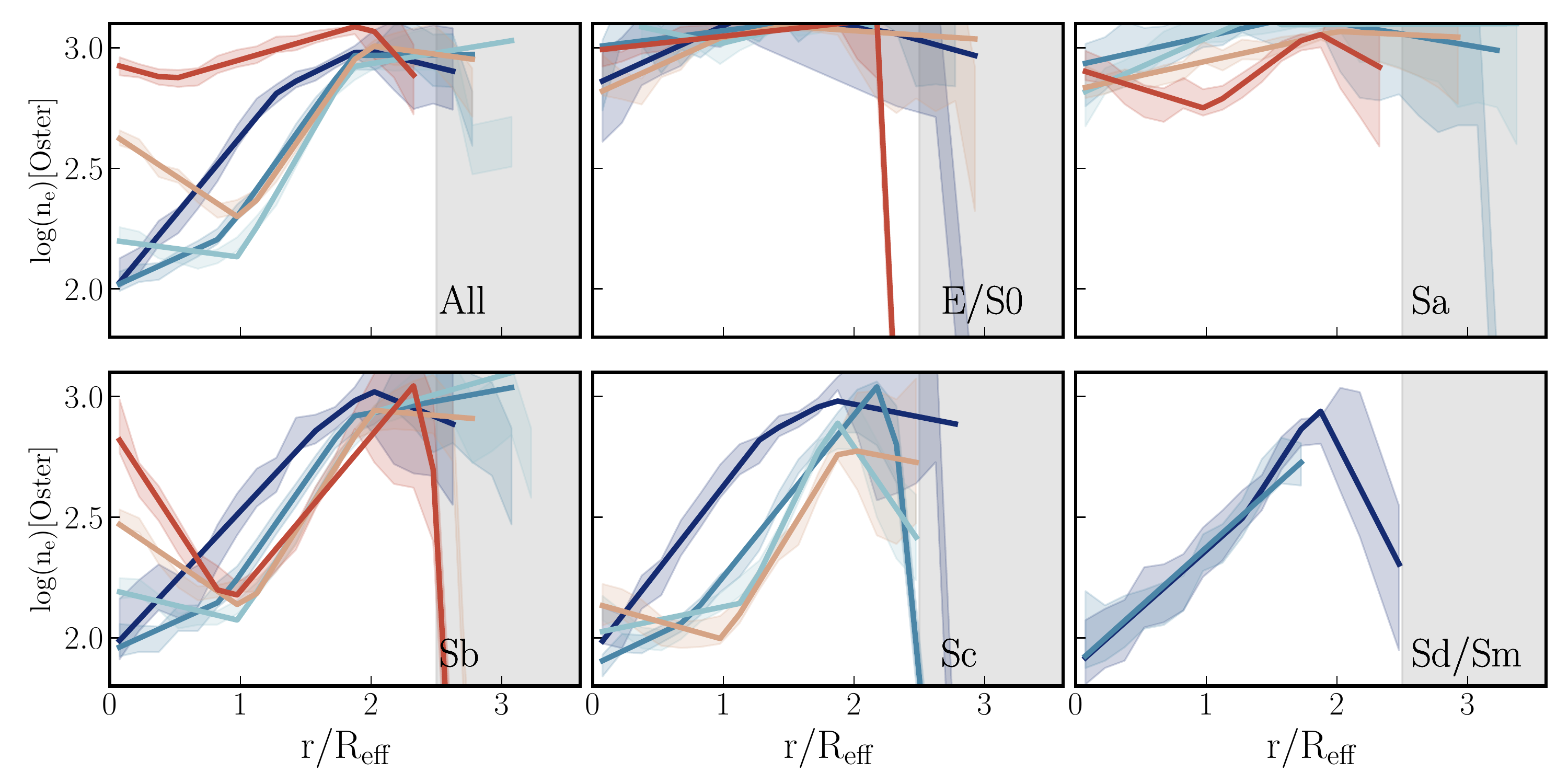}
\caption{Similar to Fig.~\ref{fig:Sstar_rad} for the electron density.}
\label{fig:ne_rad}    
\end{figure*}

\begin{figure*}
\includegraphics[width=\linewidth]{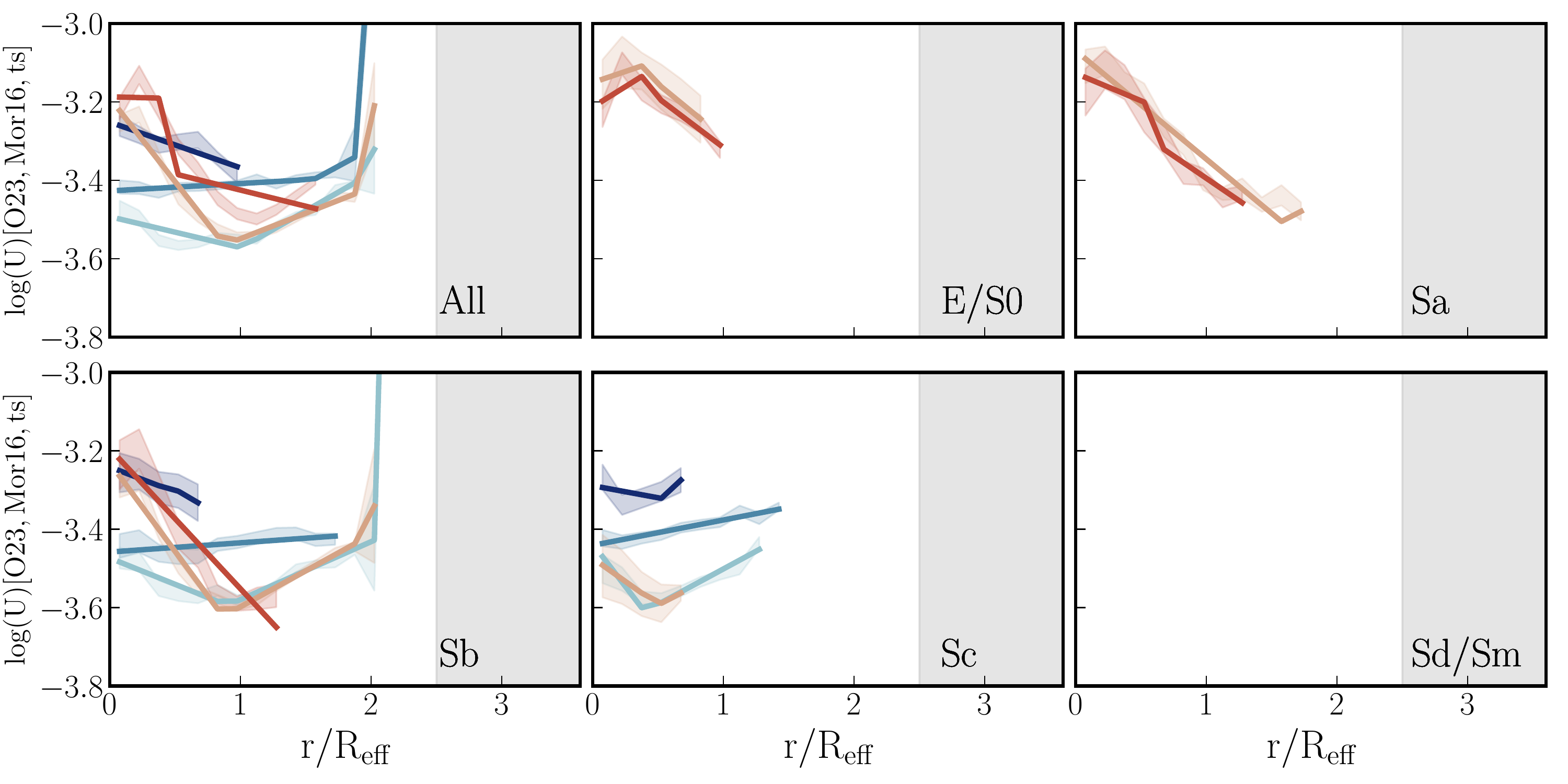}
\caption{Similar to Fig.~\ref{fig:Sstar_rad} for the ionization parameter.}
\label{fig:U_rad}    
\end{figure*}

\section{Oxygen abundances using different calibrators}
\label{app:OH_calib}
\begin{figure*}
\includegraphics[width=\linewidth]{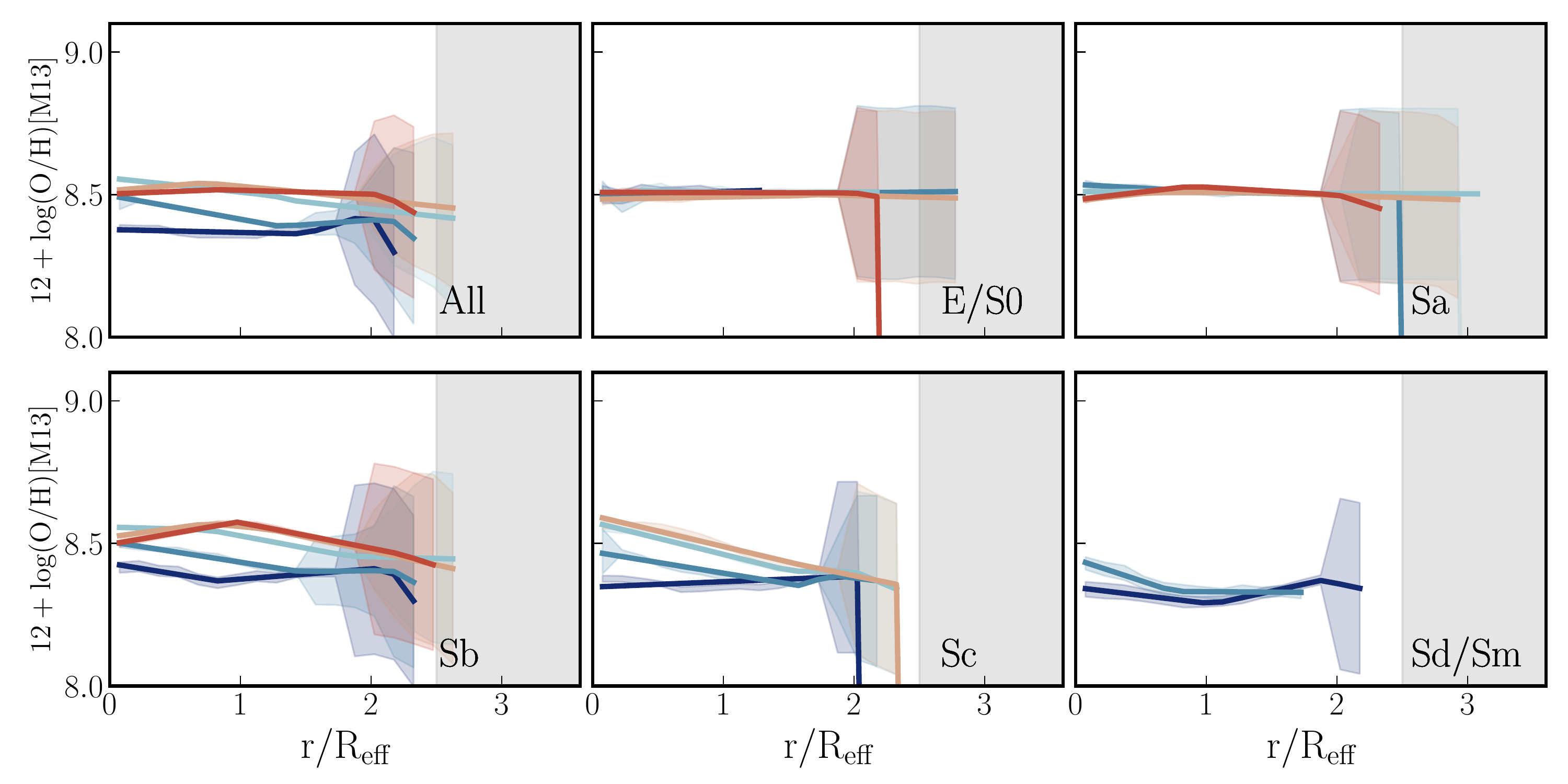}
\caption{Similar to Fig.~\ref{fig:Sstar_rad} for the oxygen abundance derived using the calibrator presented in \citet{Marino_2013}}
\label{fig:OH_M13_rad}    
\end{figure*}

\begin{figure*}
\includegraphics[width=\linewidth]{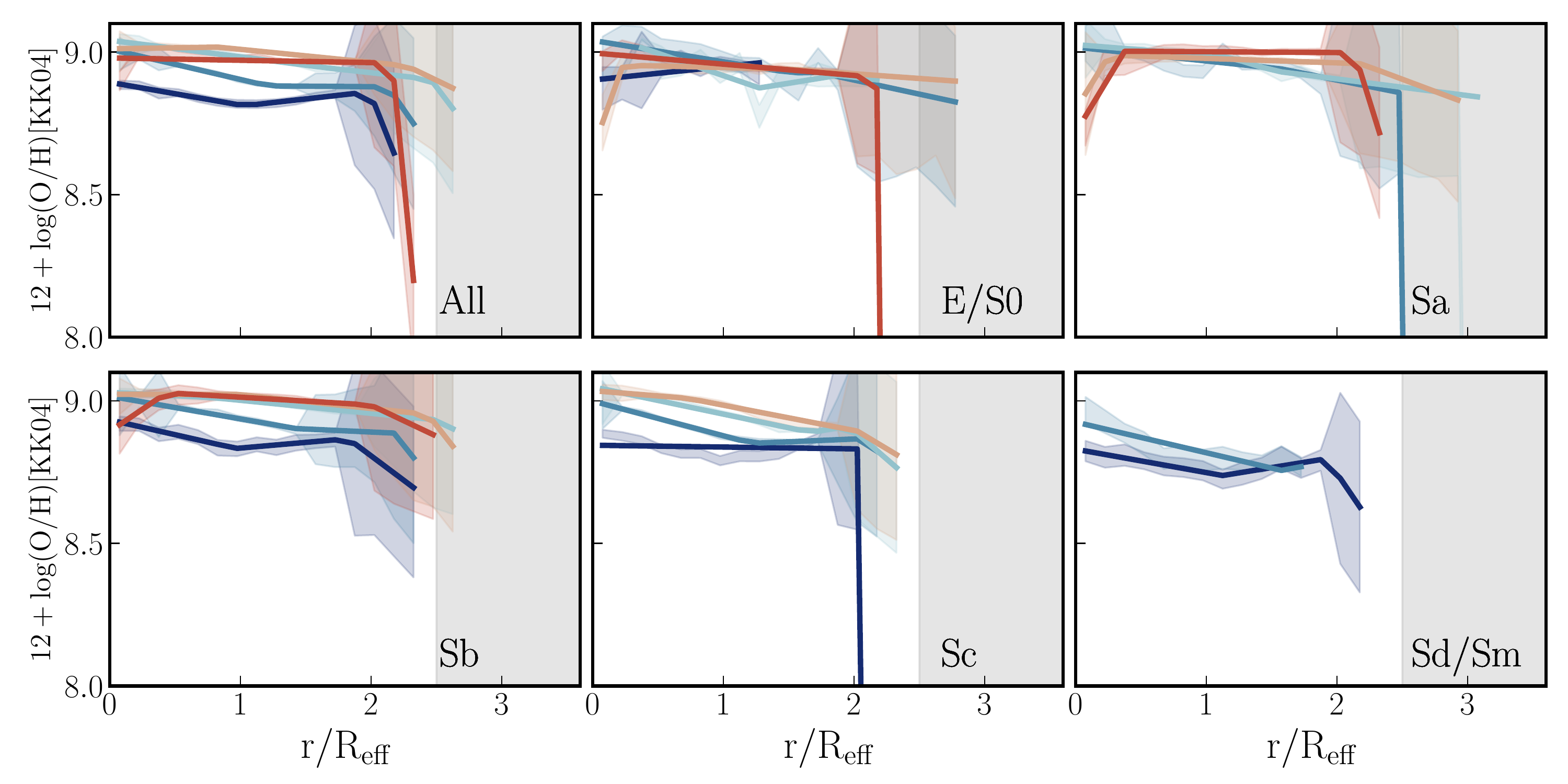}
\caption{Similar to Fig.~\ref{fig:Sstar_rad} for the oxygen abundance derived using the calibrator presented in \citet{KK_2004}}
\label{fig:OH_KK04_rad}    
\end{figure*}

\begin{figure}
\includegraphics[width=\linewidth]{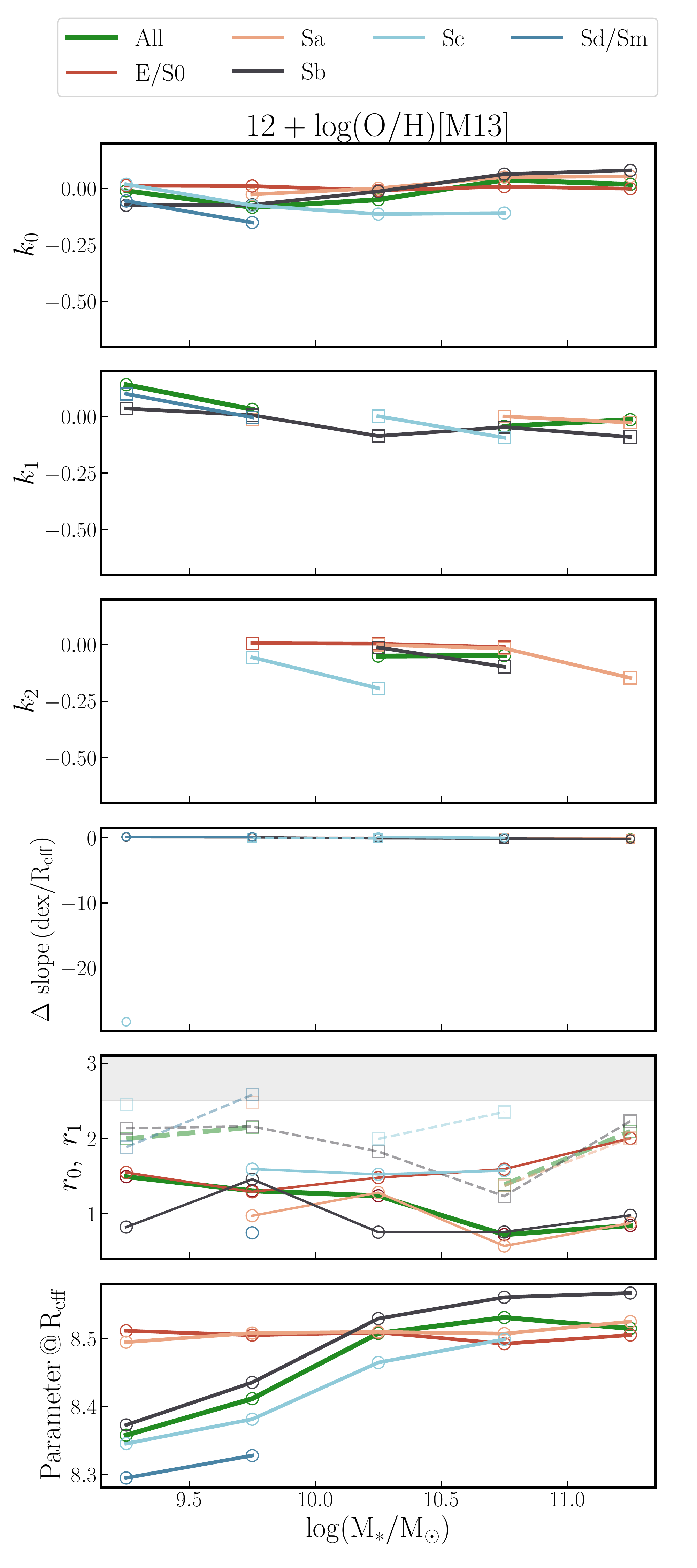}
\caption{Similar to Fig.~\ref{fig:Sstar_rad} for the oxygen abundance derived using the calibrator presented in \citet{Ho_2019}}
\label{fig:OH_M13}    
\end{figure}

\begin{figure}
\includegraphics[width=\linewidth]{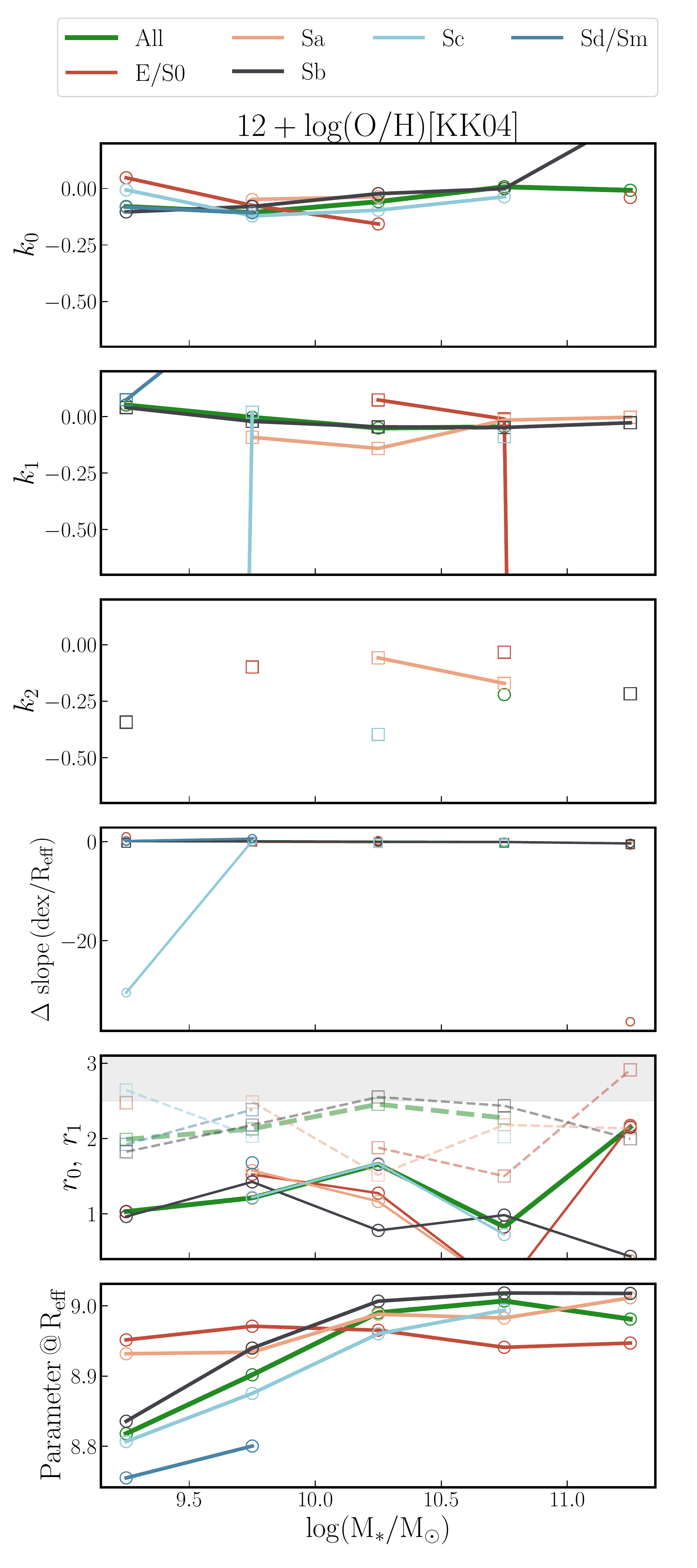}
\caption{Similar to Fig.~\ref{fig:Sstar_rad} for the oxygen abundance derived using the calibrator presented in \citet{Ho_2019}}
\label{fig:OH_KK04}    
\end{figure}

As we mention in Sec.~\ref{sec:OH}, thanks to the \texttt{pyOxy} script, we are able to estimate the oxygen abundance using a large suite of strong-line calibrators. Although the script allows an estimation of the oxygen abundance for more than 20 calibrators, in this appendix we present the radial distribution and the piece-wise analysis using two of them that we consider are representative of this large set of calibrators: the empirical calibrator using the O3N2 ratio from \citet{Marino_2013}, and the theoretical calibrator using different line ratios proposed by \citet{KK_2004}.

In Fig.~\ref{fig:OH_M13_rad} we present the radial distribution of the oxygen abundance using the O3N2 abundance calibrator derived in \citet{Marino_2013}, whereas in Fig.~\ref{fig:OH_M13}, we show the result from the piece-wise analysis of these radial profiles. In comparison to the results derived using the calibrator derived by \citep[][ see Fig.~\ref{fig:OH_Ho}]{Ho_2019}, we find that the central gradients ($k_0$) using this O3N2 calibrator are flatter (and in some cases of opposite sign). The trend from $k_0$ for different bins of \Mstar\ for the Sb galaxies using the O3N2 calibrator is similar as those described by \citet{Belfiore_2017}: $k_0$ increases as \Mstar\ increases (except for the lowest-mass bin). However, for Sc galaxies we find that $k_0$ decreases with \Mstar. A similar trend is observed for Sd/Sm galaxies. For early-type galaxies we find flat gradients regardless \Mstar. For the outskirts of galaxies the piece-wise analysis shows that the gradients ($k1$) change from positive to negative as \Mstar\ increases. This piece-wise analysis provides a more complete description of the radial gradients in comparison to assume a single gradient to the entire radial distribution of the oxygen abundance \citep[e.g.,][]{Sanchez-Menguiano_2018}. Finally, we find similar trends of the characteristic abundance using the O3N2 calibrator than those derive using the calibrator derived by \citep{Ho_2019}. For late-type galaxies this abundance increases with \Mstar -- albait some variation among galaxies from this type: For a given stellar mass bin the oxygen abundance increases from Sd/Sm galaxies to Sb ones. For early-type galaxies the characteristic value of the oxygen abundance remains constant regardless the morphological type (E/S0 or Sa) and \Mstar (\mbox{$12+\log({\rm O/H}) \sim$ -8.5}). 

In Fig.~\ref{fig:OH_KK04_rad} we present the radial distribution of the oxygen abundance using the O3N2 abundance calibrator derived in \citet{KK_2004}, whereas in Fig.~\ref{fig:OH_KK04}, we show the result from the piece-wise analysis of these radial profiles. In general, we find similar results as those derived using the O3N2 calibrator.

\end{document}